\documentclass[pss]{wiley2sp} 
\usepackage{amsmath}
\usepackage{tabularx}

\usepackage{epstopdf}
\begin{document}

\title{Nanostructured clathrates and clathrate-based nanocomposites}

\titlerunning{Nanostructured clathrates ...}

\author{%
  R. Christian\textsuperscript{\textsf{\bfseries 1}},
  M. Ikeda\textsuperscript{\textsf{\bfseries 1}},
  G. Lientschnig\textsuperscript{\textsf{\bfseries 1,2}},
  L. Prochaska\textsuperscript{\textsf{\bfseries 1}},
  A. Prokofiev\textsuperscript{\textsf{\bfseries 1}},
  P. Tome\v{s}\textsuperscript{\textsf{\bfseries 1}},
  X. Yan\textsuperscript{\textsf{\bfseries 1}},
  A. Zolriasatein\textsuperscript{\textsf{\bfseries 1,3}},
  J. Bernardi\textsuperscript{\textsf{\bfseries 4}},
  T. Schachinger\textsuperscript{\textsf{\bfseries 4}},
  S. Schwarz\textsuperscript{\textsf{\bfseries 4}}, 
  A. Steiger-Thirsfeld\textsuperscript{\textsf{\bfseries 4}},
  P. Rogl\textsuperscript{\textsf{\bfseries 5}},
  S. Populoh\textsuperscript{\textsf{\bfseries 6}},
  A. Weidenkaff\textsuperscript{\textsf{\bfseries 6,7}}, and
  S. Paschen\textsuperscript{\Ast,\textsf{\bfseries 1}}}

\authorrunning{R.\ Christian et al.}

\mail{e-mail
  \textsf{paschen@ifp.tuwien.ac.at}, Phone:
  +43-1-58801-13716, Fax: +43-1-58801-13899}

\institute{%
  \textsuperscript{1}\,Institute of Solid State Physics, TU Wien, Wiedner Hauptstr. 8-10, 1040 Vienna, Austria\\
  \textsuperscript{2}\,Center for Micro- and Nanostructures, TU Wien, Floragasse 7, A-1040 Vienna, Austria\\
  \textsuperscript{3}\,Faculty of Materials Science and Engineering, K. N. Toosi University of Technology, Tehran, Iran\\
  \textsuperscript{4}\,Universit\"are Service-Einrichtung f\"ur Transmissions-Elektronenmikroskopie, TU Wien, Wiedner Hauptstr. 8-10, 1040 Vienna, Austria\\
  \textsuperscript{5}\,Institute of Physical Chemistry, University of Vienna, W\"ahringer Strasse 42, 1090 Vienna, Austria\\  
  \textsuperscript{6}\,Laboratory for Solid State Chemistry and Catalysis, EMPA,  \"Uberlandstrasse 129, 8600 D\"ubendorf, Switzerland\\
    \textsuperscript{7}\,Institut f\"ur Materialwissenschaft, Universit\"at Stuttgart, Heisenbergstr.\ 3, 70569 Stuttgart, Germany\\
}


\received{XXXX, revised XXXX, accepted XXXX} 
\published{XXXX} 

\keywords{clathrates, hot pressing, nanostructuring, melt spinning, thermoelectric and transport measurements, thermoelectric materials }

\abstract{%
%
%
%
\abstcol{%
Intermetallic clathrates are candidate materials for thermoelectric applications above room temperature. Here we explore whether their intrinsically low lattice thermal conductivities can be further reduced by nanostructuring and whether this can further enhance their thermoelectric performance.}{As bulk nanostructuring routes we studied melt spinning and ball milling. To optimize the compaction process and/or stabilize the nanostructure we varied the process parameters, used additives, and studied clathrate-based composites. Initial results on clathate nanowires as simpler model nanostructures are also presented.}}

%
%

\maketitle   

\section{Introduction}

Nanostructuring has in recent years been one of the most extensively explored
optimization strategies for thermoelectric materials. It has been shown to
enhance the thermoelectric efficiency of traditional thermoelectric materials
both for model systems \cite{Bou08.1,Hoc08.1} and for bulk materials
\cite{2008Jos,2008Ma,Pou08.1,Wan08.2,Min09.1}. On intermetallic clathrates
comparatively little has been done. These cage-like materials have intrinsically
low lattice thermal conductivities and thus the nanostructuring route, which
aims primarily at reducing the lattice thermal conductivity without degrading
the electronic properties to much, is less obvious to succeed. Here we present
such a study. We have investigated melt spinning as bottom-up and ball milling
as top-down technique for the production of nanostructured bulk clathrates. To
stabilize the nanostructures we have also explored clathrate-based
nanocomposites. Finally we present our nanowire measurement setup to study
model nanostructures.

\section{Melt spinning}\label{MS}

Melt spinning comprises injecting a metal melt onto the surface of a rapidly
rotating copper wheel, which results in very high cooling rates of the order
10$^5$ - 10$^6$\,K/s. This process was originally developed for the preparation
of amorphous materials and has been the main technique for the production of metallic glasses. Metal alloys with lower glass forming ability usually
form nanometer-sized polycrystalline materials. The advantage of the technique
for preparation of nanocrystalline materials is its simplicity and short process
duration resulting in a lower probability of oxidation. Moreover, clathrate
phases typically have broader composition/homogeneity ranges which allows to
optimize their thermoelectric properties in a broader range
\cite{Lau11.1,Lau12.1}. Melt spinning has shown to be an appropriate technique
for nanostructuring of some thermoelectric materials
\cite{Li08.2,Li09.1,Su12.1}. In particular, the technique was used to prepare
Bi$_2$Te$_3$ with nano-layered microstructure with layer thicknesses of 10 -
40\,nm \cite{Tan07}.

\begin{figure}
\includegraphics[width=\columnwidth]{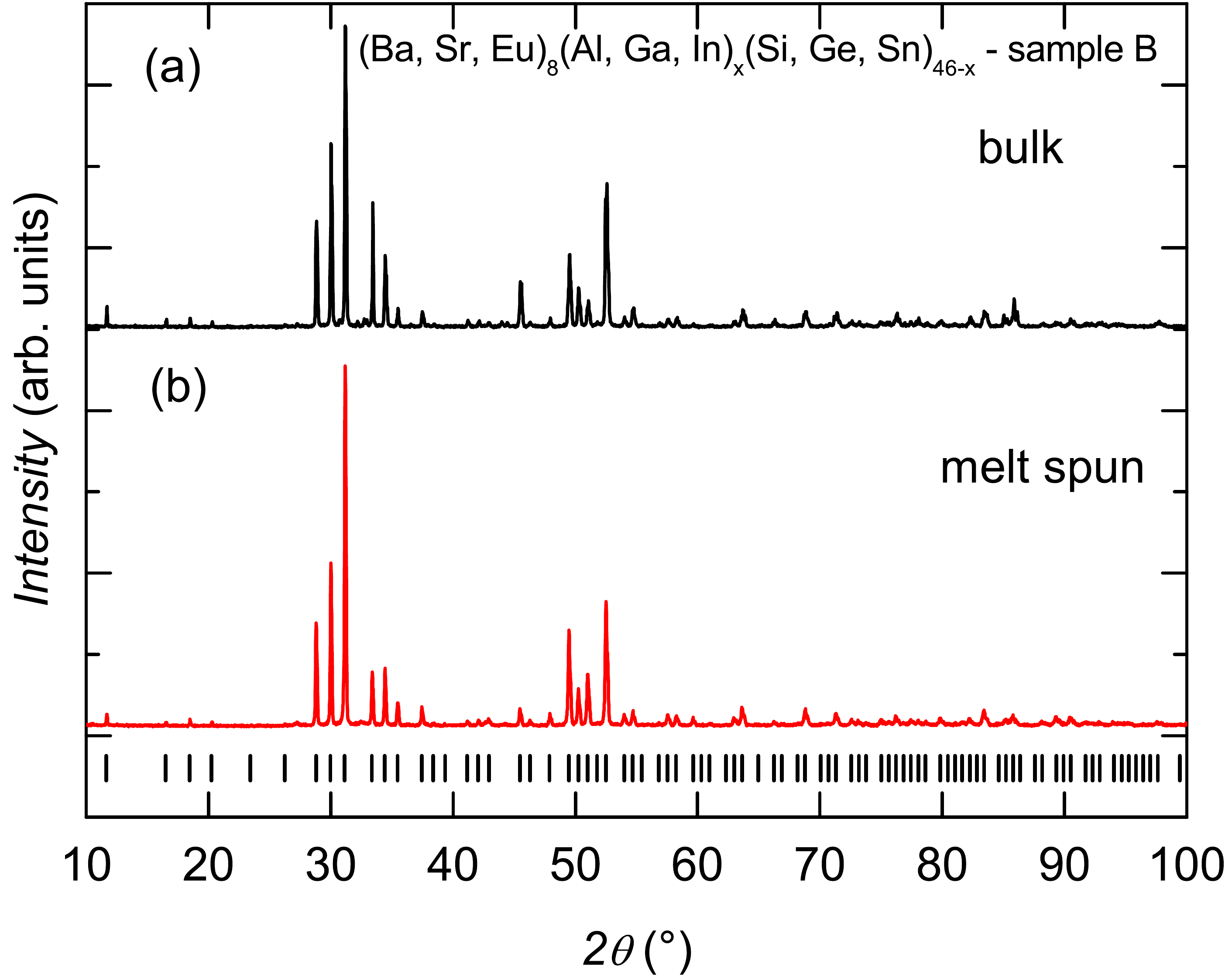}
\caption{XRD pattern of sample B before (a) and after (b) melt spinning. From \cite{Pro13}.}
\label{fig:diffractogram}
\end{figure}

However, our first experiments on the melt spinning of clathrates yielded
samples with grain sizes of at least 1\,$\mu$m \cite{Lau11.1,Lau12.1,Pro08}.
This might point to a very high crystallite growth rate of the clathrate phase.
For this reason, the formation of foreign phases appeared to be kinetically
suppressed. On the one hand this is a great advantage for producing bulk
clathrates since the typically needed long-term annealing to remove foreign
phases can be omitted. On the other hand this hampers the direct application of
melt spinning for the production of nanostructured clathrates. Thus, a profound 
investigation of the applicability of melt spinning to nanostructure clathrates
appeared necessary to understand the specifics of clathrates in rapid cooling
processes.

Clathrates have very high crystal as opposed to glass forming ability. Thus all
measures which usually lead to glass formation should reduce the clathrate grain
size. We describe two approaches. Firstly, we adjusted the parameters of the
melt spinning process. We varied the temperature of the melt (melt overheating)
and the cooling rate (Sect.\,\ref{app1}). Secondly, we studied the effect of the
melt composition on the grain size (Sect.\,\ref{app2}).

\subsection{Experimental details}\label{synth}

The polycrystalline starting materials were prepared by melting elementary
metals in a water cooled copper boat in argon atmosphere using high-frequency
heating (purity of metals better than 99.95\%). At least three re-meltings were
done for sample homogenization. For melt spinning the inductively heated melt
was sprayed from a fused silica nozzle onto the surface of a quickly rotating
(600 - 3000 revolutions per minute, rpm) copper wheel, with a diameter of
30\,cm. This method provided a cooling rate of about 10$^5$ - 10$^6$\,K/s. Both
the slow and rapid crystallizations were performed in argon atmosphere.

Scanning electron microscopy (SEM) and high-resolution transmission electron
microscopy (TEM) investigations were performed on Philips XL30 ESEM (with an
EDAX New XL-30 135-10 UTW+ detector) and FEI TECNAI F20 microscopes,
respectively. The data reduction of the spectra and quantitative analyses were
done by the EDX Control Software (from EDAX Inc.). X-ray powder diffraction
(XRD) data were collected using a Siemens D5000 diffractometer with
Cu-K$_{\alpha1,2}$ radiation. Electrical resistivity and Hall effect
measurements were done in a Physical Property Measurement System (PPMS) of
Quantum Design. The typical sample size for electrical resistivity measurements
was $3 \times 1\times 0.5$\,mm$^3$.

To measure the thermal conductivity, we used either thermal diffusivity
measurements (Flash technique) or the 3$\omega$ method. The latter is an ac
technique which heats the sample locally and thus reduces errors caused by
radiation at room temperature and below to a negligibly small level
\cite{Cah90.1}. A narrow metal line (20\,$\mu$m wide and 1\,mm long) serves as
both the heater and the thermometer. To avoid electrical contact between heater
and sample the polished sample surfaces were covered with a thin layer of
SiO$_2$ by chemical vapor deposition. The heater structures were made by
standard optical lithography techniques using a Karl Suess MJB4 mask aligner.
Sputtering of a 4\,nm thick titanium sticking layer and the 64\,nm gold film was
done in an Ardenne LS 320 S sputter system. The metal line was heated by an
oscillating current at a circular frequency $\omega$, which thus leads to a
2$\omega$ temperature oscillation of both the heater and the sample. Due to the
linear temperature dependence of the metallic heater, the 2$\omega$ temperature
oscillation translates to a 3$\omega$ voltage oscillation, which is detected
using a lock-in amplifier \mbox{(7265, Signal Recovery)}. Prior to the thermal
3$\omega$ voltage detection, the first harmonic and all related higher harmonics
are subtracted from the signal using an active filter design based on the
technique suggested by Cahill et al. \cite{Cah87.1,Cah90.1}. More details on our
3$\omega$ measurements are given in Ref.\,\cite{Ike15.1}.

The phonon contribution $\kappa_{ph}$ was evaluated by subtracting the electronic part $\kappa_{el}$ from the total thermal conductivity $\kappa$, using the Wiedemann-Franz law $\kappa_{el}=L_{0}T/\rho$ with $L_{0}$ = 2.44 x 10$^{-8}$ V$^{2}$K$^{-2}$. 

\subsection{Process parameters for melt spinning of clathrates}\label{app1}

The transition metal clathrate phase Ba$_8$Au$_x$Si$_{46-x}$ was chosen for the
melt spinning experiments because of its rather wide homogeneity range (about $4
< x < 6$) \cite{Ayd11.1,Zei12.1}, which lowers the risk for foreign phases
formation. Also, as a rule, a difference in atomic size $\geq 10$\,\%
strengthens the glass forming ability of metallic phases. The atomic size
difference of Au and Si is just 10\,\%.

\begin{figure}
\subfloat{\label{fig:9elements}\includegraphics[width=\columnwidth]{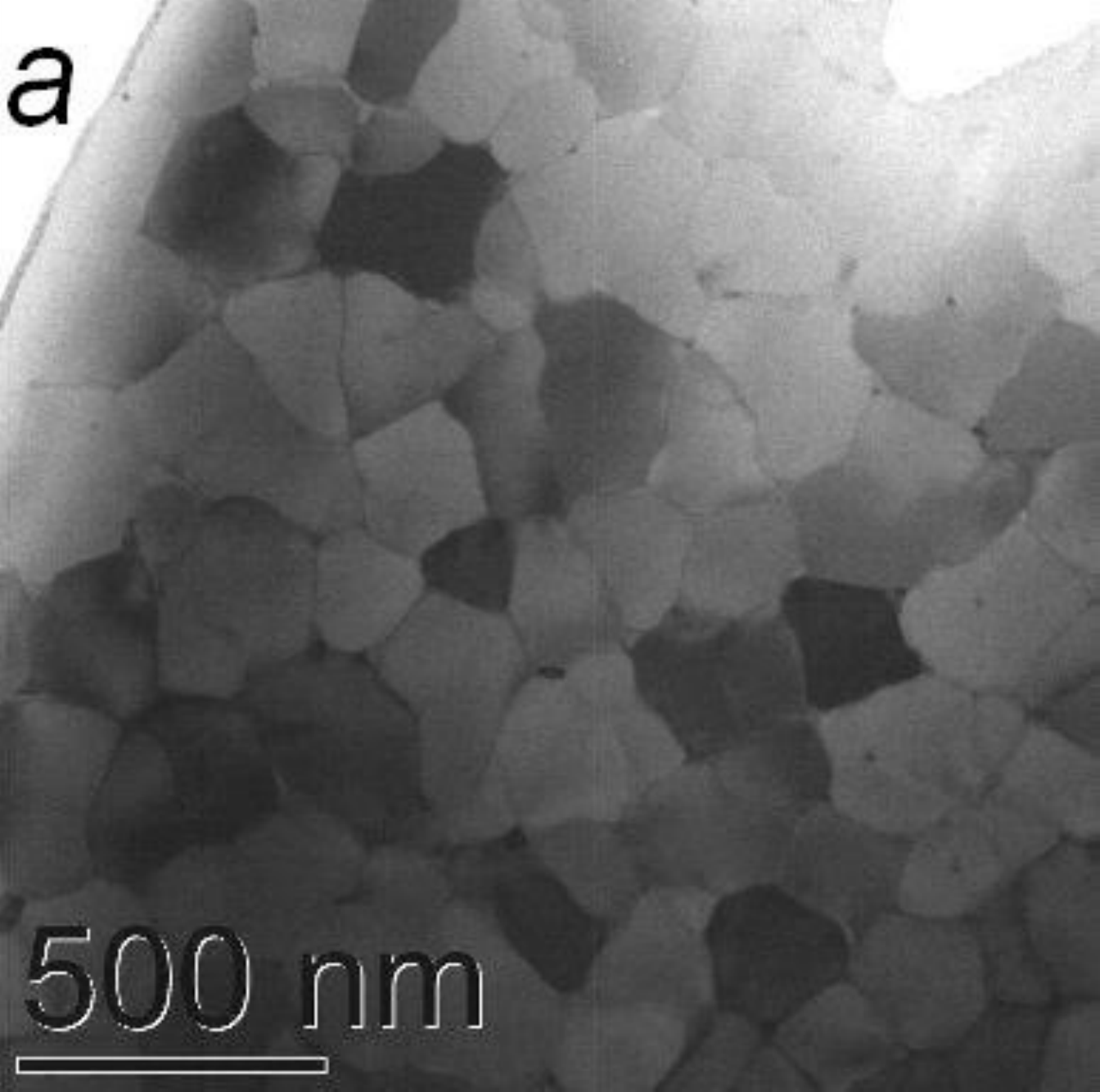}}\hfill   \subfloat{\label{fig:3elements}\includegraphics[width=\columnwidth]{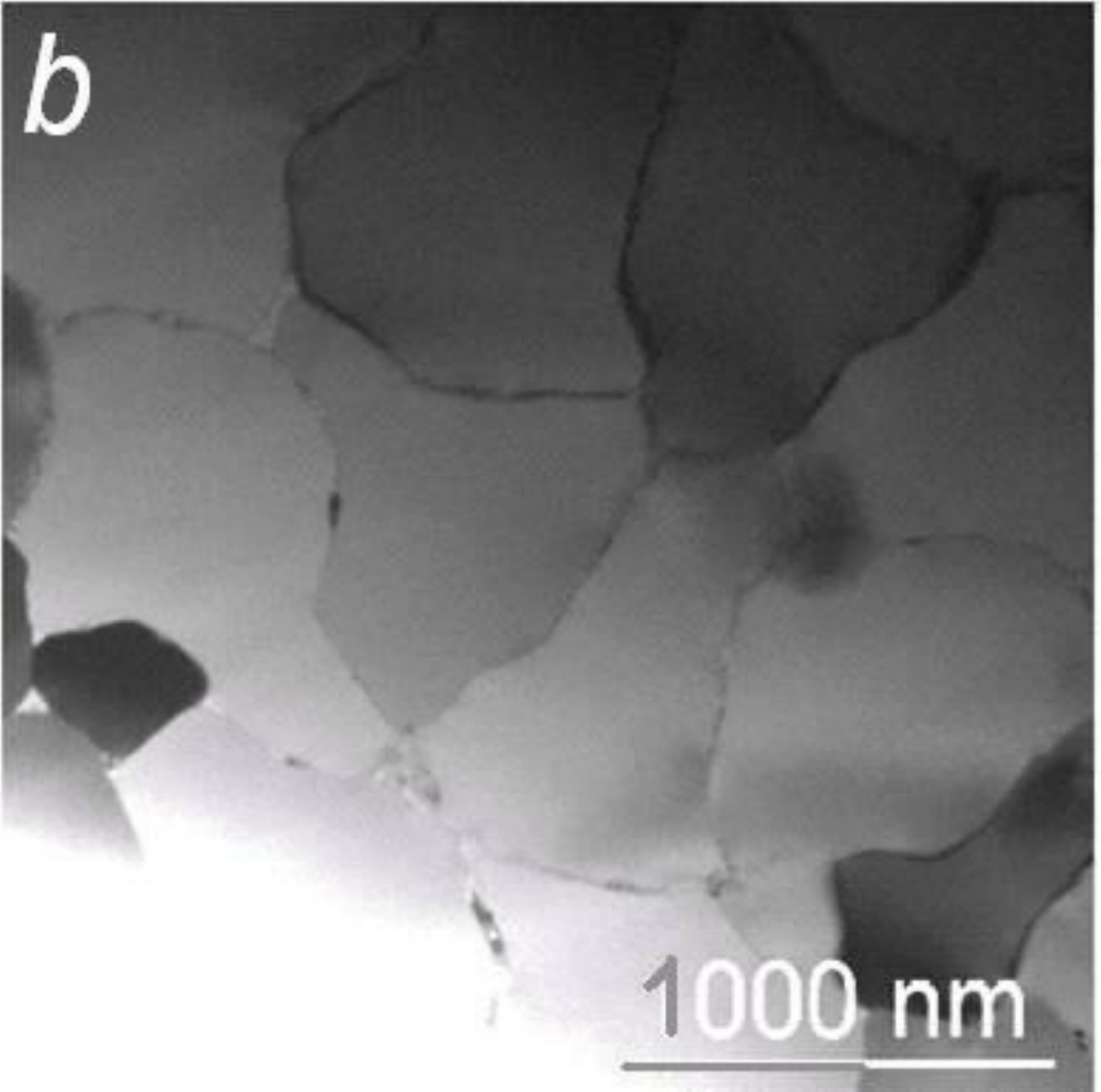}}
     \caption{TEM images of the meltspun nine element sample B (a) and of Ba$_8$Ga$_{16}$Ge$_{30}$ for comparison (b). From \cite{Pro13}.}
  \label{fig:TEM2}
\end{figure}

We used $x = 5$ as starting Au content. It is interesting to note that it was a
Au-Si alloy that was first obtained as metal glass by rapid quenching
\cite{Kle60}. Three different melt spinning experiments were performed (Table
\ref{tab:table1}). Samples 1 and 2 were quenched with a wheel rotation speed of
1500\,rpm from two different melt temperatures, 1300$^{\circ}$C (just above the
melting point of 1260$^{\circ}$C) and 1600$^{\circ}$C, respectively. Sample 3
was quenched at a higher rotation speed of 2500\,rpm from 1300$^{\circ}$C.
Usually melt spinning is carried out in inert gas atmosphere. However, very fast
wheel rotation leads to a gas turbulence near the wheel surface, which destroys
the melt jet and leads to the formation of small droplets before the material
reaches the cooling surface. Therefore sample 3 was quenched in vacuum.

\begin{table}[ht!]
\caption{\label{tab:table1} Melt spinning parameters for Ba$_8$Au$_5$Si$_{41}$.}
\begin{tabular}{cccc}
\hline \hline
   & Quenching   & Wheel rotation  & Atmosphere\\
      & temperature ($^{\circ}$C) &  speed (rpm) &  \\
      \hline
  sample 1 & 1300 & 1500 & 300\,mbar Ar\\
  sample 2 & 1600 & 1500 & 300\,mbar Ar\\
  sample 3 & 1300 & 2500 & Vacuum \\
  \hline
  \hline
\end{tabular}
\end{table}

The SEM investigation of the obtained samples revealed micrometer or
submicrometer sized clathrate grains. Neither higher rotating (cooling) rate
(2500 vs 1500\,rpm) nor higher starting melt temperature (1600 vs
1300$^{\circ}$C) led to a reduction of the grain size. A detailed microscopy
investigation of the obtained materials is described in Sect.\,\ref{app3}.

\subsection{Composition effect in meltspun Ba$_8$Ga$_{16}$Ge$_{30}$ clathrate}\label{app2}

A well-known strategy to promote glass formation is to increase the number of
constituting elements, preferably of considerably different atomic size
\cite{Gre95.1}. Here we used the clathrate Ba$_8$Ga$_{16}$Ge$_{30}$ as starting
material and partially substitute all three elements (Ba, Ga, Ge) with isovalent
elements to obtain a clathrate phase (Ba, Sr, Eu)$_8$(Ga, Al, In)$_x$(Ge, Si,
Sn)$_{46-x}$. The In and Sn substitution would be especially important because
of the much larger atomic radii (1.66 and 1.62$\AA$, respectively) compared to
the ones of Ga, Al, Ge, and Si (atomic radii of 1.32 - 1.43\,\AA). Contrary to
the case of an amorphous final product the element ratio in the quenched
clathrate melt cannot be varied arbitrarily. The melt composition for quenching
should correspond to a stable crystalline clathrate phase as otherwise foreign
phases would form during crystallization.

In the search for a stable clathrate phase the nominal composition
(Ba$_{1/3}$Sr$_{1/3}$Eu$_{1/3}$)$_8$(Al$_6$Ga$_6$In$_4$)(Si$_5$Ge$_{15}$Sn$_{10}$)
was prepared by melting the elements followed by fast cooling in a cold boat.
The sample was a mixture of a number of phases, whose individual compositions
were investigated by SEM/EDX. Among them two phases had element ratios close to
the prototypal M$^{\rm{II}}_8$A$^{\rm{III}}_{16}$B$^{\rm{IV}}_{30}$ phase. XRD
confirmed the presence of at least two clathrate phases. According to their
measured compositions
(Ba$_6$Sr$_{1.4}$Eu$_{0.9}$)(Ga$_{6.6}$Al$_{7.7}$In$_{0.3}$)(Ge$_{16.8}$Si$_{13.5}$Sn$_{0.5}$)
and
(Ba$_{5.3}$Sr$_{2}$Eu$_{1.0}$)(Ga$_{9.9}$Al$_{4.8}$In$_{0.3}$)(Ge$_{25}$Si$_{4.9}$Sn$_{0.5}$)
the phases were prepared in pure form (hereafter referred to as sample A and
sample B, respectively). From the small contents (0.6 - 1.0\,at.\% In and Sn) we
see that the large sized atoms In and Sn are reluctant to enter the clathrate
structure.

Both samples were meltspun with a wheel rotation speed of 1500\,rpm. In sample
A a small amount of foreign phase was detected. TEM investigations show that
they form thin intergrain layers. Sample B was phase pure according to XRD
(Fig.~\ref{fig:diffractogram}). TEM reveals only a tiny amount of foreign phases
which are concentrated at the junctions of three grains, the boundaries between
two grains being free of any intergrain phase. This is why we focussed our
further investigations on sample B. Most importantly, the meltspun sample B has
a much smaller grain size (about 200\,nm) than the meltspun parent phase
Ba$_8$Ga$_{16}$Ge$_{30}$ (about 1000\,nm), as shown by TEM images
(Fig.~\ref{fig:TEM2}) \cite{Pro13}.

\begin{figure}
\subfloat{\label{fig:rho}\includegraphics[width=\columnwidth]{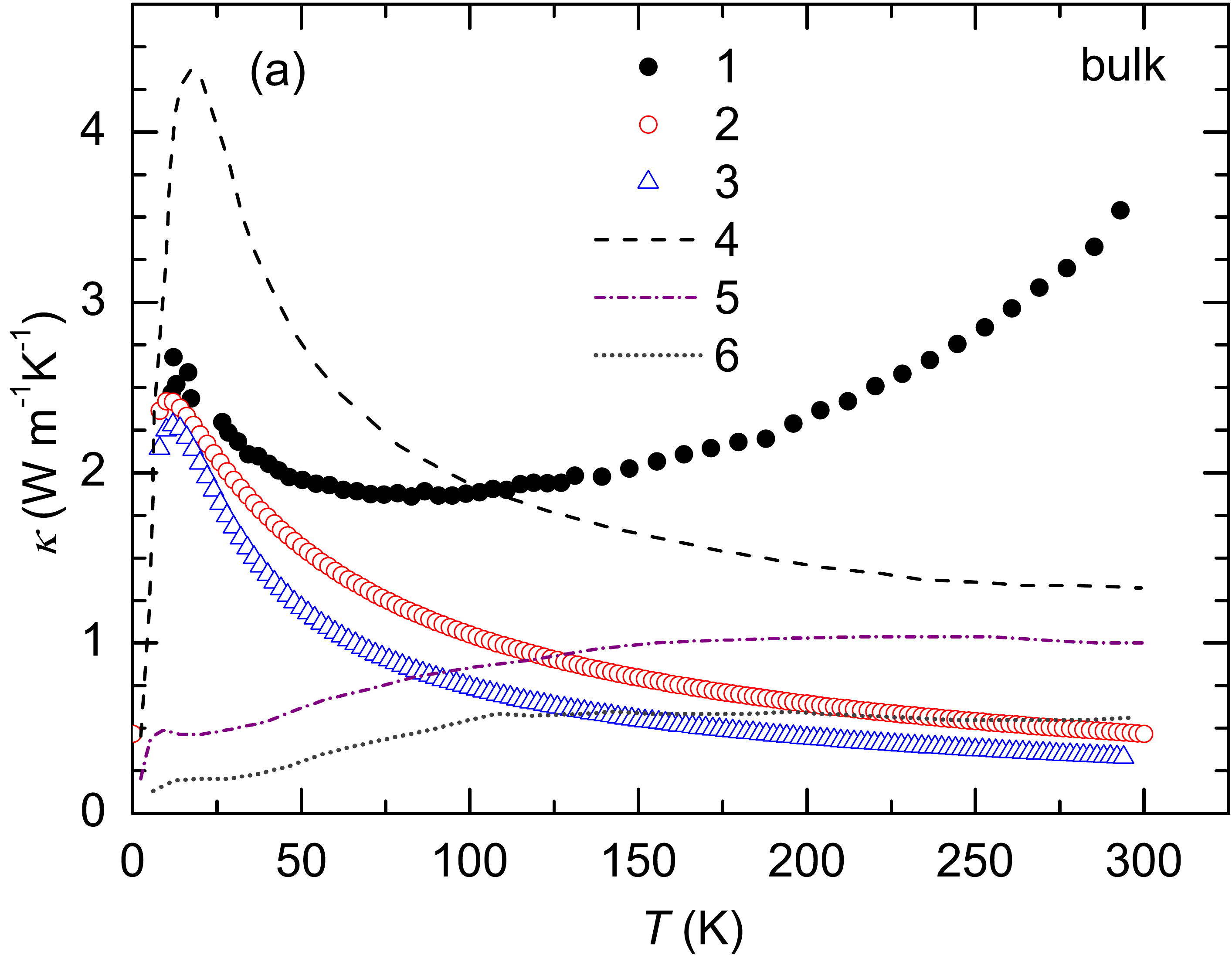}}\hfill  \subfloat{\label{fig:powerfactor}\includegraphics[width=\columnwidth]{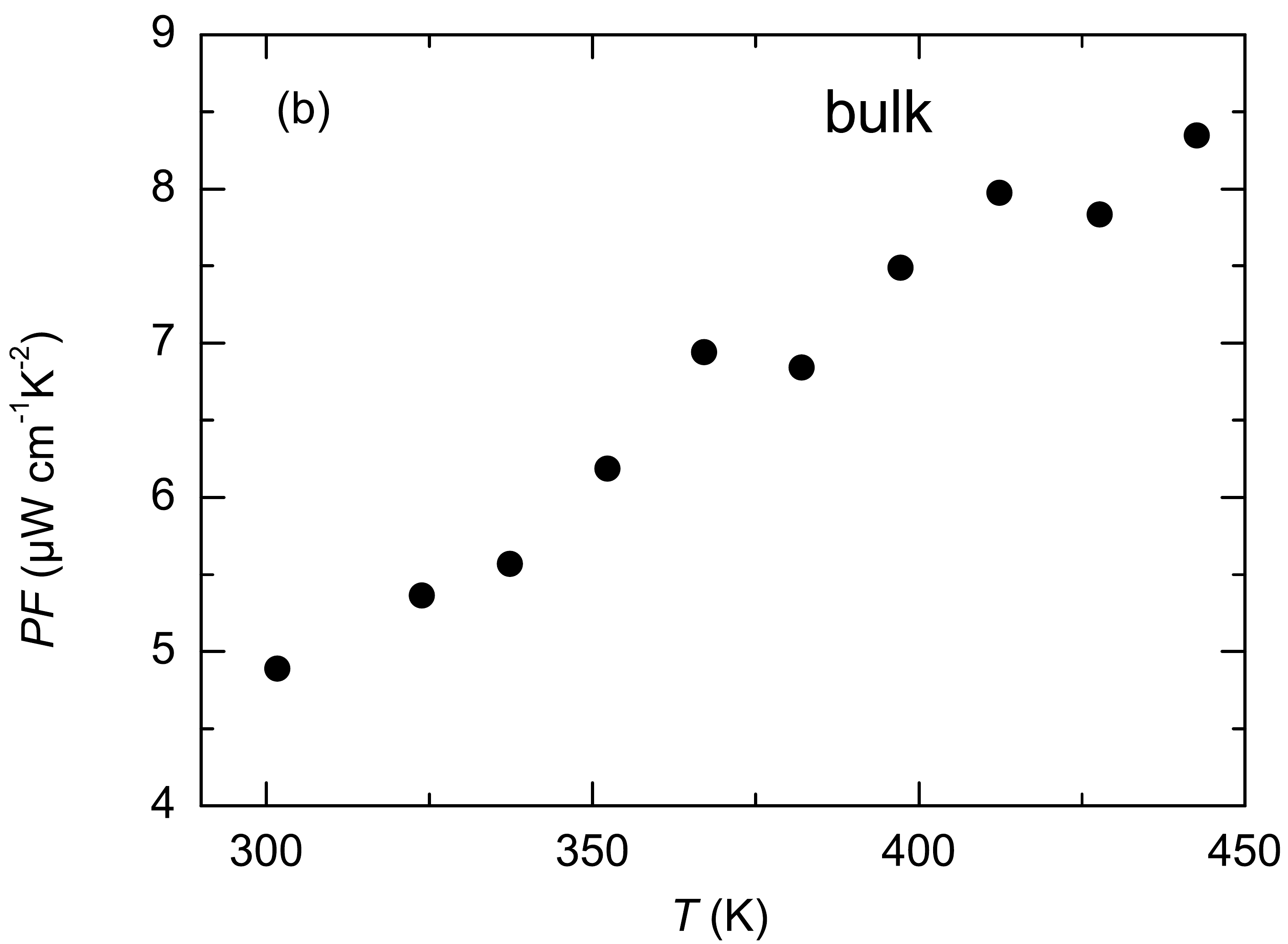}}\hfill
\subfloat{\label{fig:kappa}\includegraphics[width=\columnwidth]{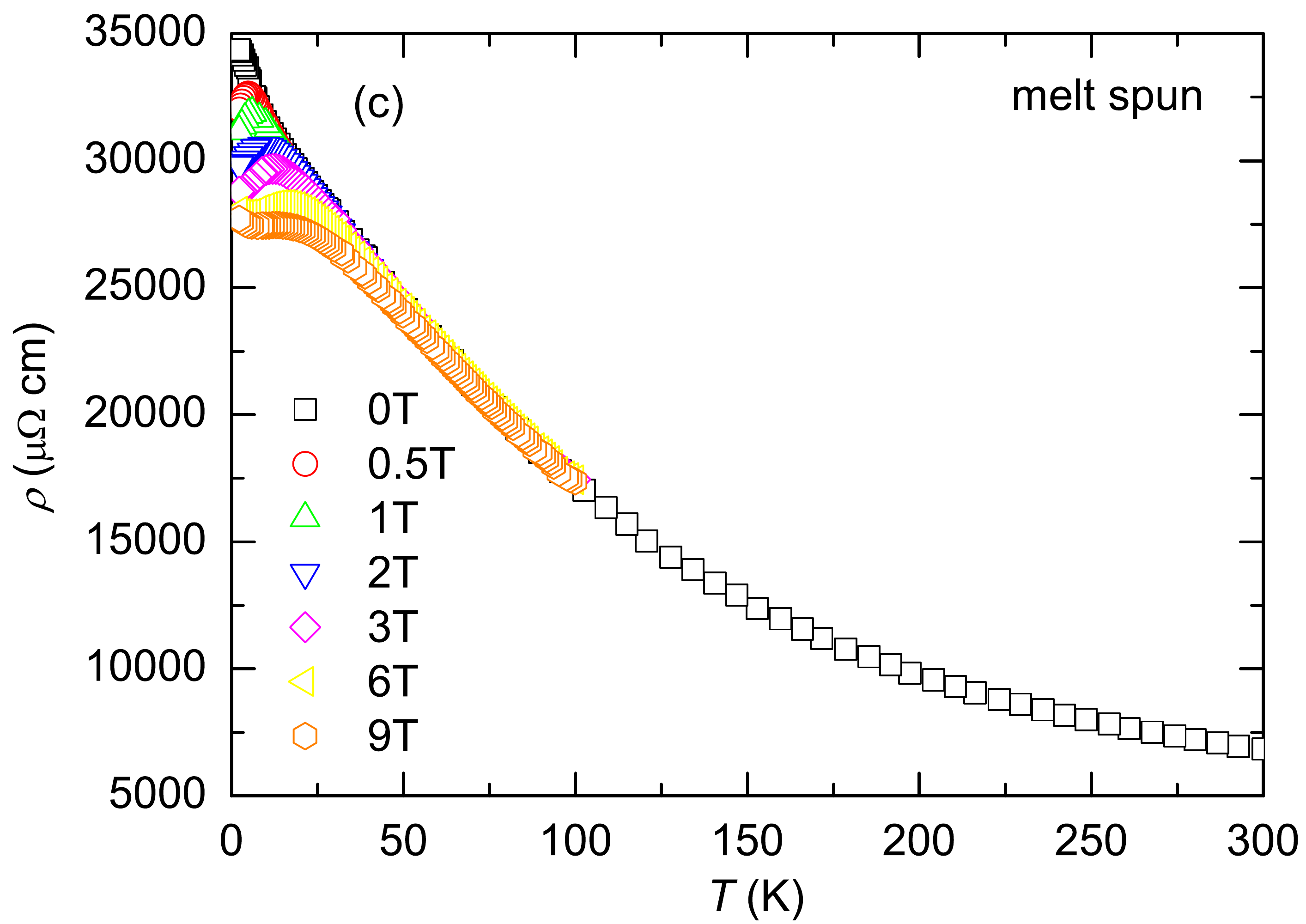}}  \caption{Thermal conductivity of the bulk clathrate alloy phase (Ba, Sr, Eu)$_8$(Ga, Al, In)$_x$(Ge, Si, Sn)$_{46-x}$ before melt spinning (a): raw data (1), lattice thermal conductivity after subtraction of the radiation and electron contributions according to the WF-C (2) and the CW (3) analysis (see Section II); the lattice thermal conductivities of Ba$_8$Ga$_{16}$Ge$_{30}$ (4), Sr$_8$Ga$_{16}$Ge$_{30}$ (5), and Eu$_8$Ga$_{16}$Ge$_{30}$ (6) \cite{Sal01.1} are shown for comparison. Power factor of the bulk sample B (b). Electrical resistivity of the meltspun sample B (c). From \cite{Pro13}.}
\label{fig:prop}
\end{figure}

In the obtained 9-element clathrate, three ``phonon engineering'' approaches are
believed to be combined: (i) the intrinsically low thermal conductivity of
clathrates \cite{Nol98.1,Coh99.1}, (ii) the alloying effect, and (iii) the grain
size reduction. Unfortunately we could not yet perform reliable thermal
conductivity measurements on the rather brittle meltspun 9-element clathrate
flakes to evidence the effect of this triple-phonon engineering approach. Here
we present only the thermal conductivity of the bulk nine-element clathrate
phase prior to melt spinning. Thus only the alloying effect on the lattice
thermal conductivity is demonstrated (Fig.\,\ref{fig:prop}, a). The lattice
thermal conductivity obtained after substraction of the electronic contribution
and a radiation term is extremely low. For comparison, the lattice thermal
conductivities of Ba$_8$Ga$_{16}$Ge$_{30}$, Sr$_8$Ga$_{16}$Ge$_{30}$, and
Eu$_8$Ga$_{16}$Ge$_{30}$ adopted from \cite{Sal01.1} are shown. In
Fig.\,\ref{fig:prop}, b the power factor of the bulk sample is presented in the
temperature range 300 - 440\,K. The room temperature $ZT$ of sample B amounts to
about 0.07.

By contrast, we did succeed to measure the electrical resistivity $\rho(T)$ of
the meltspun 9-element clathrate sample B. It shows semiconducting behavior
(Fig.\,\ref{fig:prop}, c). Below 30\,K, $\rho$ decreases with increasing
magnetic field, presumably due to a reduction of electron scattering from
Eu$^{2+}$ magnetic moments which order ferromagnetically at about 20\,K, as
evidenced by measurements of the magnetic susceptibility (not shown).

\subsection{Electron microscopy investigation of meltspun
clathrates}\label{app3}

As described above, our efforts to reduce the grain size of clathrates by melt
spinning were only moderately successful. To understand this unusual
crystallization behavior of clathrates we performed a more detailed
investigation. We studied the rapid crystallization of clathrates with the
starting compositions Ba$_8$Au$_{5}$Si$_{41}$ and Ba$_8$Ni$_{3.5}$Si$_{42.5}$.
The meltspun samples were brittle thin ribbons of the typical size of $20 \times
4 \times 0.02$\,mm$^3$ (flakes). The microstructure of the polished flake
surfaces and of the flake cross-sections was investigated by SEM and TEM.

\begin{figure}
\includegraphics[width=\columnwidth]{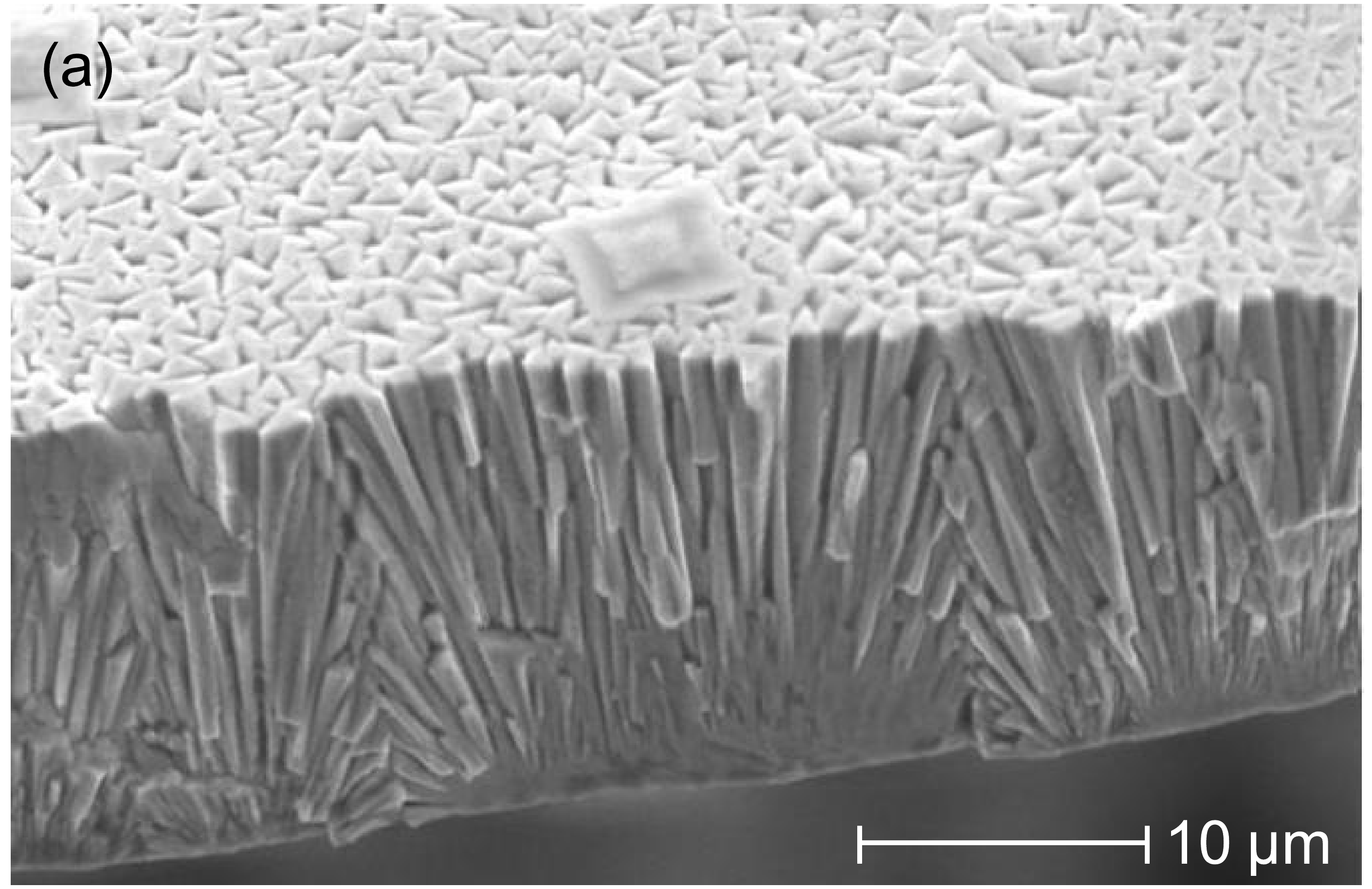}
\includegraphics[width=\columnwidth]{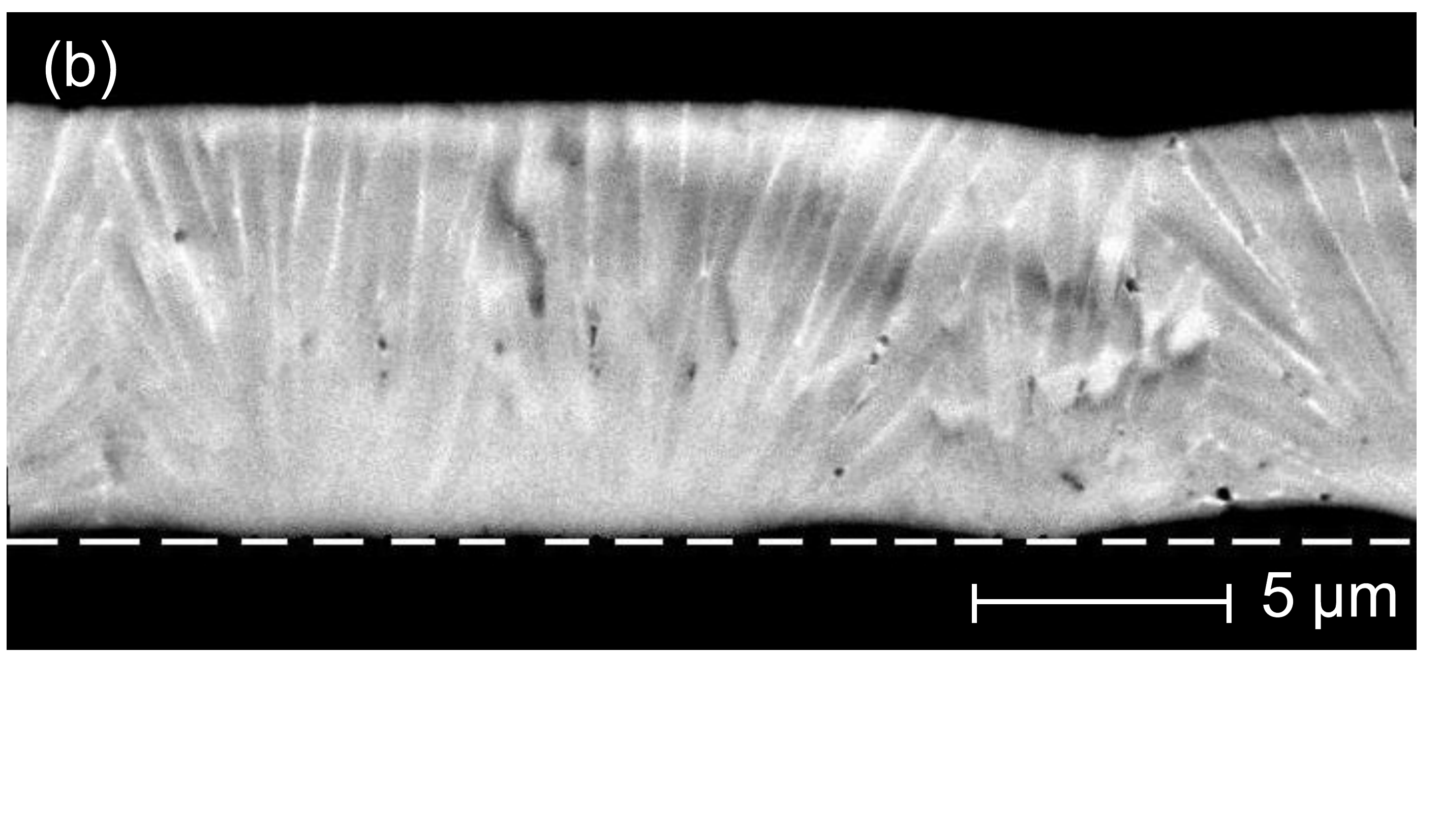}
\includegraphics[width=\columnwidth]{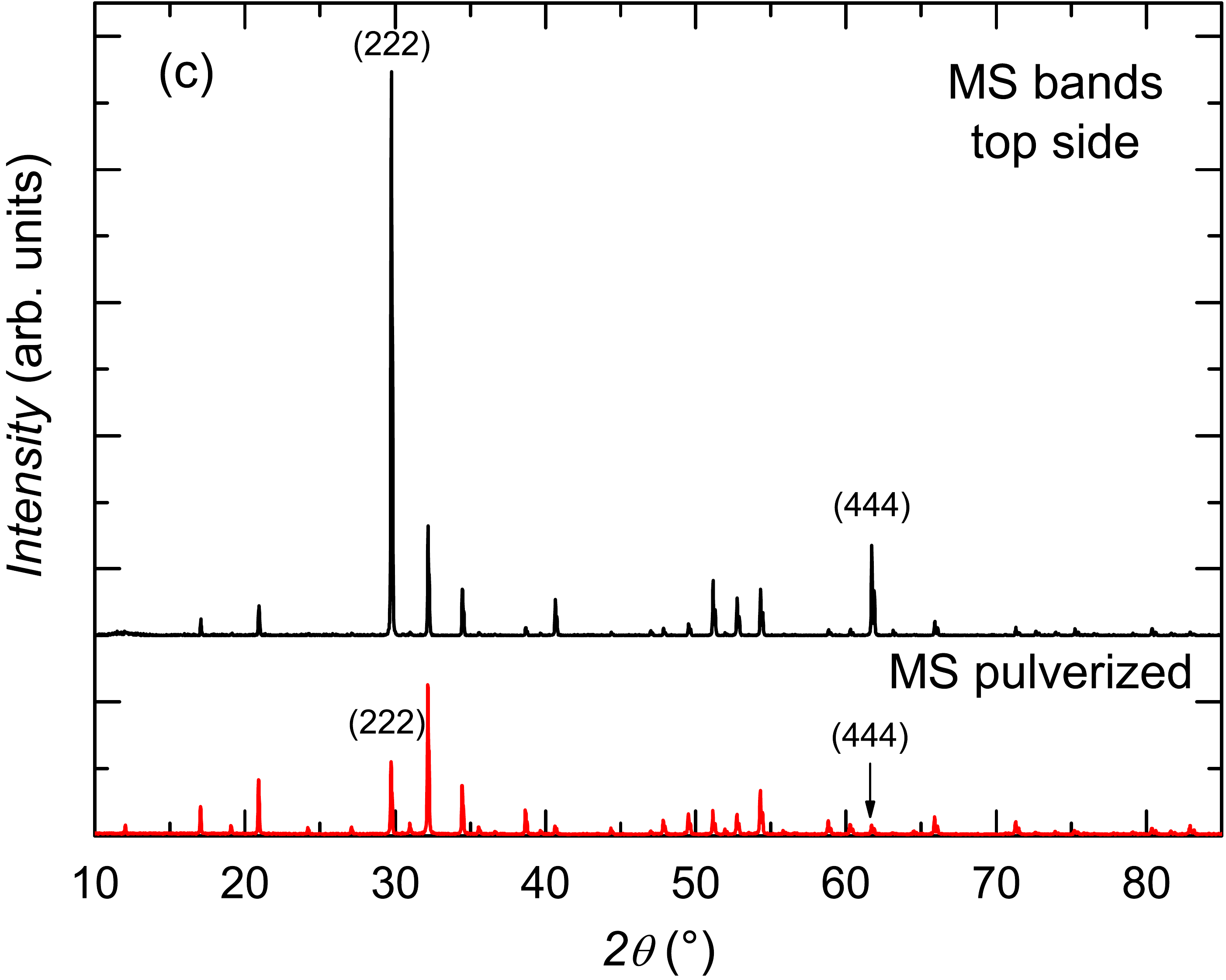}
\caption{SEM images of Ba$_8$Au$_{5.0}$Si$_{41}$ flake cross-sections: broken surface (a), polished surface (b). The dashed line shows the contact surface with the copper wheel. XRD of meltspun Ba$_8$Au$_{5.0}$Si$_{41}$ (c). Anomalously high (222) and (444) reflexions of the intact as-cast meltspun bands in comparison with those of the pulverized bands indicate that the fastest growth direction is [111]. This feature is also observed for the Ba$_8$Ni$_{3.5}$Si$_{42.5}$ flakes. From \cite{Pro14.1}.}
\label{fig:SEM}
\end{figure}

\begin{figure}
\includegraphics[width=\columnwidth]{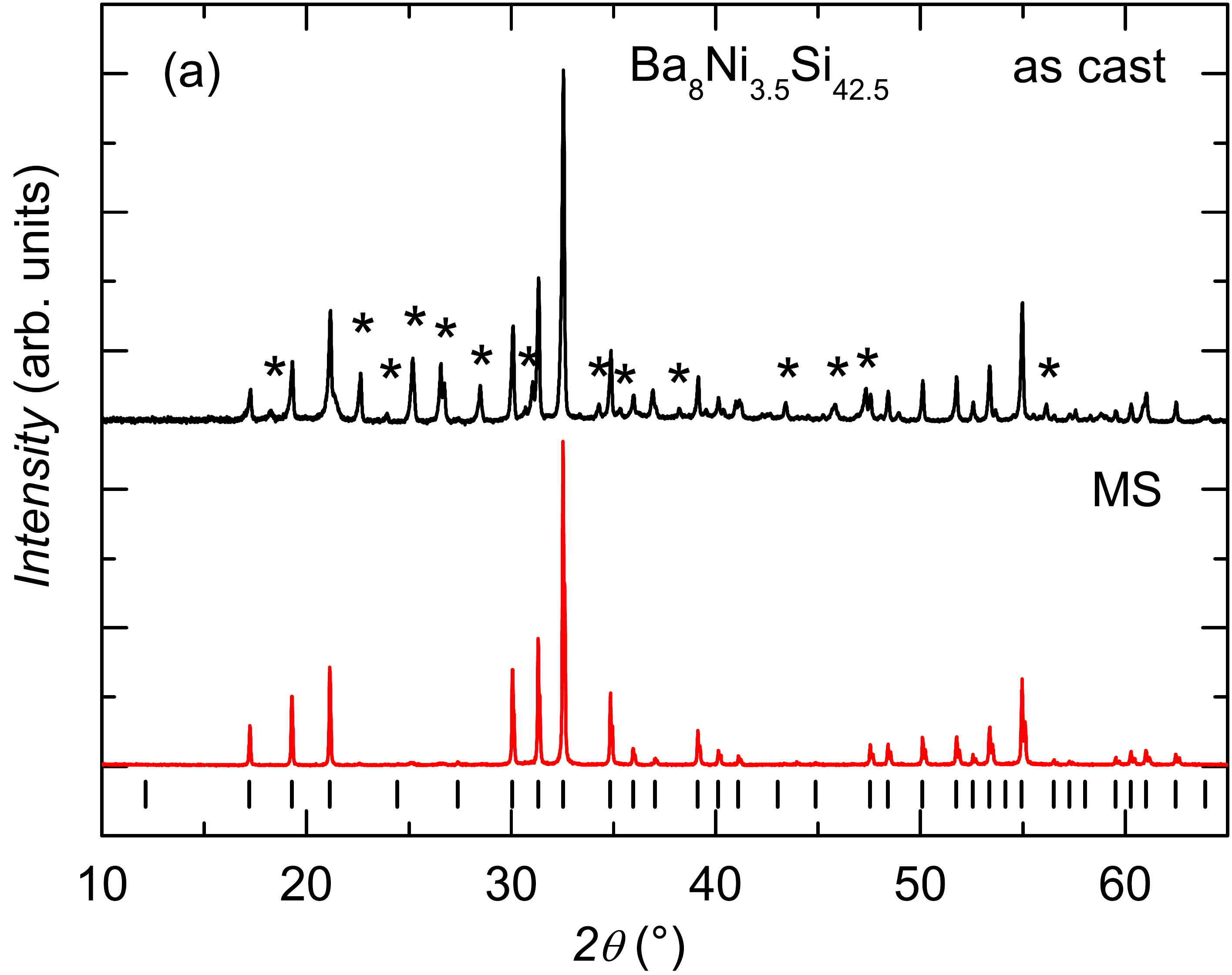}
\includegraphics[width=\columnwidth]{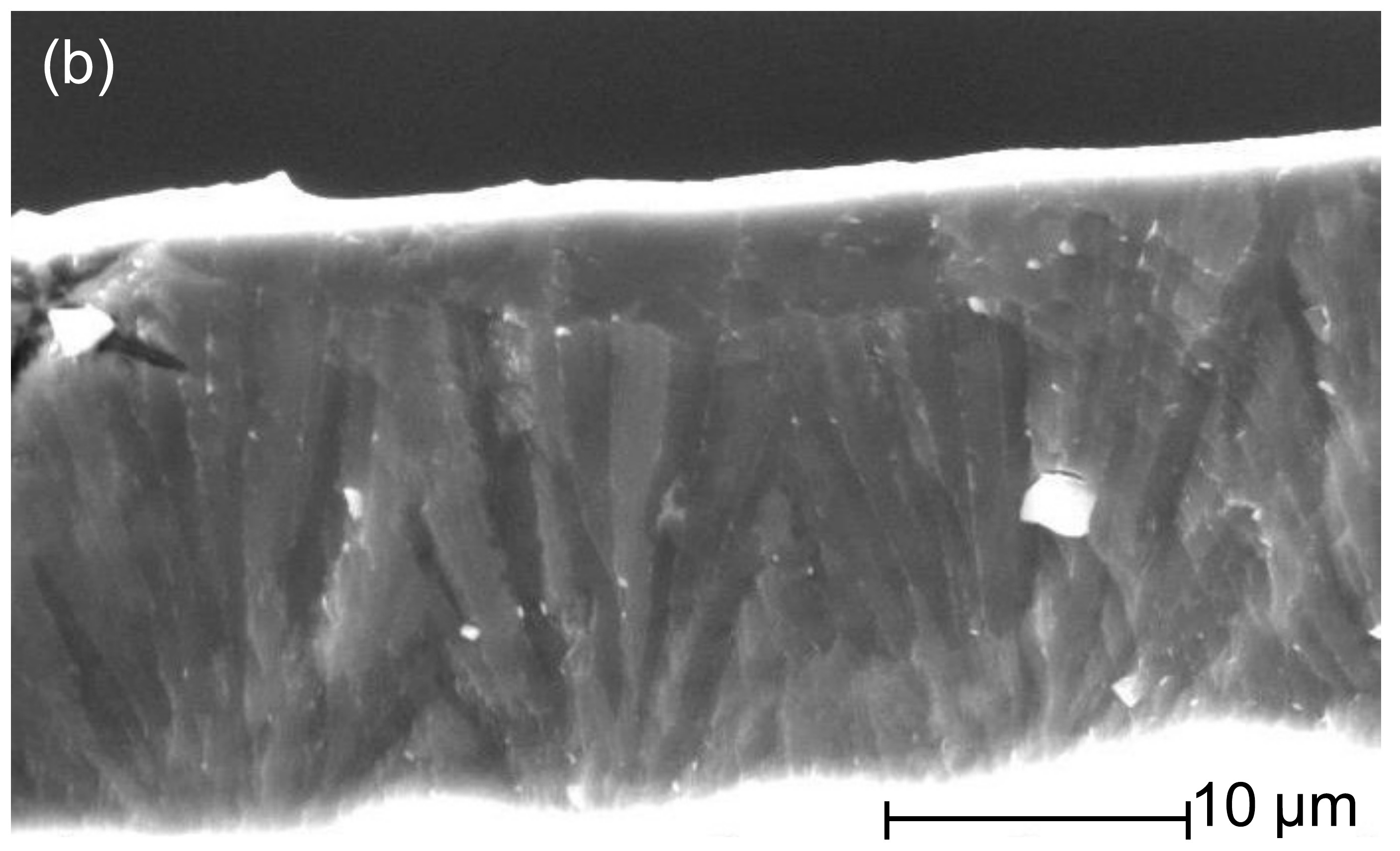}
\includegraphics[width=\columnwidth]{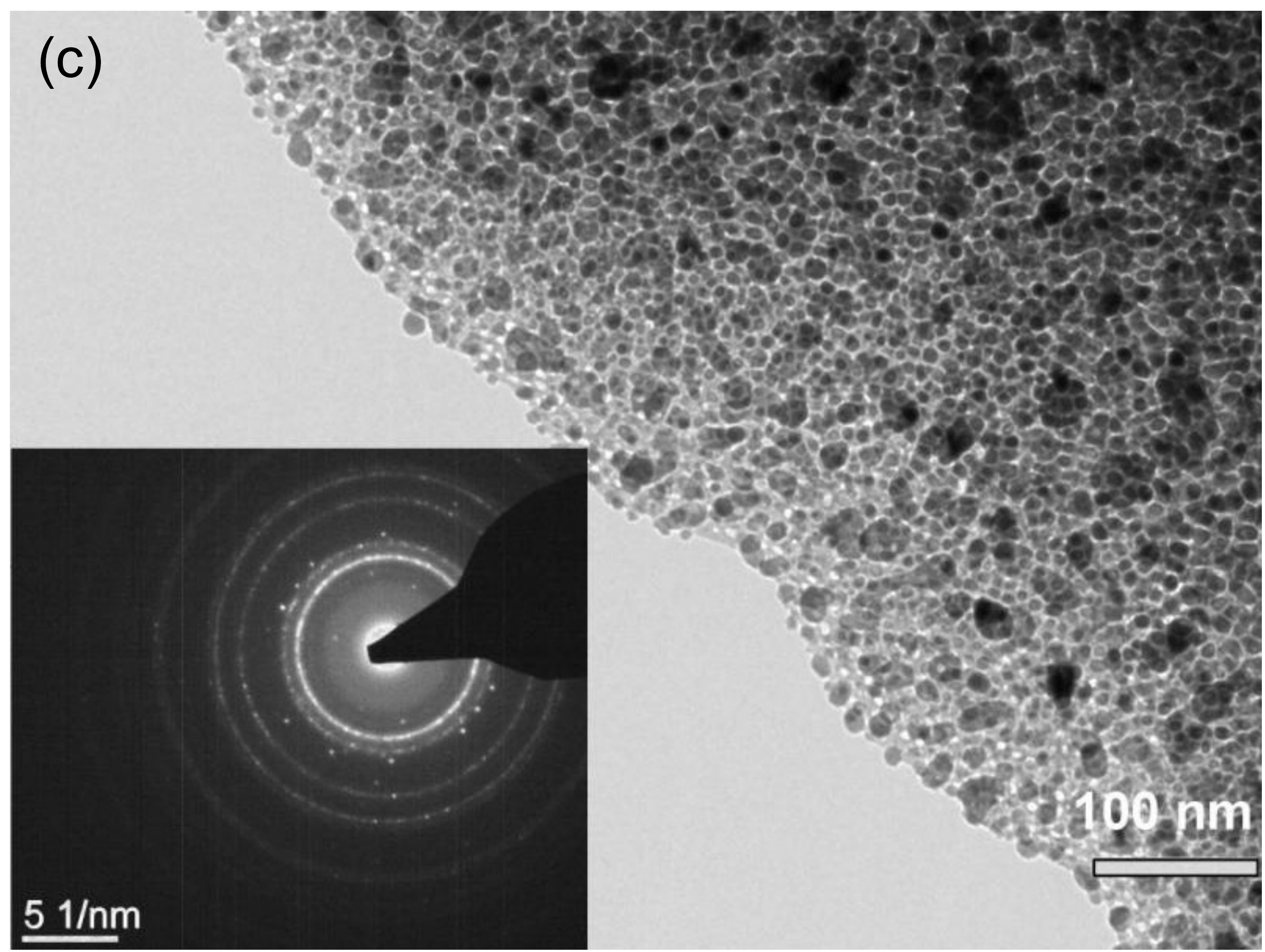}
\caption{XRD patterns of as-cast and meltspun Ba$_8$Ni$_{3.5}$Si$_{42.5}$ (a). The foreign phases (mainly BaSi$_2$) are marked by stars. SEM image of the broken cross-section of Ba$_8$Ni$_{3.5}$Si$_{42.5}$ revealing micrometer-sized crystals (b). TEM image of the thin contact layer of Ba$_8$Ni$_{3.5}$Si$_{42.5}$ (c). Inset: Selected area electron diffraction (SAED) image showing the crystallinity of the material; the structure agrees with the clathrate structure. From \cite{Pro14.1}.}
\label{fig:TEM3}
\end{figure}

Inspecting flake cross-sections helps us to understand much of the time-resolved
growth process. Both as-broken and polished surfaces reveal columnar crystallite
growth (Fig.\,\ref{fig:SEM}\,a,\,b and \ref{fig:TEM3}\,b). The preferential
crystal growth direction is [111], as seen from the comparison of the XRD
patterns of the intact as-spun and the pulverized flakes
(Fig.~\ref{fig:SEM}\,c). Despite the overall micro-sized grain picture of the
meltspun samples, the very bottom (contact surface) part contains a thin
diffuse-looking layer of material which is featureless in SEM. A high-resolution
TEM investigation of this layer in meltspun Ba$_8$Ni$_{3.5}$Si$_{42.5}$
supported its crystallinity and nanogranular structure with grain sizes of 10 -
40\,nm (Fig.\,\ref{fig:TEM3}\,c). These features of the microstructure shed
light on one of the reasons for the coarse microstructure of the meltspun
clathrate. Crystalline clathrates have an extremely low thermal conductivity. As
soon as a thin layer near the cooling surface crystallizes it impedes further
heat flow, and the effective cooling rate drops for the remaining layer of melt.

However, one more reason may underlie the formation of large grains
\cite{Pro14.1}. In complex polar liquids such as water solutions of ionic
compounds the formation of pentagonal dodecahedron clusters was revealed
\cite{Sob99}. Despite the very different chemical nature of water solutions and
``clathrate melts'' their structural similarities are obvious -- both water
molecules and group-IV elements, which dominate the melt, have a strong tendency
to form tetrahedrally bonded networks. This similarity is supported by the
existence of solid water clathrates, e.g.\ the methane hydrate
(CH$_4$)$_8$(H$_2$O)$_{46}$. In water solutions of ionic compounds such clusters
are formed around the positive ion (``ionic clathrates'') \cite{Sob99}. The
small cage of a type-I clathrate structure is a pentagonal dodecahedron too.
Thus a fragment of the clathrate structure is already present intrinsically in
the clathrate melt if we assume its similarity with water. This may be the
reason why the clathrate phase is preselected in a rapid solidification process
among all other competing phases with crystallographic types of local symmetry.
Since the pentagonal dodecahedron which is responsible for this process is also
present in clathrates of types I to III and IX \cite{Rog05.1} similar behavior
is expected also for these clathrates. It is the early nucleation that possibly
also leads to large grains: if the nucleation was strongly retarded to a lower
temperature when the crystal growth was already suppressed due to a low
diffusion rate, nanoscale grains would form \cite{Pro14.1}. 

\subsection{Melt spinning of
Ba$_{8}$Cu$_{5}$Si$_{6}$Ge$_{35-x}$Sn$_{x}$}\label{app4}

Even though melt spinning does not readily produce nanostructured clathrates, it
is advantageous to produce clathrate phases with broader composition ranges than
what is possible by conventional synthesis techniques \cite{Lau11.1,Lau12.1}.
The main motivation of this study was to increase the solubility limit of Sn in
Ba$_{8}$Cu$_{5}$Si$_{6}$Ge$_{35-x}$Sn$_{x}$ clathrates to improve their
thermoelectric properties. We expected a partial Ge substitution by heavier Sn
to lower the thermal conductivity, which was indeed realized. In addition, the
change of the electronic structure shifted the maximum of $ZT$ to lower
temperatures \cite{Yan13.2}.

\begin{figure}
\includegraphics[width=\columnwidth]{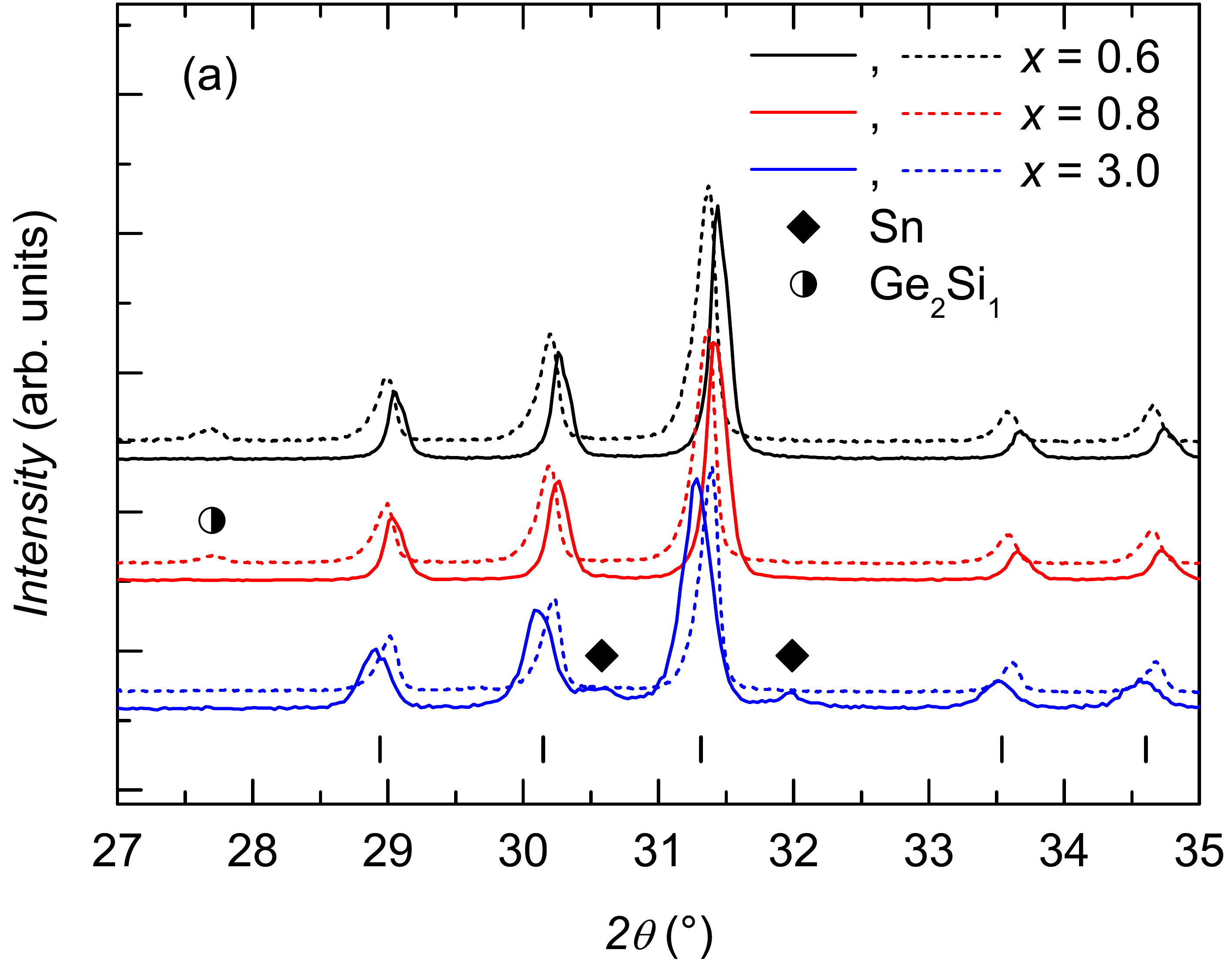}
\includegraphics[width=\columnwidth]{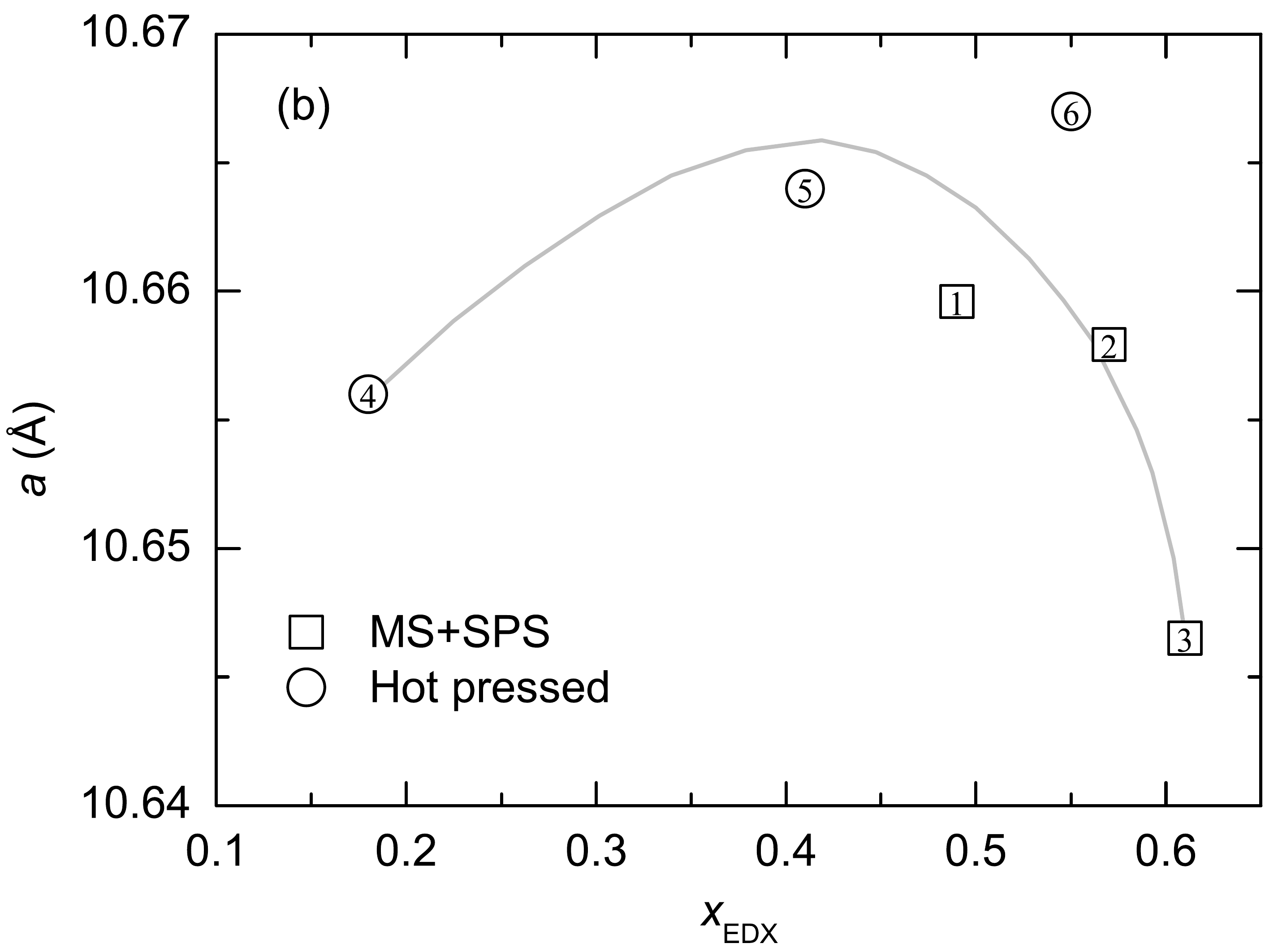}
\caption{XRD patterns, normalized to the highest peak of the clathrate
phase, of Ba$_{8}$Cu$_{5}$Si$_{6}$Ge$_{35-x}$Sn$_{x}$ ($x$ = 0.6 - 3.0) meltspun
(MS) and MS+SPS samples (a). Lattice parameters of the clathrate phase vs Sn
content derived from EDX, $x_{\rm{EDX}}$ (b). Data for hot pressed samples are
added from \cite{Yan13.2}. The line is a guide to the eye and the numbers are
the sample codes from Table~\ref{tab:table1_Tomes}.}
\label{Graph1_Tomes}
\end{figure}

\begin{figure}
\includegraphics[width=\columnwidth]{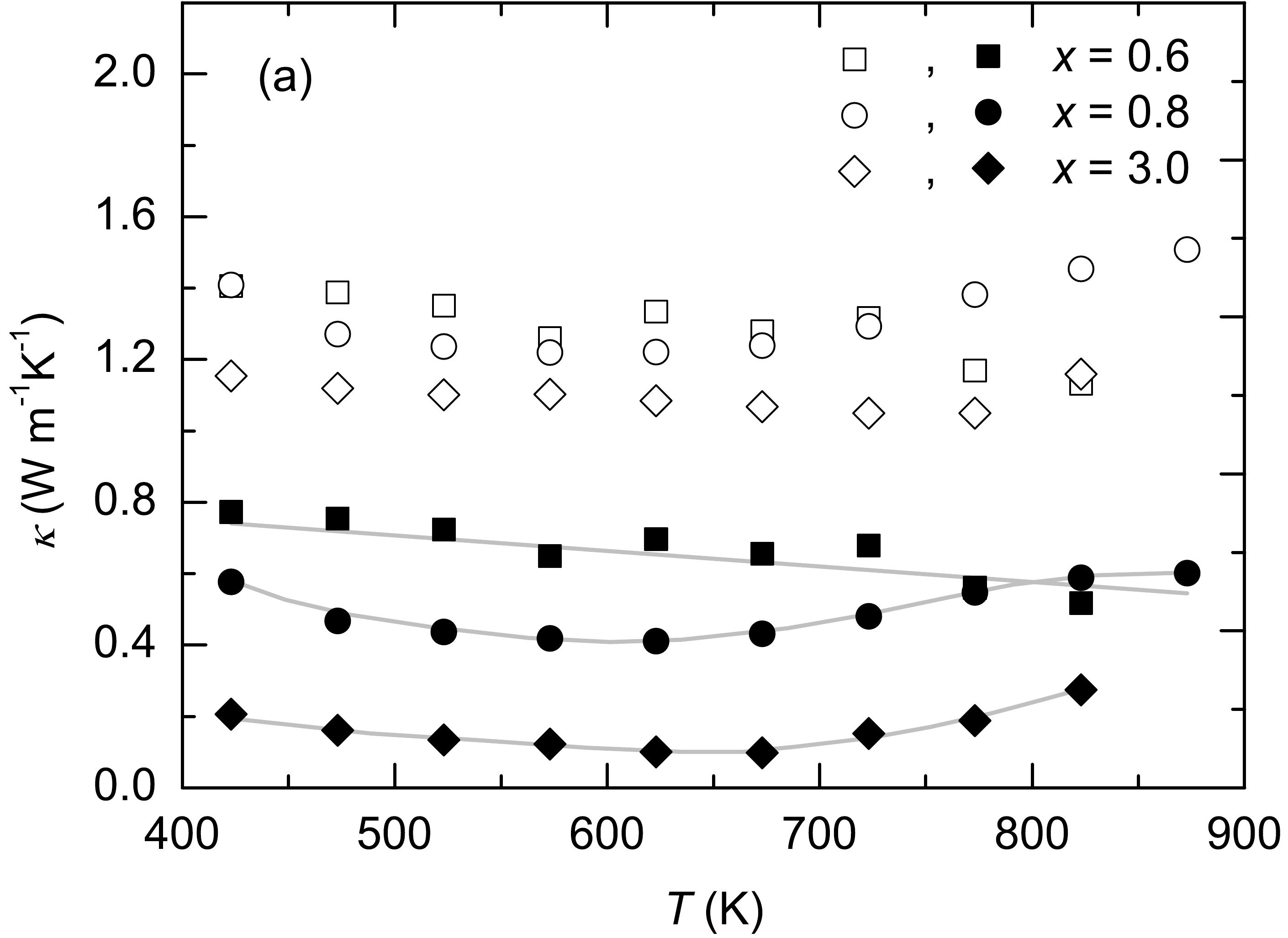}
\includegraphics[width=\columnwidth]{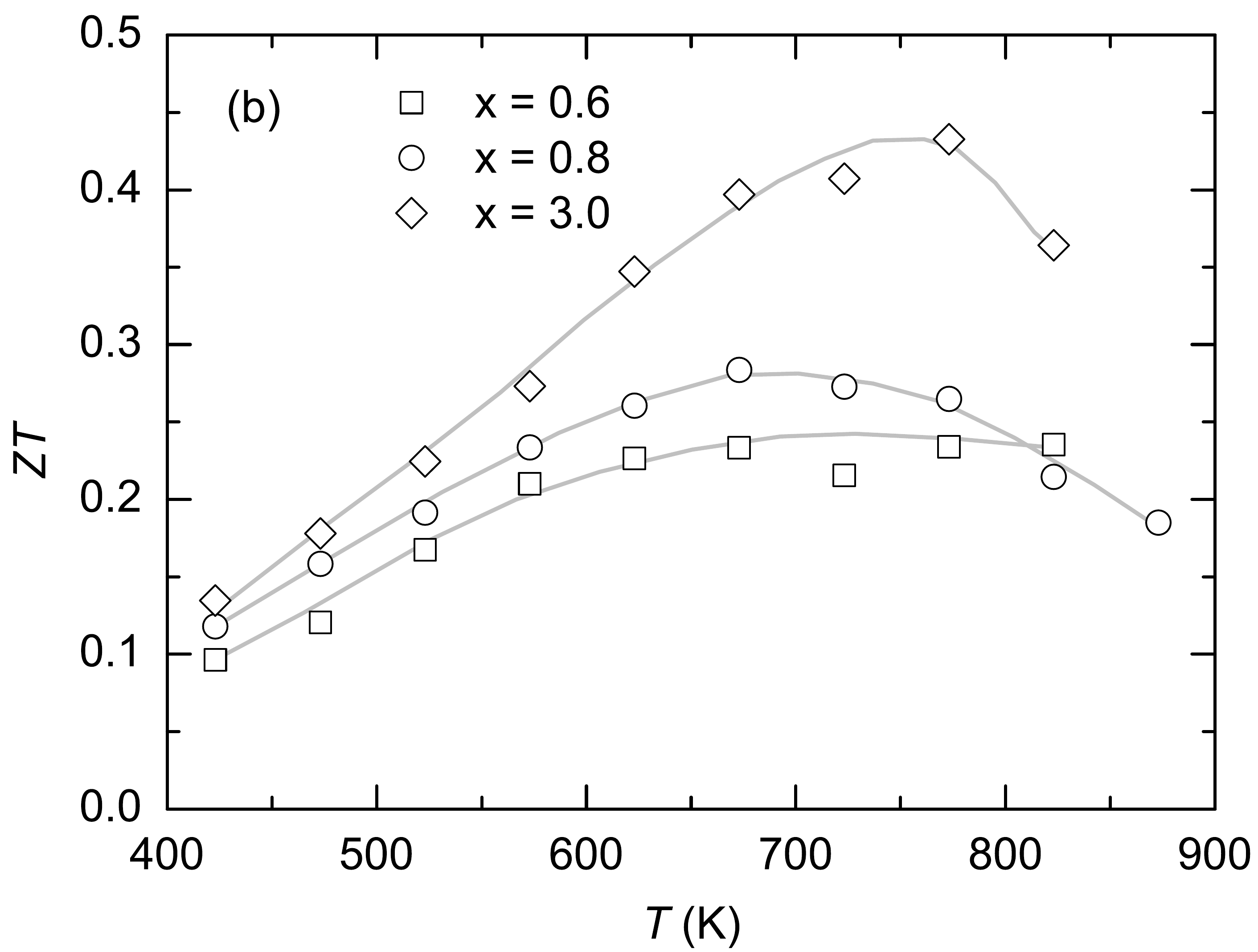}
\caption{Temperature dependence of the total thermal conductivity $\kappa$ (open
symbols) and phonon thermal conductivity $\kappa_{ph}$ (solid symbols) (a) and
the dimensionless figure of merit $ZT$  (b) of
Ba$_{8}$Cu$_{5}$Si$_{6}$Ge$_{35-x}$Sn$_{x}$ (\textit{x} = 0.6 - 3.0). The lines
are guides to the eye.}
\label{Graph2_Tomes}
\end{figure}

Polycrystalline samples with the nominal compositions
Ba$_{8}$Cu$_{5}$Si$_{6}$Ge$_{35-x}$Sn$_{x}$ (\textit{x} = 0.6 $-$ 3.0) were
prepared using melt spinning and spark plasma sintering (SPS). According to XRD,
all meltspun (MS) samples are type-I clathrate phases with a small amount of Sn
as impurity phase (Fig.\,\ref{Graph1_Tomes}\,a). During the SPS process a small
amount of diamond phase Ge$_{2}$Si forms while the Sn impurity phase disappears.
The composition of the MS+SPS samples evaluated from EDX, their relative
densities, and their lattice parameters are shown in
Table\,\ref{tab:table1_Tomes}. The maximum amount of Sn in the type-I clathrate
phase is approximately 0.6\,at./u.c. which, unfortunately, is only slightly
larger than the value reported in Ref.\,\cite{Yan13.2} (see also
Table\,\ref{tab:table1_Tomes}). The Cu content decreases with increasing
\textit{x}. This, together with an appearance of vacancies, influences the
electrical transport. All MS+SPS samples are well compacted, with relative
densities $d$ between 93\,\% and 98\,\%. Figure\,\ref{Graph1_Tomes}\,b shows
that the substitution of Ge by larger Sn atoms increases the lattice parameter
$a$ in the hotpressed samples with a low substitution level $x_{\rm{EDX}}$.
However, in the MS+SPS samples with a higher substitution level, $a$ decreases
with further increasing $x_{\rm{EDX}}$. We assume that the latter is due to the
decrease of the (Cu+Ge)/(Si) ratio, as Cu and Ge are larger than Si.

The temperature dependence of the thermal conductivities $\kappa$ of all samples
and their phonon contributions $\kappa_{ph}$ are shown in
Fig.\,\ref{Graph2_Tomes}\,a. $\kappa$ decreases with increasing temperature up
to about 650\,K. Above this temperature a bipolar contribution is observed, as
in the case of the hotpressed samples (Ref.\,\cite{Yan13.2}). The decrease of
$\kappa_{ph}$ at low temperatures can be understood as the combined effect of
Umklapp scattering of acoustic phonons ($\kappa_{ph}\approx T^{-1}$) and alloy
scattering ($\kappa_{ph}\approx T^{-1/2}$). The decrease of $\kappa_{ph}$ with
higher $x$ is most probably due to the higher Sn concentration and a larger
vacancy content. At 750\,K, $\kappa_{ph}$ of the MS+SPS samples is only half of
the value of the hotpressed samples \cite{Yan13.2}. The dimensionless figure of
merit $ZT$ is 0.43 at 773\,K for \textit{x} = 3.0 (Fig. \ref{Graph2_Tomes}b).

\begin{table*}[h!]
  \caption{Sn content \textit{x} in the nominal composition, sample code, chemical composition, lattice parameter \textit{a}, (Cu+Ge)/(Si) ratio, and relative density $d$ of Ba$_{8}$Cu$_{5}$Si$_{6}$Ge$_{35-x}$Sn$_{x}$ (\textit{x} = 0.6 - 3.0) MS+SPS samples (codes 1-3), and of hotpressed samples (codes 4-6) from Ref.\,\cite{Yan13.2} for comparison.}
   \begin{tabularx}{\textwidth}{lllllll}
    \hline
    \textit{x} & Code & Composition & \textit{a} & (Cu+Ge)/(Si) & \textit{d}\\
    & & & (nm) & & ($\%$)\\
    \hline
    0.6 \hspace{0.5cm} &1&Ba$_{8}$Cu$_{5.23}$Si$_{6.43}$Ge$_{33.68}$Sn$_{0.48}$ \hspace{1cm} & $1.06596(1)$ \hspace{1cm} & $6.05$ \hspace{1cm}& $96.3$ \\
 0.8&2&Ba$_{8}$Cu$_{5.21}$Si$_{6.79}$Ge$_{33.34}$Sn$_{0.57}$&$1.06579(3)$&$5.68$&$94.8$ \\
 3.0&3&Ba$_{8}$Cu$_{5.06}$Si$_{7.46}$Ge$_{32.76}$Sn$_{0.60}$&$1.06465(1)$&$5.07$&$98.0$ \\ \hline
    0.2&4&Ba$_{8}$Cu$_{4.98}$Si$_{5.88}$Ge$_{34.76}$Sn$_{0.18}$&$1.0656(2)$&$6.76$&$94.6$ \\
    0.4&5&Ba$_{8}$Cu$_{4.92}$Si$_{5.92}$Ge$_{34.57}$Sn$_{0.41}$&$1.0664(1)$&$6.67$&$96.7$ \\
     0.6&6&Ba$_{8}$Cu$_{5.01}$Si$_{5.93}$Ge$_{34.28}$Sn$_{0.55}$&$1.0667(2)$&$6.63$&$95.8$ \\
    \hline
  \end{tabularx}
  \label{tab:table1_Tomes}
\end{table*}

\subsection{Melt spinning of the skutteridite CoSb$_3$}

The unusual crystallization behavior of clathrates described in
Sects.\,\ref{app1}-\ref{app3} poses the question whether it is unique for
clathrates or typical also for other cage compounds with intrinsically low
thermal conductivities. Which of the suggested mechanisms discussed in
Sect.\,\ref{app3} -- the suppression of the heat flow or the peculiarity of the
crystal structure  -- is the dominant one? To answer this question we performed
a similar investigation of melt spinning on the skutterudite CoSb$_3$. We
demonstrate that for CoSb$_3$ melt spinning can be successfully used for
nanostructuring.

We produced a series of meltspun and subsequently spark plasma sintered (SPS)
CoSb$_3$ samples, differing only by the cooling speed during the quenching
procedure. Table\,\ref{tab:table2} summarizes details for all samples studied
here. The series consists of five different CoSb$_3$ samples produced at various
rotation speeds (600 - 3000\,rpm) of the quenching copper wheel. After the melt
spinning procedure all materials were compacted using SPS at a temperature of
550$^\circ$C and a pressure of 40\,MPa for 5 minutes. The samples are named
according to the rotation speed of the copper wheel during the melt spinning
procedure (e.g., 600RPM).

\begin{table*}[ht!]
\caption{\label{tab:table2} Sample code, impurity phase, lattice parameter $a$, grain size (GS) determined from XRD data and TEM images, and relative density $\rho_{rel}$ of the meltspun CoSb$_3$ samples after the SPS process. From Ref.\,\cite{Ike15.1}.}
\begin{tabularx}{\textwidth}{ccccccc}
\hline
\multicolumn{1}{c}{Code}&\multicolumn{1}{c}{Impurities}&\multicolumn{1}{c}{$a$}&\multicolumn{1}{c}{GS (XRD)}&\multicolumn{1}{c}{GS (TEM)}&\multicolumn{1}{c}{$\rho_{rel}$}\\
 (rpm)&&(\AA)&(nm)&(nm)&($\%$)\\ \hline
 600RPM&$8\%$     Sb   &$9.0358 (3)$&$343.5$&$500-1000$&$95.6$ \\
 1200RPM&$5\%$ CoSb$_2$&$9.0355(2)$&$336.3$&$250-800$&$97.3$ \\
 1800RPM&$5\%$ CoSb$_2$&$9.0360(2)$&$-$&$150-400$&$88.6$ \\
 2400RPM&$8\%$ CoSb$_2$&$9.0370(3)$&$157.1$&$<50$&$87.7$ \\
 3000RPM&$20\%$CoSb$_2$&$9.0350(6)$&$-$&$-$&$98.3$ \\
 \hline
\end{tabularx}
\end{table*}

XRD on the meltspun CoSb$_3$ samples resulted in a phase mixture with CoSb$_3$
as the majority phase and Sb and CoSb$_2$ impurity phases in all meltspun
samples \cite{Ike15.1}. This is consistent with the fact that CoSb$_3$ melts
incongruently. In contrast to type-I clathrates, the crystallization rate of
CoSb$_3$ is too small for the kinetic suppression of foreign phases. In line
with studies on meltspun filled skutterudites \cite{Li08.2,Li09.1,Su12.1}, the
content of foreign phases is readily reduced by the heat treatment during the
SPS procedure. A small amount of Sb in 600RPM and CoSb$_2$ (see
Tab.\,\ref{tab:table1}) was, however, detected by XRD investigations even in
the sintered samples. Due to the relatively large amount of CoSb$_2$ in the
sample 3000RPM, this material is not considered in the discussion presented
below.

\begin{figure}[ht!]
\centering
\includegraphics[width=\columnwidth]{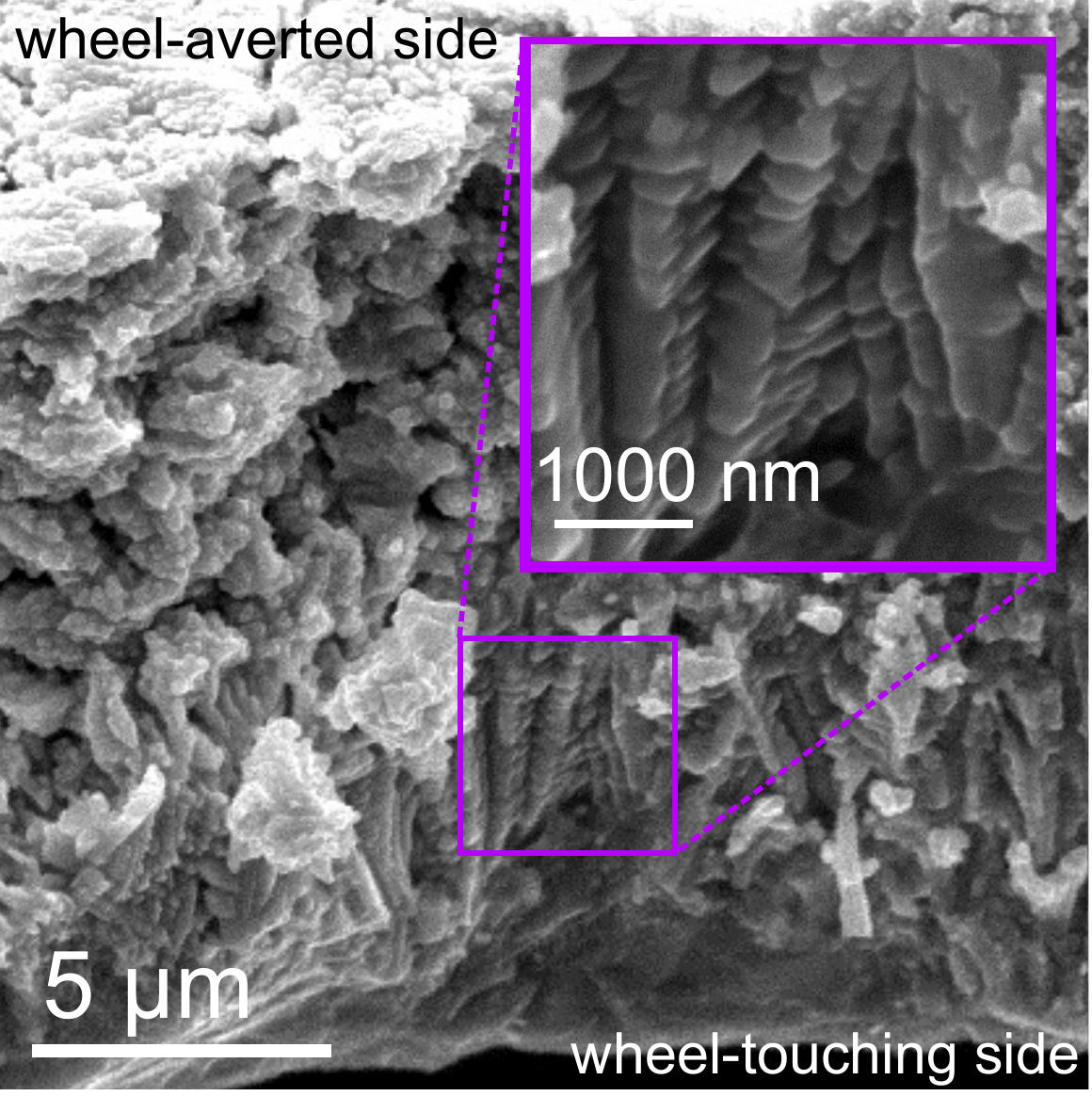}
\caption{SEM image of a fractured 1200RPM meltspun CoSb$_3$ flake. The inset
shows a magnified image of the area indicated by the violet
box. Adapted from Ref.\,\cite{Ike15.1}.}\label{fig:CoSb3SEMcross}
\end{figure}

Figure\,\ref{fig:CoSb3SEMcross} shows the broken cross-section of a meltspun
CoSb$_3$ flake of the batch 1200RPM. In contrast to type-I clathrates
\cite{Pro13.0,Pro14.1}, the meltspun CoSb$_3$ samples consist of nearly
spherical grains. This is the case even for the seemingly columnar areas on the
wheel-touching side (Fig.\,\ref{fig:CoSb3SEMcross} inset). In contrast to type-I
clathrates, which show columnar grain growth, the rate of crystallization and
the speed of grain growth is thus found to be moderate.

\begin{figure}[!tbp]
	\centering
  \includegraphics[width=\columnwidth]{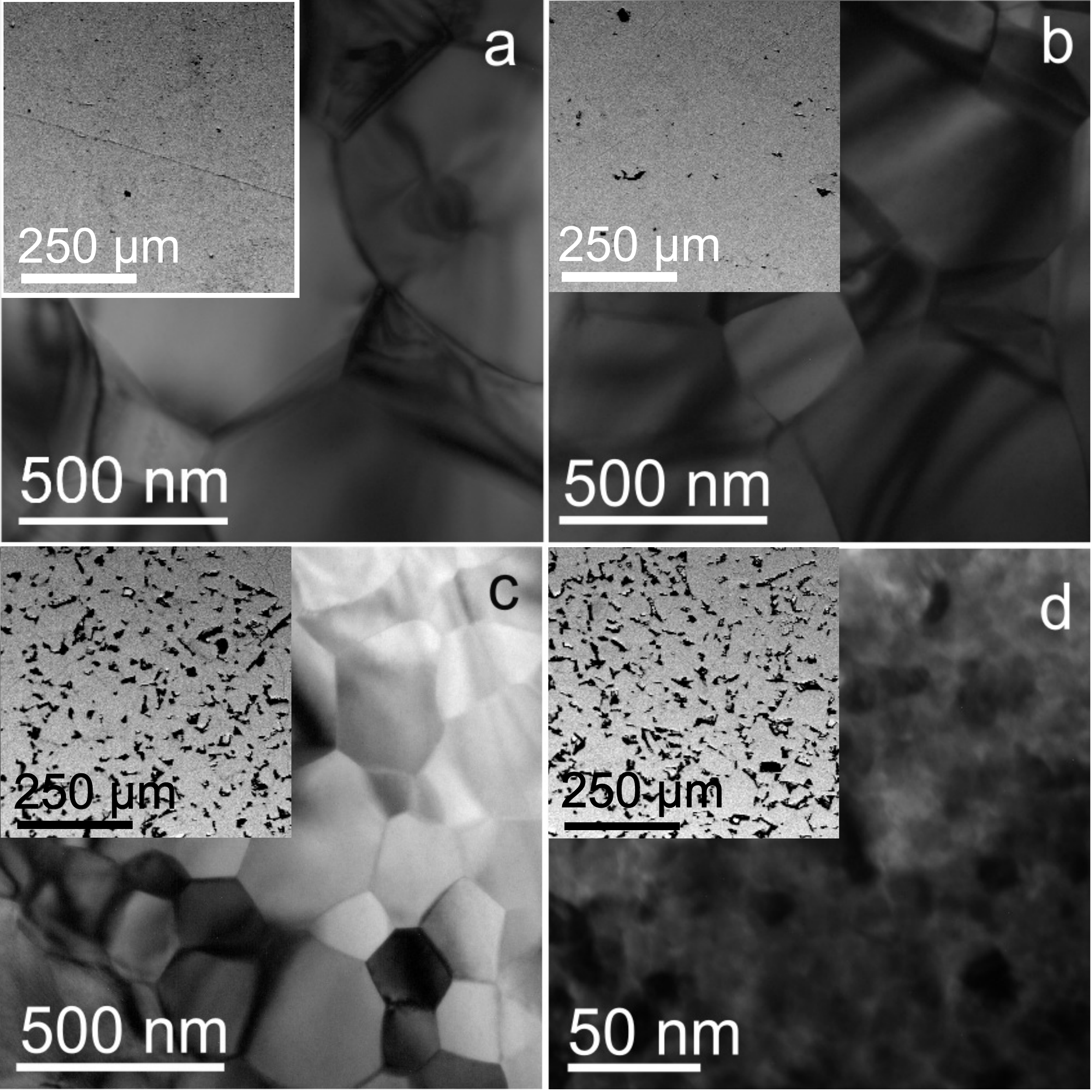}
	\caption{TEM bright field images of the meltspun and SPS treated CoSb$_3$ samples, produced at 600\,rpm (a), 1200\,rpm (b), 1800\,rpm (c), and 2400\,rpm (d). The insets show SEM images of the corresponding materials. Adapted from Ref.\,\cite{Ike15.1}.}
	\label{fig:CoSb3TEM}
\end{figure}

\begin{figure*}
\centering	\subfloat{\includegraphics[height=0.36\textwidth]{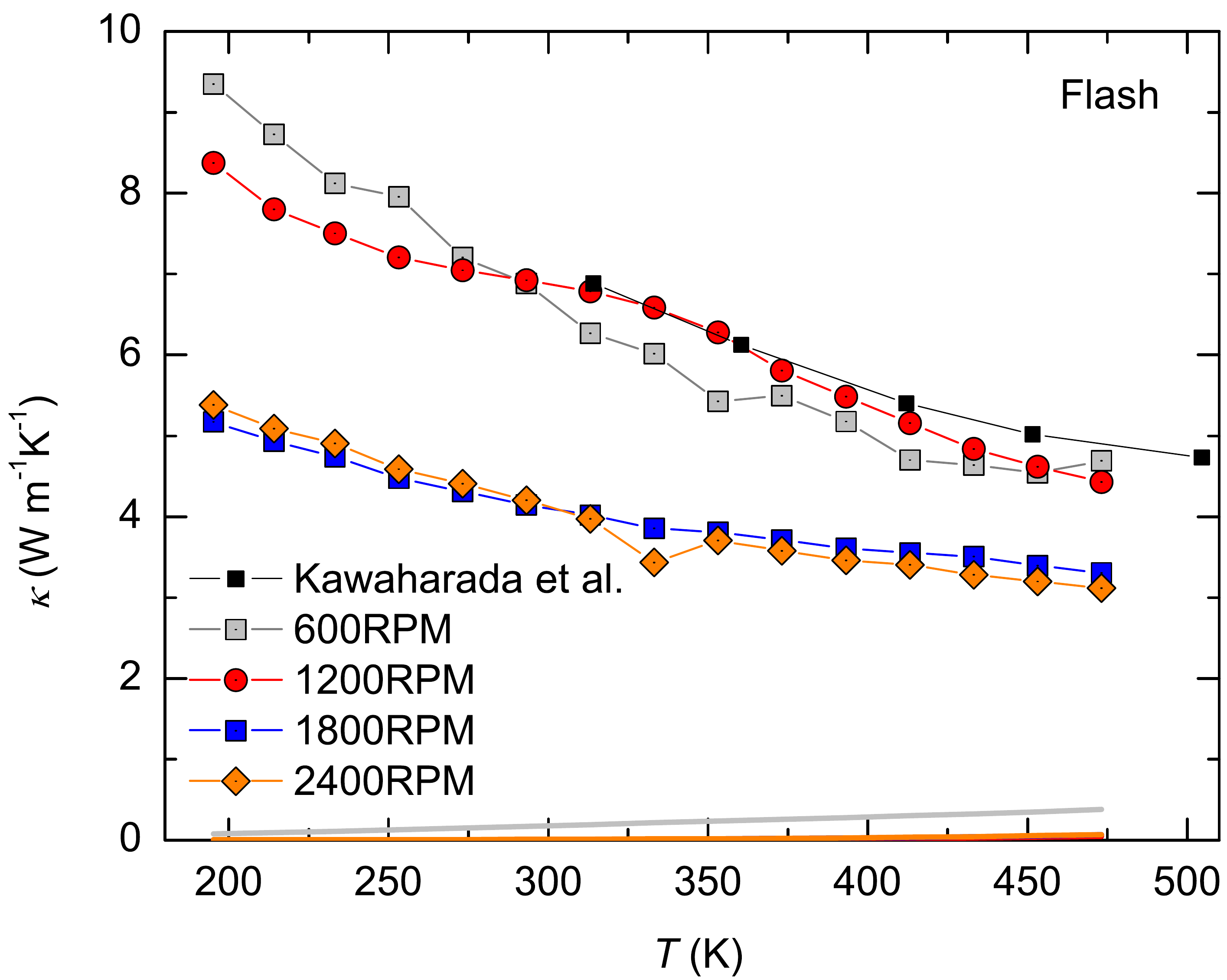}}\hfill
\subfloat{\includegraphics[height=0.36\textwidth]{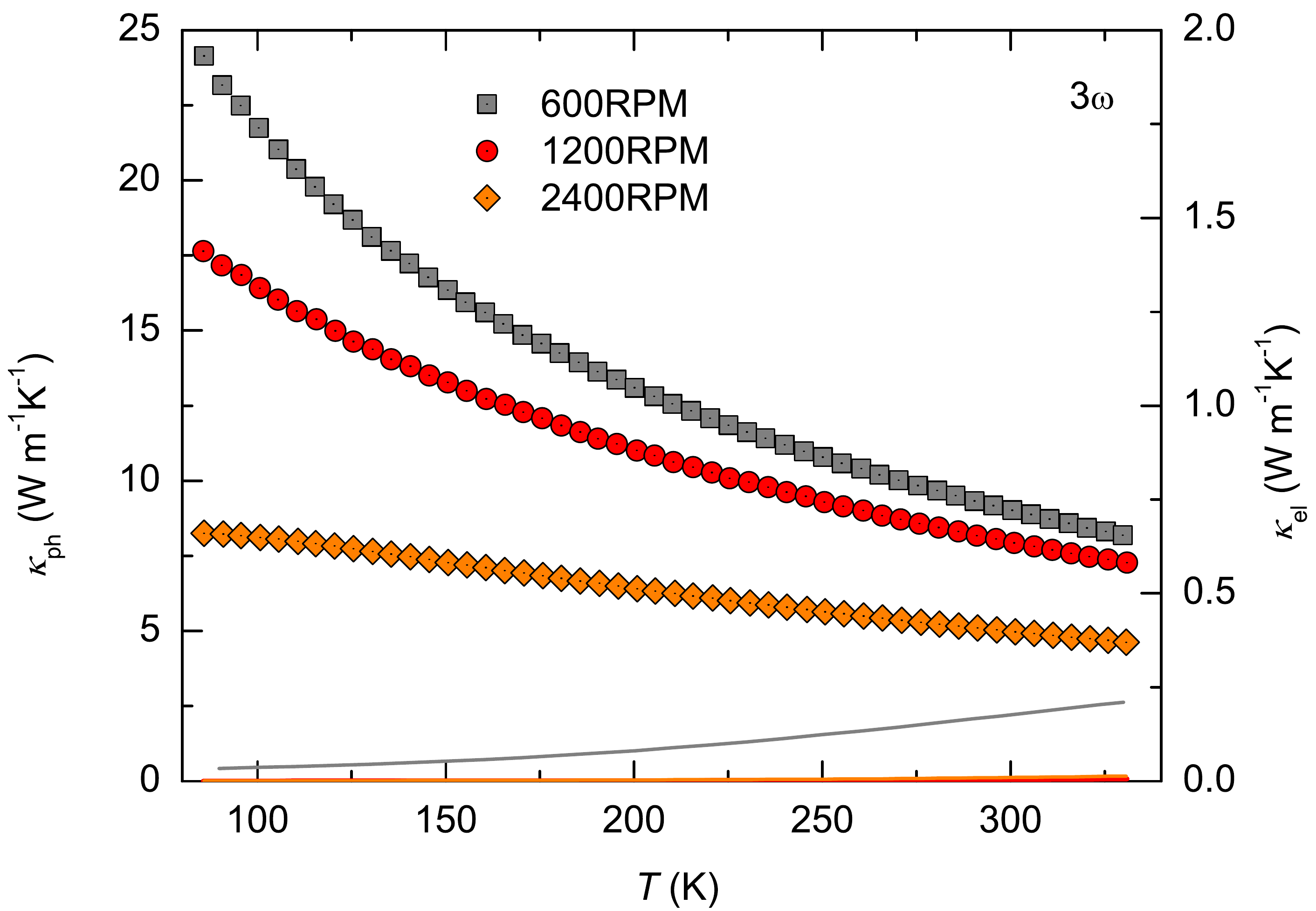}}
\caption{Temperature-dependent phonon (symbols) and electronic thermal
conductivity (lines, calculated using the Wiedemann-Franz law) of samples
produced at 600 - 2400\,rpm. The black squares show literature data
\cite{Kaw01.2} on polycrystalline CoSb$_3$. The results shown left are measured
with the Flash technique, the right panel shows 3$\omega$ data. Adapted from Ref.\,\cite{Ike15.1}.}\label{fig:kappa1}
\end{figure*}

TEM investigations on the samples 600RPM, 1200RPM, 1800RPM, and 2400RPM after
the SPS procedure clearly evidence that the grain size of the materials is
dependent on the cooling rate during melt spinning. The grain size determined by
TEM and, alternatively, by XRD is summarized in Tab.\,\ref{tab:table2}. The
discrepancy observed for 2400RPM is most likely explained by the presence of
regions with different grain sizes. The insets of Fig.\,\ref{fig:CoSb3TEM} show
SEM pictures of the materials 600RPM to 2400RPM. The incomplete compaction of
1800RPM and 2400RPM, indicated by the lower measured relative densities
(Tab.\,\ref{tab:table2}), is also clearly observed in the corresponding SEM
pictures.

The thermal conductivity of the meltspun and SPS samples was studied by two
different techniques. First, the thermal conductivity was determined from the
measured room temperature density $\rho_{\text{RT}}$, the thermal diffusivity
$\alpha$, and the specific heat $C_{\text{p}}$ (Ref.\,\cite{Ike15.1}) according
to $\kappa = \alpha \cdot C_{\text{p}} \cdot \rho_{\text{RT}}$. The result is
shown in Fig.\,\ref{fig:kappa1} (left). Within the resolution of our Flash
experiment, the phonon contribution $\kappa_{\text{ph}}$ of 600RPM and 1200RPM
is similar. 1800RPM and 2400RPM show a significantly reduced thermal
conductivity. Unfortunately these samples not only possess a smaller grain size
compared to 600RPM and 1200RPM, but also an approximately 10\,\% lower relative
density (Tab.\,\ref{tab:table1}). Thus, the mechanism of the
$\kappa_{\text{ph}}$ reduction cannot be pinned down from these results.

Therefore, we studied the thermal conductivity of the meltspun and SPS samples
in more detail using additional 3$\omega$ thermal conductivity measurements. As
long as the sample surface is sufficiently flat, the heater geometry can be
determined with high accuracy. For porous materials on the other hand, the
length of the heater can considerably deviate from its nominal value. This
introduces an error in the measured thermal conductivity, because the line power
density is a multiplicative quantity for calculating the in-phase temperature
oscillation. As observed for the sample 1800RPM, moreover the insulating layers
between heaters and porous samples are likely to show pinholes. In such cases,
accurate 3$\omega$ measurements cannot be performed. The overall trend of our
3$\omega$ phonon thermal conductivity data is similar to the Flash thermal
conductivity results but the absolute values are systematically higher. Since
the absolute values of the 3$\omega$ measurements on the dense samples 600RPM
and 1200RPM can be considered as accurate the Flash data appear to underestimate
$\kappa$. In agreement with the decreasing grain size the 3$\omega$
$\kappa_{\text{ph}}$ values are found to decrease with increasing rotation
speed. Due to the incomplete compaction of 2400RPM, its low $\kappa_{\text{ph}}$
values can, however, not be fully attributed to scattering on additional grain
boundaries. The difference in the $\kappa_{ph}$ values among all meltspun and
SPS treated CoSb$_3$ samples increases towards lower temperatures. This is in
agreement with the increase of the phonon mean-free-path associated with the
freezing of Umklapp processes upon cooling.

\begin{figure}
\centering
\includegraphics[width=\columnwidth]{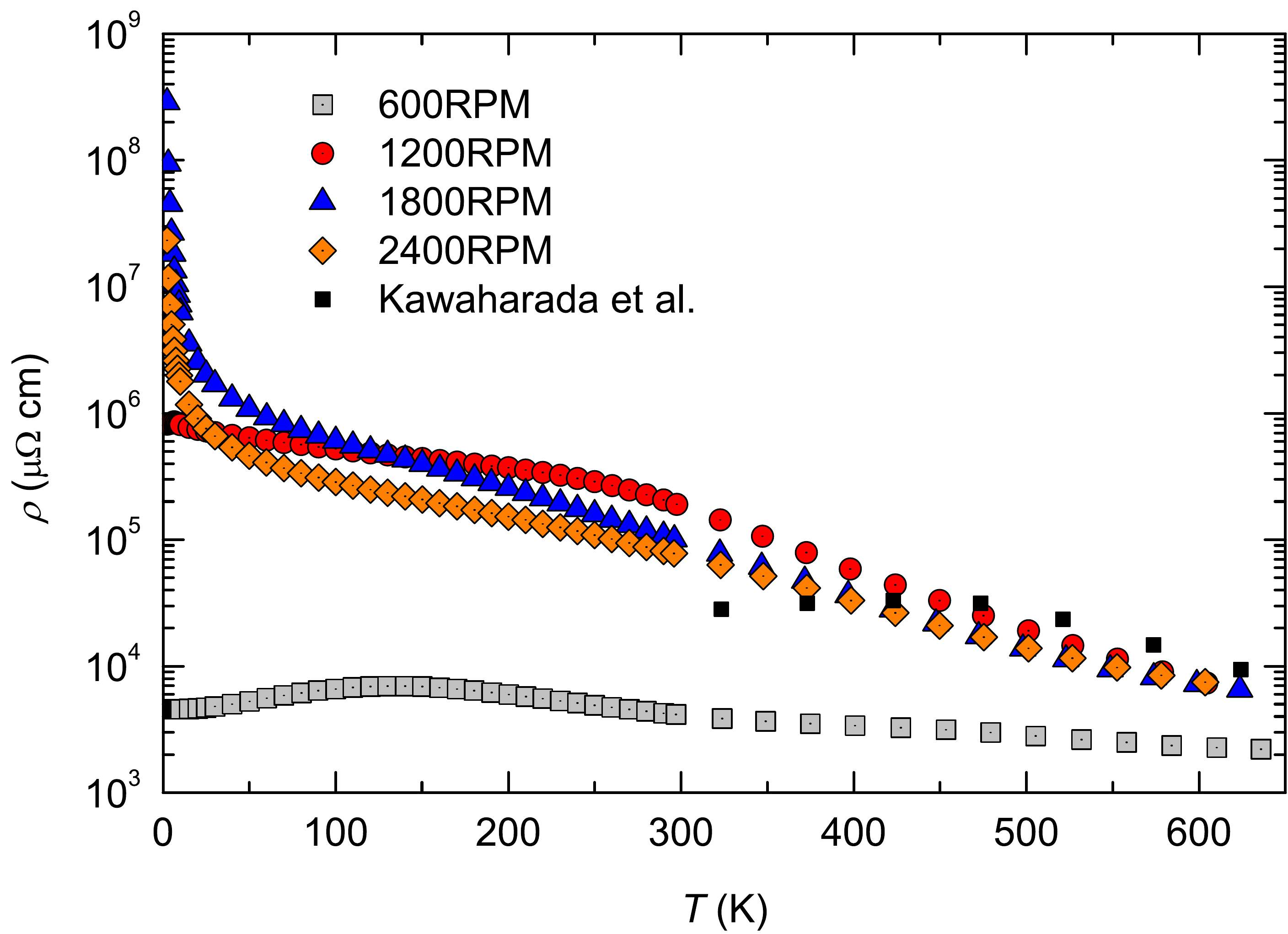}
\caption{Electrical resistivity of all meltspun and SPS treated CoSb$_3$ samples as a function of temperature. The black squares are literature data for polycrystalline CoSb$_3$ from \cite{Kaw01.2}. Adapted from Ref.\,\cite{Ike15.1}.}
\label{fig:CoSb3rho}
\end{figure}

Figure\,\ref{fig:CoSb3rho} shows the electrical resistivity of all investigated
CoSb$_3$ samples as a function of temperature. 1200RPM, 1800RPM, and 2400RPM
show semiconducting behavior in the entire temperature range. 600RPM, however,
shows features of weak metallicity at low temperatures. By studying 
magnetoresistance and Hall effect data of the presented sample series in terms
of a multiband model, we attributed the decrease of the electrical resistivity
of 600RPM to additional charge carriers with high carrier mobility introduced by
the Sb foreign phase \cite{Ike15.1}. A surprising result is that the poorly
compacted materials 1800RPM and 2400RPM show similar resistivities as the
well compacted sample 1200RPM. A reason for this behavior might be the SPS
mechanism: the compaction is achieved by pressure and heat generated by an
electrical current flowing through the material. Thus, even in samples with low
densities there are continuous low-resistance current paths through the sample.

\section{Ball milling and hot pressing}

\subsection{Nanostructuring by ball milling}\label{ballmilling}

Nanostructured bulk materials can also be obtained by ball milling. This top-down nanostructuring technique is generally considered as a simple, economic, and efficient method that has been used frequently in different intermetallic systems \cite{Koc03.1,Fec06.1}. The fundamental mechanism for powder refinement during this process is to use the kinetic energy of the mill system to fragmentate large particles to small ones \cite{Koc03.1}. The particle size in powders is determined by the milling energy, milling time, and other process parameters (e.g., the atmosphere and choice of additives) as well as the mechanic  properties of the raw materials \cite{Koc03.1,Fec06.1,Eck92.1,Koc93.1,Koc97.1}. 

Studies of crystallite size vs milling time were carried out on clathrates in the Ba-Cu-Ge-Si system. Our previous investigation on Ba$_8$Cu$_{5}$Si$_x$Ge$_{35-x}$ showed that the best thermoelectric performance was found in compounds with low $x$ \cite{Yan13.2}. Thus, we chose the composition Ba$_8$Cu$_{4.5}$Si$_6$Ge$_{35}$ for further investigations \cite{Yan15.1}.

The ball milling was performed in a Fritsch planetary mill. Pre-synthesized and
annealed ($800^\circ\mathrm{C}$, 20 days) large samples (4 g each) were used for
ball milling. The process parameters were as follows: speed of the main disk =
120 rpm, weight ratio of balls to powders = 82, planetary ratio = -2.5, ball
diameter $\O = 1$ cm, material of containers and balls = WC. The milling time is
denoted by $t_{\rm m}$. 

\begin{figure}[h]%
\vspace{-1.2cm}

\hspace{-1.4cm}\includegraphics*[width=1.35\linewidth]{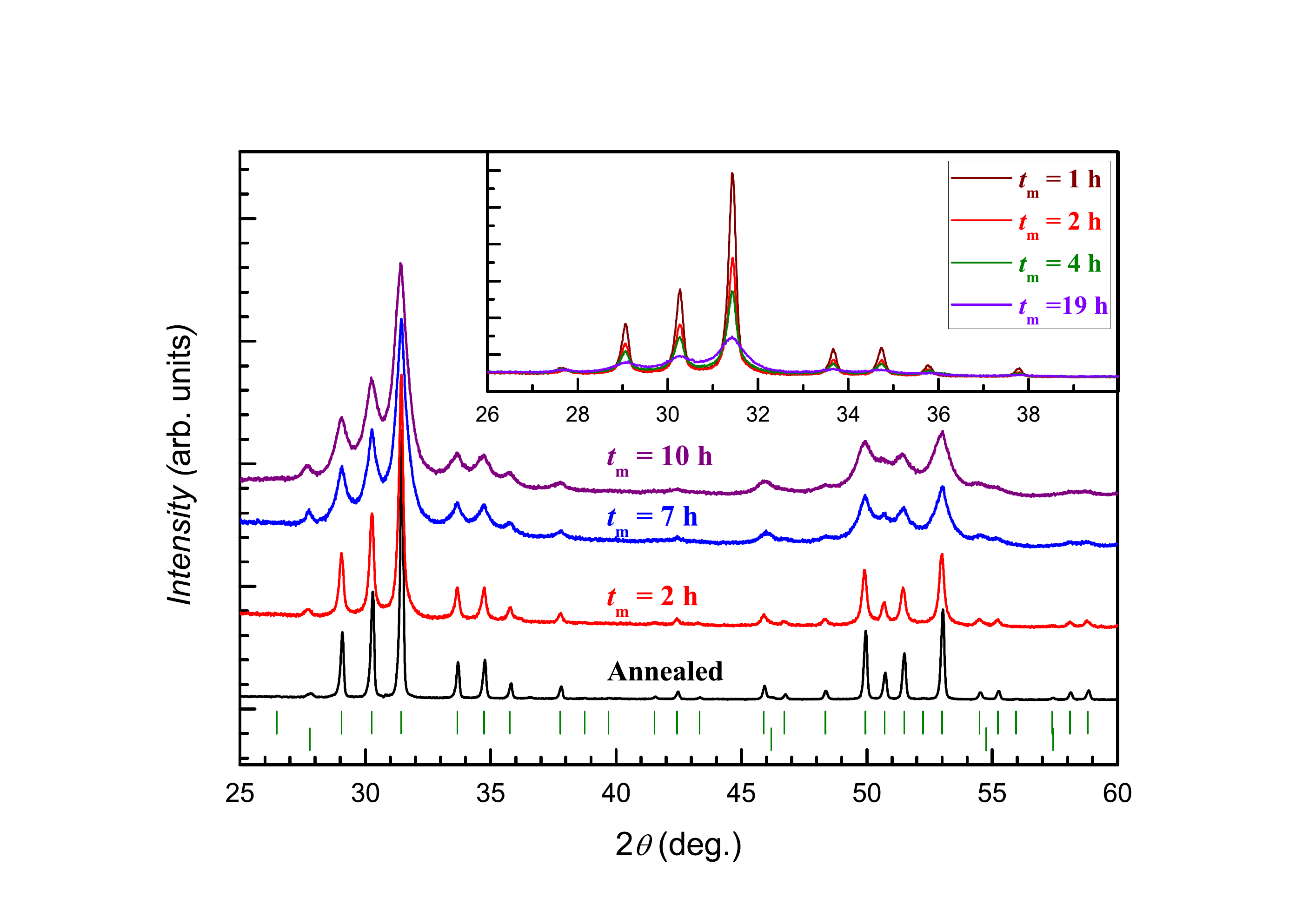}
\vspace{-0.8cm}

\caption{%
  XRD patterns of an annealed sample and selected samples milled with different $t_{\rm m}$.
The patterns were normalized to the highest peak of the type-I clathrate phase. The
vertical lines (dark green) at the bottom denote the Bragg positions of the clathrate and diamond phase. The inset shows the non-normalized patterns of the powders recorded with the same X-ray exposure time.}
\label{Fig:fig1}
\end{figure}

The phase analysis of the as-milled powders was performed using XRD. The XRD patterns of an annealed sample and several as-milled powders are shown in Fig.\,\ref{Fig:fig1}. The broadening of the peaks in the patterns of the as-milled samples arises from the reduced grain/crystallite size and increased residual strain after milling. No peak shift or new peaks appear in the patterns after milling, indicating that the phase composition is stable during the milling process. 

\begin{figure}[h]%
\centering
\includegraphics*[width=\columnwidth]{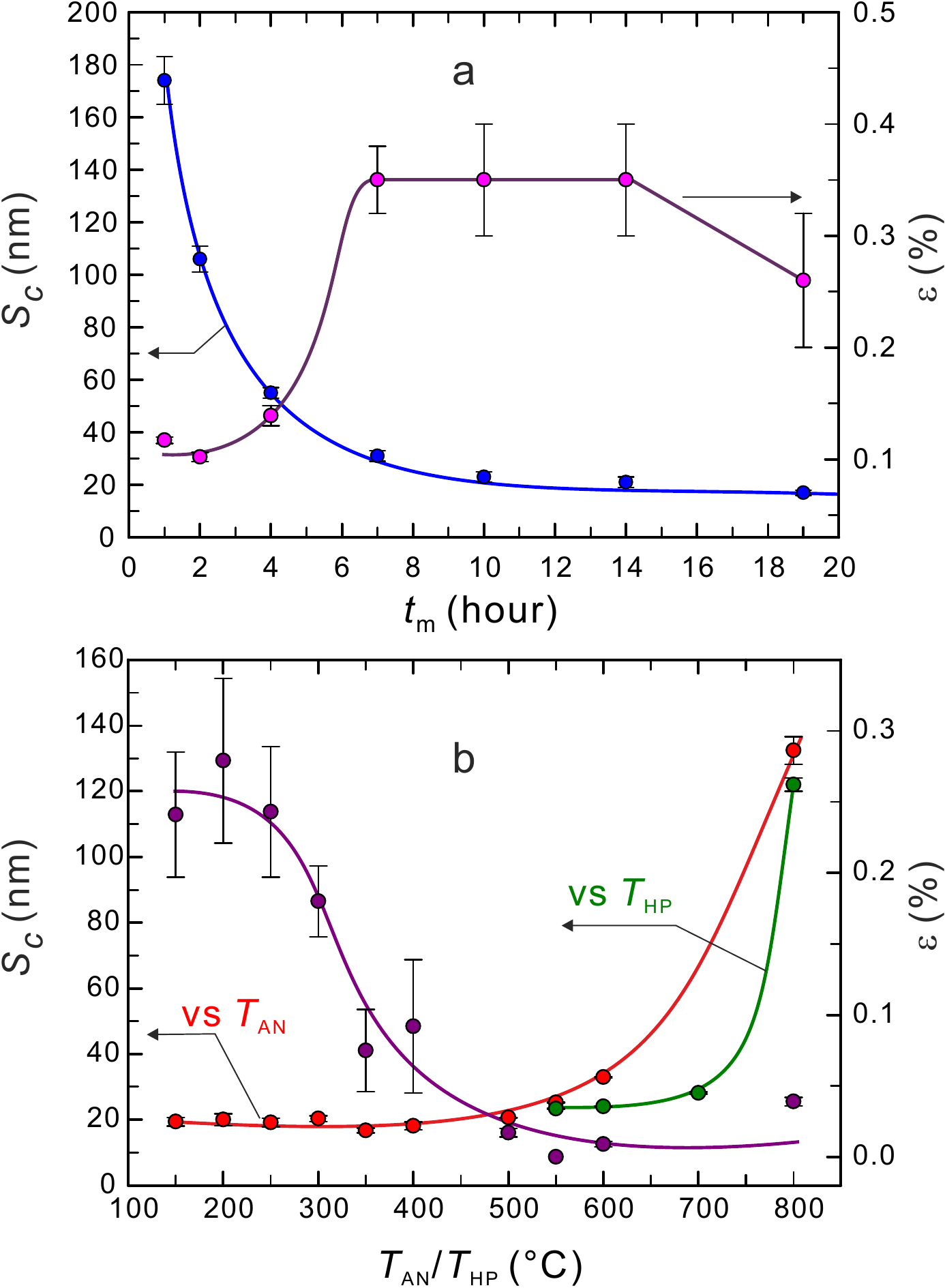}
\caption{%
Crystallite size $S_{\rm c}$ and residual strain $\varepsilon$ vs milling time $t_{\rm m}$ (a). At each opening, the powder is separated from the container and the balls. $S_{\rm c}$ and $\varepsilon$ vs $T_{\rm AN}$ of this powder (batch II) (b). $S_{\rm c}$ of sintered samples from batch I vs pressing temperature $T_{\rm HP}$ is also included for comparison. The solid lines are guides to the eye. 
}
\label{Fig:fig2}
\end{figure}

The crystallite size $S_{\rm c}$ decreases rapidly in the first 4 h of the milling treatment, and  approaches saturation slowly thereafter  (Fig.\,\ref{Fig:fig2} (a)). The minimum size after 19 h milling was $18(1)$ nm. The residual strain $\varepsilon$, on the other hand, increases in the beginning, reaches a maximum of 0.35\% after 7\,h, and decreases slightly after 14 h. Our study indicates that a repeated powder separation from both the container and the balls is necessary during the ball milling process; without this separation the minimum crystallite size after 14\,h reaches only 45(5)\,nm. The powders without and with powder separation are referred to as batch I and batch II in what follows.
  
To determine the best hot pressing temperature, a study of crystallite size 
growth/strain reduction was carried out by annealing the nanopowders in vacuum
at different temperatures $T_{\rm AN}$ for 2\,h. The crystallite size of powder
from batch II remains essentially unchanged when the annealing temperature is
below $600^\circ\mathrm{C}$ but the residual strain is released
(Fig.\,\ref{Fig:fig2}\,b). Above $600^\circ\mathrm{C}$, the crystallite size
increases rapidly. This suggests that the pressing temperature should not exceed
$600^\circ\mathrm{C}$ to keep the grains small. 

\subsection{Consolidation by hot pressing} \label{hotpressing}
  
Powder consolidation is required to obtain bulk samples. Conventional pressureless sintering is often unable to provide dense materials with very fine grains \cite{Hun09.1}. Pressure-assisted sintering methods such as hot pressing, hot isostatic pressing, or spark plasma sintering have thus been at focus \cite{Fec06.1,Gro06.1}. The process parameters are pressure, pressing temperature, holding times, heating rate, and atmosphere for sintering. They determine the microstructure and the physical and mechanical properties of the sintered samples. 

\begin{figure}[h]%
\centering
\includegraphics*[width=\columnwidth]{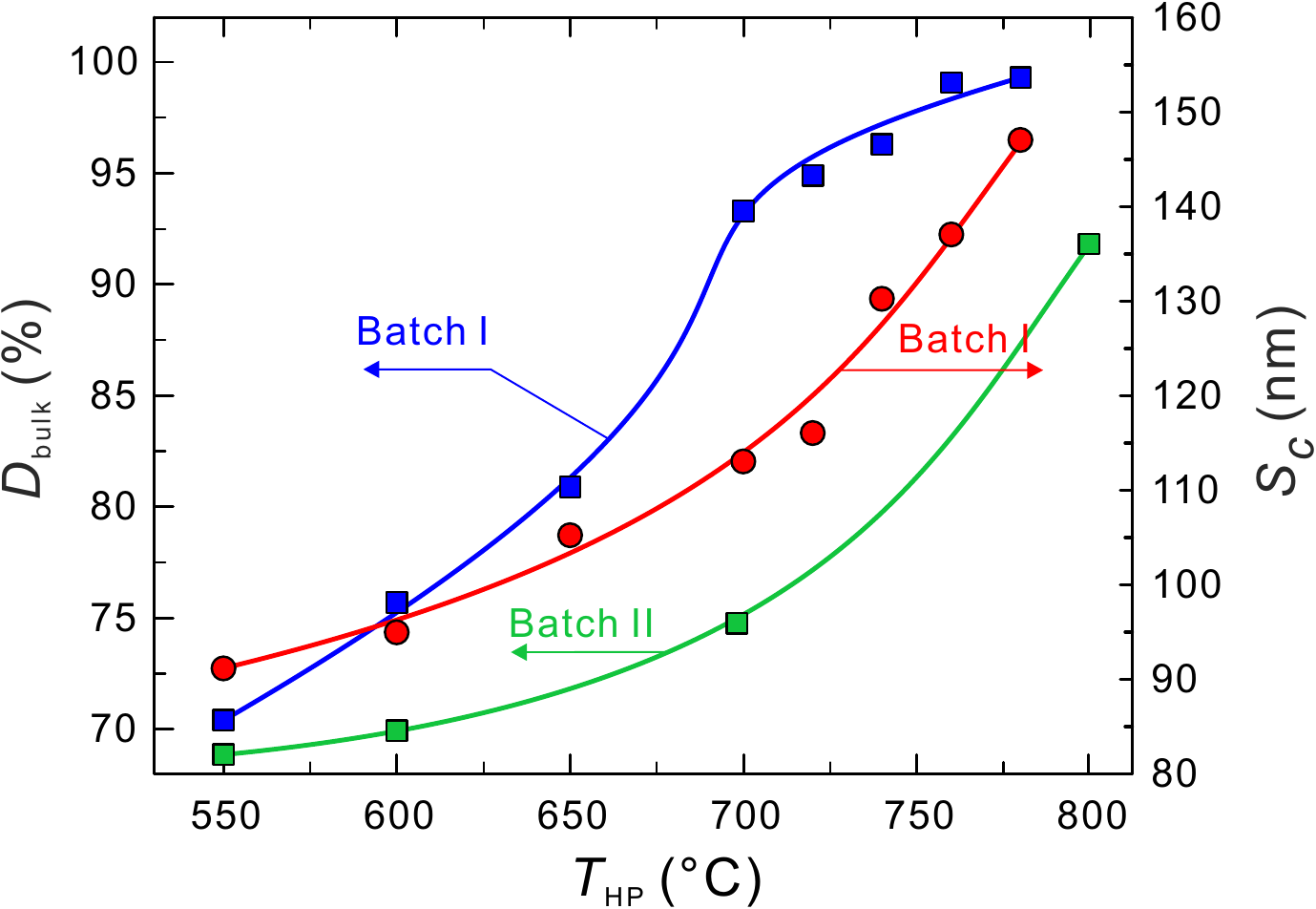}
\caption{%
Bulk density $D_{\rm bulk}$ and crystallite size $S_{\rm c}$ vs hot pressing temperature $T_{\rm HP}$ for sintered samples from batch I and $D_{\rm bulk}$ vs $T_{\rm HP}$ for the samples of batch II.
}
\label{Fig:fig3}
\end{figure}

We focused on the effect of the hot pressing temperature $T_{\rm HP}$ on the
microstructure and the thermoelectric properties of sintered samples
\cite{Yan14.1}. We fixed the pressure $p=56$ MPa, holding time $t=2$ h, and
heating rate $r_T=5^\circ\mathrm{C}$/min during the hot pressing process. Since
the properties of the hotpressed samples are also determined by the properties
of the nanopowders for sintering, we studied the two above-discussed batches of
nanopowders which have different crystallite size (45(5)\,nm for batch I,
18(1)\,nm for batch II). We chose pressing temperatures between 550 and
$800^\circ\mathrm{C}$ (Fig.\,\ref{Fig:fig2}\,b).

Figure\,\ref{Fig:fig3} shows the crystallite size $S_{\rm c}$ and relative bulk density $D_{\rm bulk}$ of samples sintered from nanopowders of batch I vs the pressing temperature $T_{\rm HP}$. Both $S_{\rm c}$ and $D_{\rm bulk}$ increase with $T_{\rm HP}$. To obtain dense samples with small grain size, $T_{\rm HP}$ should be between $700^\circ\mathrm{C}$ and $760^\circ\mathrm{C}$. The qualitative behaviour of batch II is similar (Figs.\,\ref{Fig:fig2}\,b and \ref{Fig:fig3}).

\begin{figure}[ht]%
\centering
\includegraphics*[width=\columnwidth]{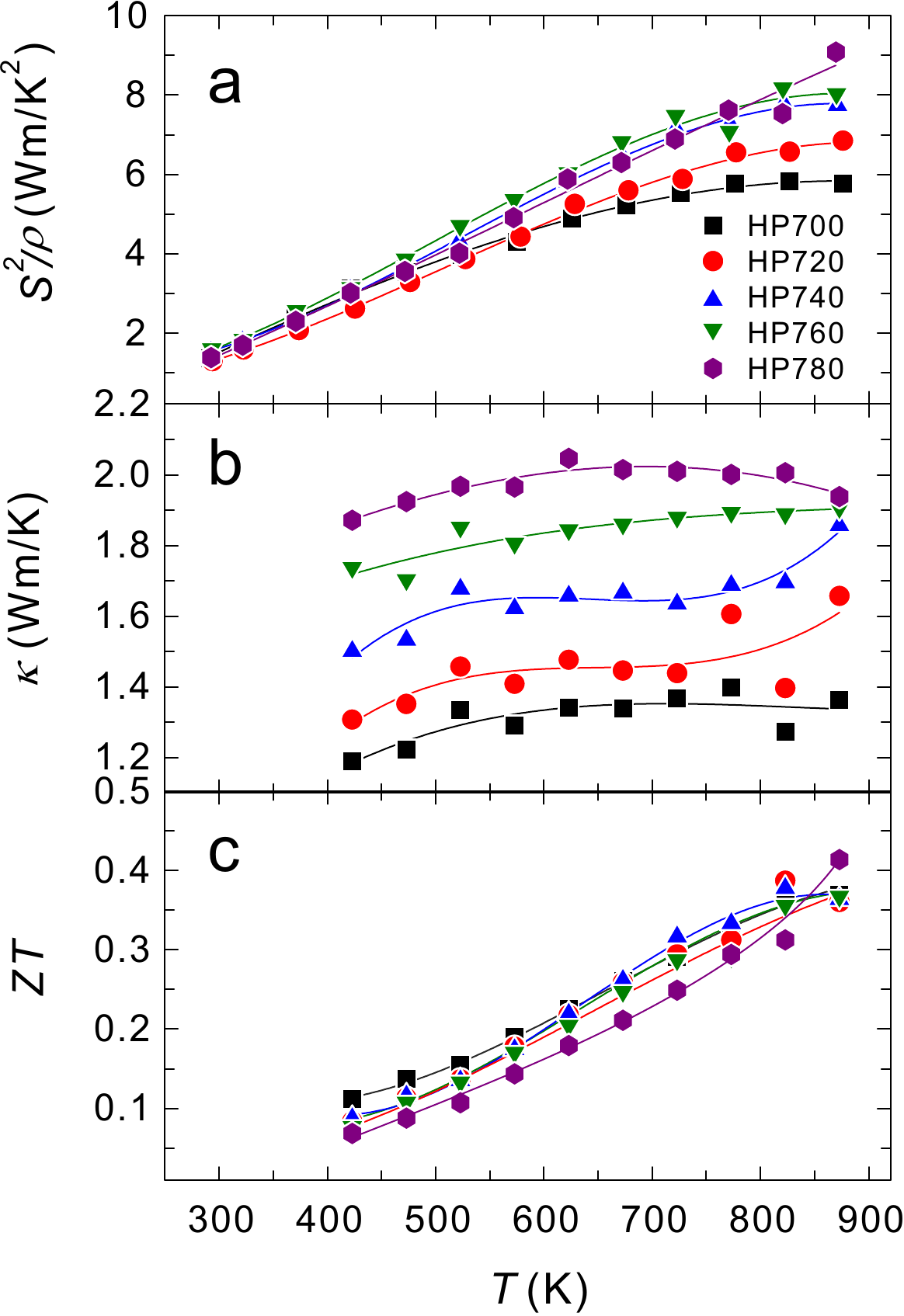}
\caption{%
Thermoelectric properties of sintered samples of batch I. Power factor $S^2/\rho$ vs temperature (a), thermal conductivity $\kappa$ vs temperature (b), and $ZT$ values (c). The samples are labelled by HP$T_{\rm HP}$.
}
\label{Fig:fig5}
\end{figure} 

\begin{figure}[ht]%
\centering
\includegraphics*[width=\columnwidth]{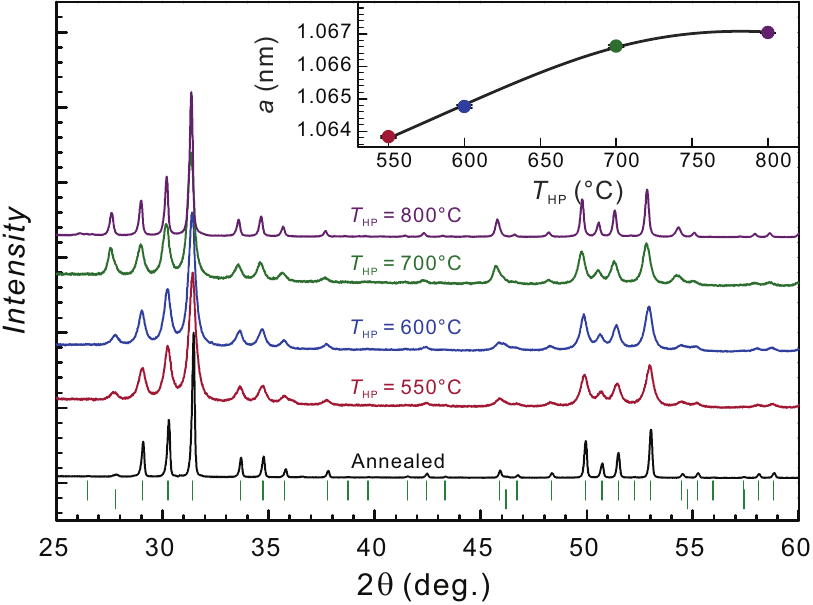}
\caption{XRD patterns of sintered samples from the nanopowders of batch II with different $T_{\rm HP}$. The $T_{\rm HP}$ dependence of the lattice parameters of the type-I clathrate is shown in the inset. The patterns were normalized according to the strongest peak. The intensity of the peak at around $27.5^\circ$, which belongs to the diamond phase, increases with increasing $T_{\rm HP}$. 
}
\label{Fig:fig6}
\end{figure}

Below 850 K, the $ZT$ values of the samples sintered with $700^\circ\mathrm{C}
\leq T_{\rm HP} \leq 760^\circ\mathrm{C}$ are higher than for the sample
sintered with $T_{\rm HP} = 780^\circ\mathrm{C}$. As all samples have similar
power factor this is mainly attributed to the lower thermal conductivity caused
by the smaller grain size (Fig.\,\ref{Fig:fig5}).

hotpressed samples using powder from batch II behave quite differently in terms
of stability and thermoelectric properties \cite{Yan15.1}. Firstly, the phase composition is
not stable during the hot pressing process, as evidenced by the change of the
lattice parameter of the clathrate phase and the increase of the peak intensity
of the diamond phase with $T_{\rm HP}$ (Fig.\,\ref{Fig:fig6}). Secondly, all
samples show semiconducting behavior with high electrical resistivities between
0.5 and 4\,$\Omega$cm at room temperature. The $ZT$ values are below 0.05 for
all samples (not shown). Additionally, the samples with the lowest bulk
densities are unstable during the transport measurements at high temperatures.
All this may at least in part be attributed to the composition instability.

\begin{figure}[h]%
\centering
\includegraphics*[width=\columnwidth]{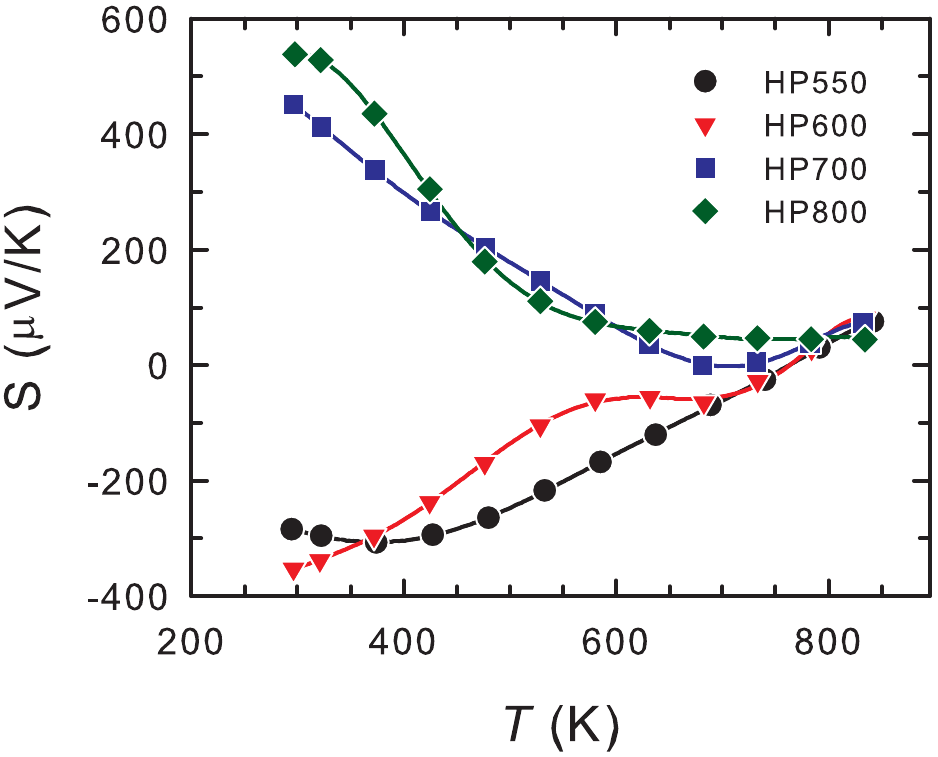}
\caption{Seebeck coefficient of the samples sintered from batch II.
}
\label{Fig:fig7}
\end{figure}

The change of composition during hot pressing most likely leads to an increase
of the Cu content and a decrease of the charge carrier concentration, leading to
the observed increased electrical resistivity and absolute Seebeck coefficient
values \cite{Yan12.1}. Within a rigid-band picture, this would shift the Fermi
level first downwards to the lower edge of the conduction band, and then further
downwards across the band gap to the top of the valence band. Indeed the Seebeck
coefficient changes from negative to positive with increasing $T_{\rm HP}$ (Fig.\,\ref{Fig:fig7}).

Interestingly, the samples sintered from the finer powders of batch II have much
lower $D_{\rm bulk}$ than those of batch I (Fig.\,\ref{Fig:fig3}). This seems
counterintuitive since finer powders should be easier to sinter due to their
higher surface energy and larger contact areas between particles
\cite{Hun09.1,Fec06.1,Gro06.1}. We suggest that the formation of contaminations
during ball milling and composition changes during hot pressing consume surface
energy. On the other hand, composition gradients between the shell and the core
of small particles and contaminations on the particle surfaces prevent particle
growth \cite{Gro06.1} and bonding between particles. Together, this leads to
smaller grain sizes but lower bulk densities of the samples hotpressed from the powder of batch II.

The crystallite size $S_{\rm c}$ of the samples hotpressed from the powder of
batch II remains very small even for $T_{\rm HP}$ up to $700^\circ\mathrm{C}$
(Fig.\,\ref{Fig:fig2}\,b). During annealing, a clear increase of the grain size
sets in already at about $500^\circ\mathrm{C}$. This indicates that pressure
indeed suppresses grain growth.

Based on the above results, we considered to use a process control agent during
ball milling to inhibit powder agglomeration, to use different powder
consolidation techniques, and to form nanocomposites for clathrates. Also we
simplified the clathrate from the optimized quaternary system to the ternary
system Ba-Cu-Si.

\subsection{Influence of process control agent} \label{PCA}

Samples of the nominal composition Ba$_8$Cu$_{4.8}$Si$_{41.2}$ used for this
investigation were previously shown to have the largest Cu solid solubility (at
$800^\circ\mathrm{C}$) \cite{Yan10.1} and the best thermoelectric performance \cite{Yan12.1}
within the ternary Ba-Cu-Si system. We used stearic acid,
CH$_3$(CH$_2$)$_{16}$CO$_2$H, as a process control agent (PCA). The aim is to
study how the PCA influences the behavior of grain size reduction during ball
milling and growth during hot pressing \cite{Zol14.1,Zol15.1}. The process
parameters for ball milling and hot pressing were fixed according to Sects.\,\ref{ballmilling} and
\,\ref{hotpressing}.    

\begin{figure}[h]%
\centering
\includegraphics*[width=\columnwidth]{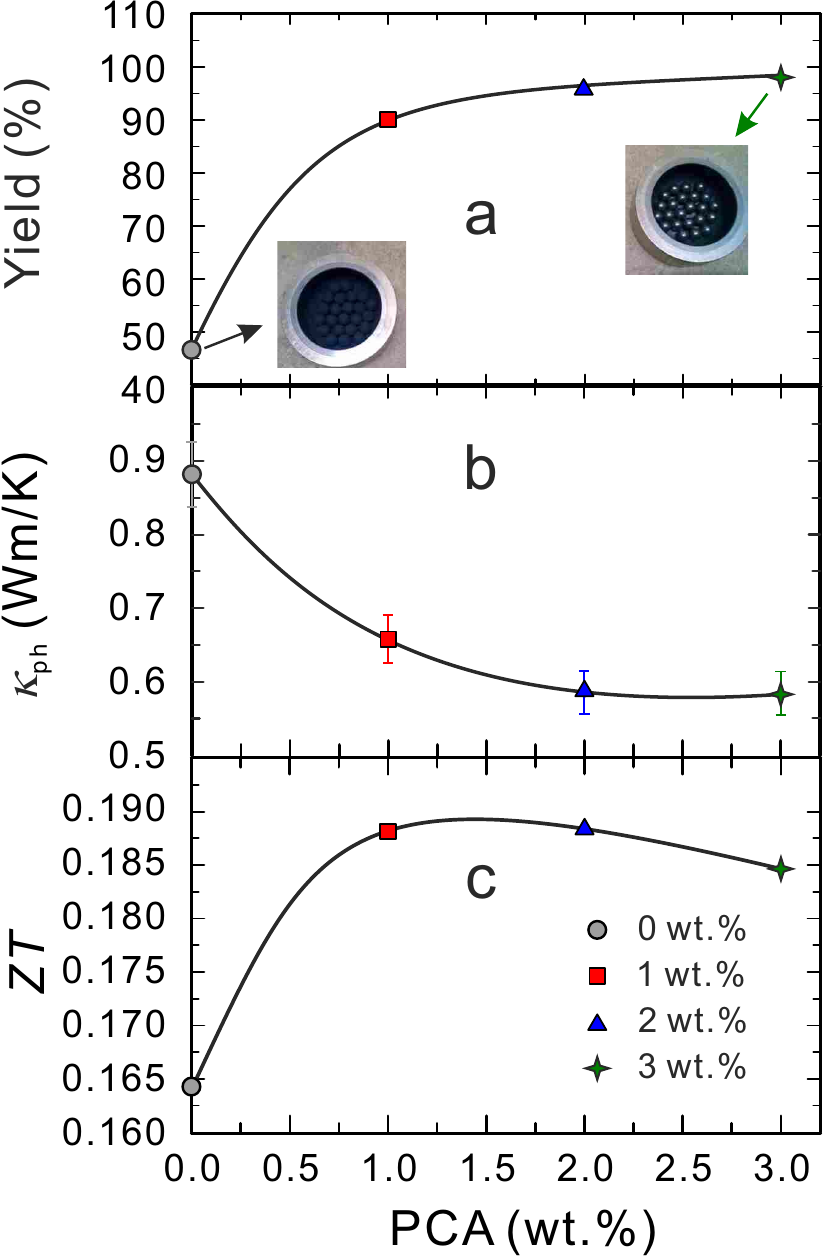}
\caption{Influence of PCA on the milling behavior and thermoelectric properties. Powder yield (a), phonon thermal conductivity $\kappa_{\rm ph}$ (b), and $ZT(450^\circ\mathrm{C})$ (c) vs PCA content. The appearance of containers and balls after removing the as-milled powders with or without PCA (inset of a).}
\label{Fig:fig8}
\end{figure}

The powder yield, which is the fraction of powder that can be collected after
the ball milling process, increases from 46\% without the PCA to 98\% with 3
wt.\% PCA, confirming that the PCA indeed reduces agglomeration effectively
(Fig.\,\ref{Fig:fig8} (a)). It also largely facilitates the removal of the
powders from the containers. 

After 5\,h milling treatment powders milled with the PCA have $S_{\rm c}\approx
50$\,nm, which is lower than for powders milled without the PCA (70\,nm) if the
powders are not repeatedly manually separated from the balls and containers
(batch I). However, it is larger than the $S_{\rm c}$ reached in batch II (with
manual powder separation). Thus the PCA is not as effective as mechanical
separation. Longer milling times might be required.

We had expected the stearic acid to evaporate during powder sintering by hot
pressing at $800^\circ\mathrm{C}$ because its boiling temperature is
$360^\circ\mathrm{C}$. However, TEM revealed very thin ($\sim$5\,nm) carbon
layers around the grains. This suggests that the stearic acid has (partially)
decomposed into carbon and gases to form these layers. They effectively prevent
grain growth during hot pressing, leading to lower thermal conductivities
(Fig.\,\ref{Fig:fig8}\,b), but are also responsible for increased electrical
resistivities. The $ZT$ values of the PCA-containing samples are overall higher
than that of the PCA-free sample in the whole temperature range (by 15\% higher
at $450^\circ\mathrm{C}$). The sample with 2 wt.\% PCA has the lowest lattice
thermal conductivity (Fig.\,\ref{Fig:fig8}\,b).  

\subsection{Powder consolidation by spark plasma sintering}

Spark plasma sintering (SPS) represents a promising powder consolidation
technique for the rapid densification at relatively low temperatures. An
electrical current is forced to flow through the sample during heating
\cite{Hun09.1,Mun06.1}. The low temperatures and short holding times are ideal
to limit grain growth. Compounds for this study \cite{Yan15.2} again have the
nominal composition Ba$_8$Cu$_{4.8}$Si$_{41.2}$. This study was performed at the
laboratory for Solid State Chemistry and Catalysis at EMPA, where annealed
samples were ballmilled and the as milled powders were sintered by SPS. In
parallel, annealed samples from the same batch were also ballmilled at the
Institute of Physical Chemistry at the University of Vienna and the as milled
powders were sintered by hot pressing. Unfortunately, because of the different
ball milling systems and different process parameters for milling, the
nanopowders had different grain/crystallite sizes: $S_{\rm c}\approx 100$\,nm
for the powders for the SPS treatment and $S_{\rm c}\approx 25(5)$\,nm for the
powders for hot pressing.

\begin{figure}[h]%
\centering
\includegraphics*[width=\columnwidth]{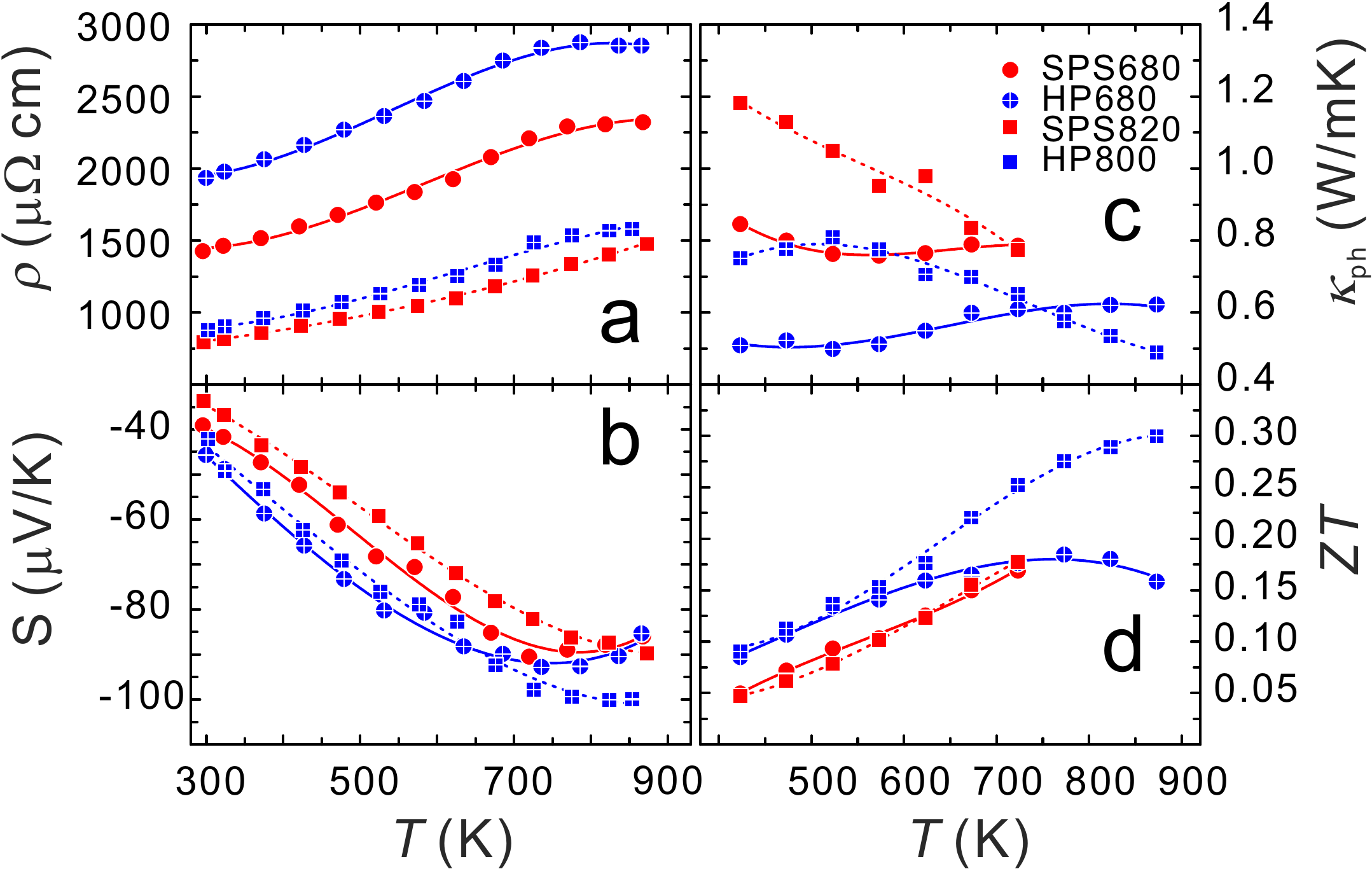}
\caption{Comparison of the thermoelectric properties of samples sintered by hot pressing and SPS. Electrical resistivity $\rho(T)$ (a), Seebeck coefficient $S(T)$ (b), lattice thermal conductivity $\kappa_{\rm ph}(T)$ (c), and figure of merit $ZT(T)$ (d). 
}
\label{Fig:fig10}
\end{figure}

Presumably due to the smaller crystallite size of the nanopowders, the samples
sintered by hot pressing have higher bulk densities and lower crystallite sizes
than the samples sintered by SPS with the same sintering temperature. In
addition, a white phase forms during the SPS process, which is rarely observed
in hotpressed samples. This white phase has a high Cu content and thus the Cu
content in the clathrate phase is reduced. The SPS treated samples have lower
$\rho(T)$ and $|S(T)|$ than the hotpressed samples (Fig.\,\ref{Fig:fig10}). The
hotpressed samples have higher $ZT$ than the SPS treated samples. Also the bulk
density plays an important role. The highest $ZT$ of 0.3 at 870\,K is reached in
a sample with a very high bulk density of 99\%. 

\section{Nanocomposites}

A limitation of bulk nanostructured thermoelectric materials is their thermal
instability with respect to coarsening. An approach to overcome this problem is
to use bulk nanostructured composites in which one of the constituting phases
serves as spacer between the nanograins of the active thermoelectric medium. To
prevent coarsening the spacer and active material should be of different chemical nature. Here we studied two different spacer materials, thermoelectric oxides and SiC.

\subsection{Clathrate-oxide nanocomposites}

Ideally, the spacer material is itself a thermoelectric material. Therefore, in
a first step, we chose thermoelectric oxides. The first task was to find a
pair of mutually chemically inert materials. Among a series of thermoelectric
oxides the system EuTiO$_{3-\delta}$ was selected because it shows no
traces of chemical reaction with the clathrate phases Ba$_8$Pd$_4$Ge$_{40}$ and Ba$_8$Ga$_{16}$Ge$_{30}$ selected for this study up to 800$^{\circ}$C.

\begin{figure}[h]%
\centering
\includegraphics*[width=\columnwidth]{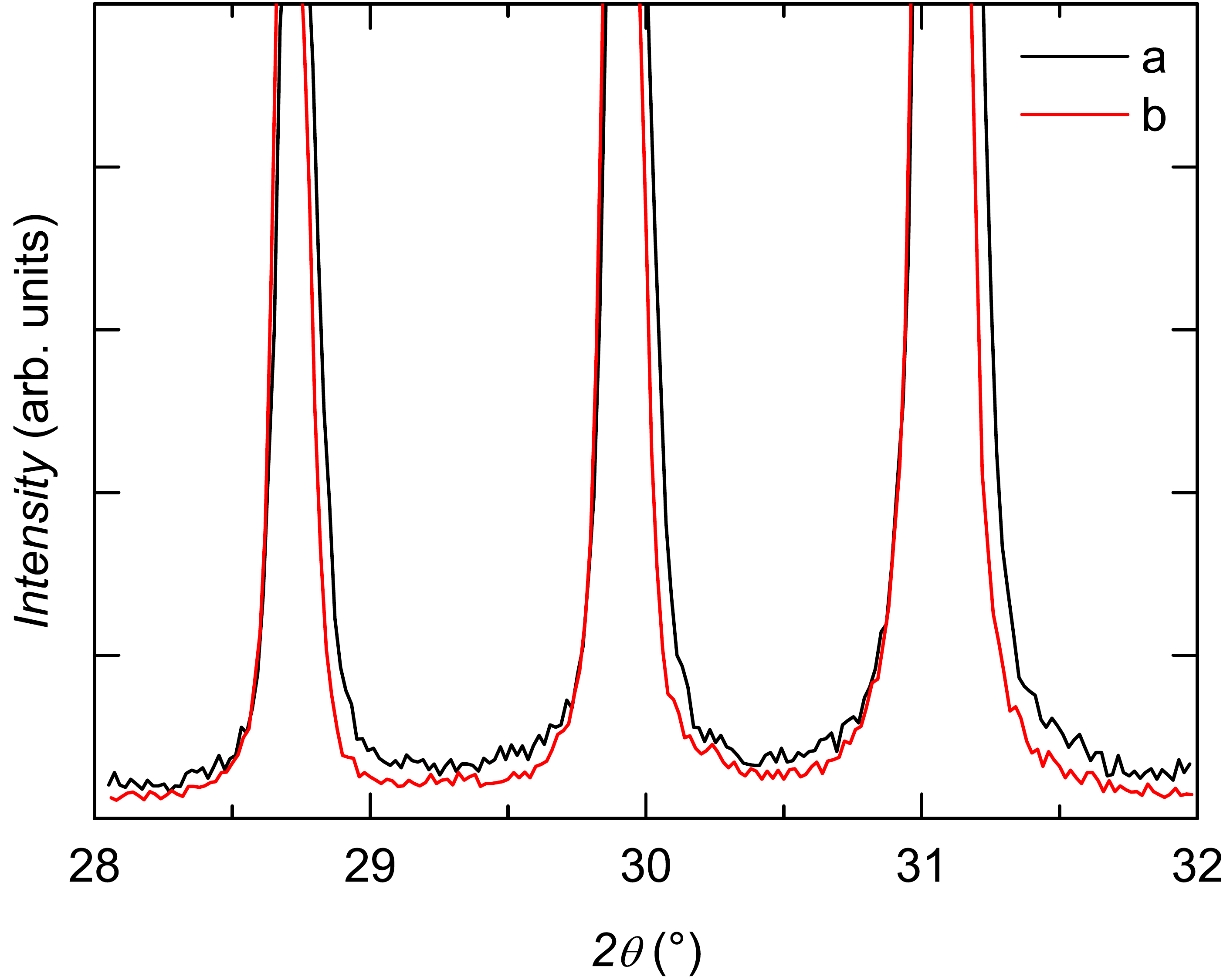}
\caption{XRD patterns of the clathrate-oxide composite (a) and the clathrate sample (b) after treatment at 500$^{\circ}$C. To show that the peaks are broader for the composite only a narrow range of angles is shown around three peaks of the clathrate structure.}
\label{prok1}
\end{figure}

For the investigation of the coarsening two series of samples were prepared: a pure Ba$_8$Ga$_{16}$Ge$_{30}$ nanostructured clathrate and the nanocomposite Ba$_8$Ga$_{16}$Ge$_{30}$ + EuTiO$_{3-\delta}$ (13\%), both ballmilled and SPS compacted at Empa. The samples were thermally treated stepwise at 500 and 650$^{\circ}$C (20 hours), and the grain size was evaluated by XRD after each step (Fig.\,\ref{prok1}). At 500$^{\circ}$C the crystallite size of the pure clathrate material and the clathrate phase in the composite were nearly equal whereas after the treatment at 650$^{\circ}$C the clathrate phase in the composite sample showed about twice smaller crystallite size.

\begin{figure}[h]%
\centering
\includegraphics*[width=\columnwidth]{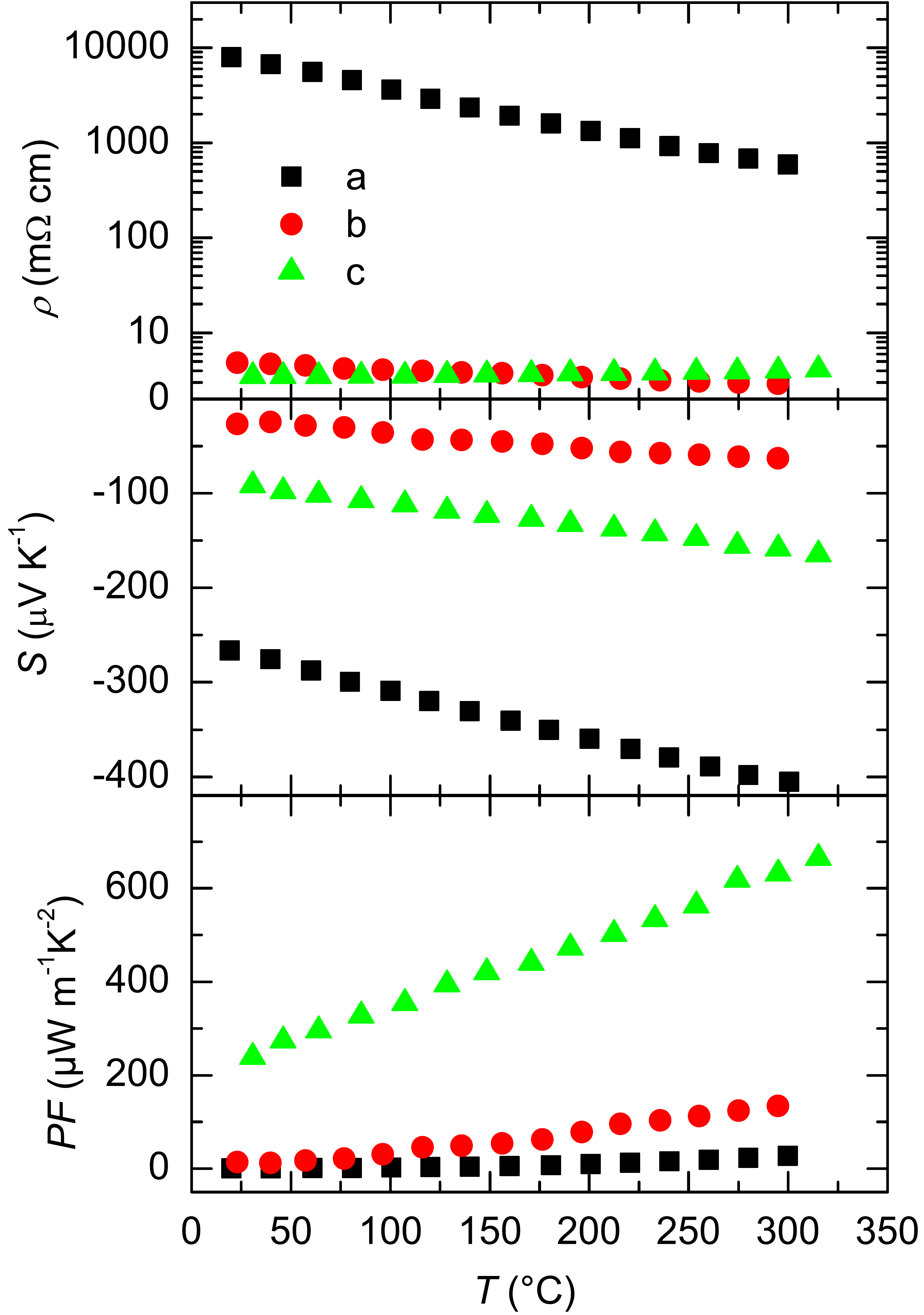}
\caption{Electrical resistivity (top), thermopower (center), and powerfactor (bottom) of the oxide (a), the clathrate-oxide nanocomposite (b), and the clathrate (c) samples (see text).}
\label{prok2}
\end{figure}

In Fig.\,\ref{prok2} we compare the thermoelectric properties of the composite
to those of the two constituents. The powerfactor $PF$ of the composite is
larger than that of the oxide, but smaller than that of the clathrate. This
disappointing result is mostly due to the unexpectedly low thermopower of the
composite. It is lower than that of both constituents, which suggests that at
least one of the two phases was degraded by the mixing process. As XRD rules out
the formation of sizable amounts of foreign phases a change in composition
within the oxide or clathrate phase should be evoked. Indeed, above
200$^{\circ}$C, the electrical resistivity of the composite is lower that that
of both constituents, suggesting that the clathrate and/or the oxide are sizably
doped in the composite. The strong effect 13\,\% oxide phase have on the
composites properties is encouraging because it indicates that the oxide is very
well distributed between clathrate grains. If the undesired doping effect can be
suppresses, this route could thus prove useful.

\subsection{Clathrate-SiC nanocomposites}

Due to the above described undesired doping effect in the clathrate-oxide
composites, the use of a passive spacer material was explored as an alternative
route. For this purpose, we mixed ballmilled Ba$_8$Cu$_{4.8}$Si$_{41.2}$
nanopowders with thermally and chemically stable SiC nanoparticles with a grain
size of 40-50\,nm. Different amounts of SiC (0, 1, and 3 wt.\%) were first mixed
with hand-ground clathrate powders and then ballmilled for 8\,h. 2 wt.\% of
stearic acid was used as a process control agent (PCA) in all samples to prevent
powder agglomeration.

\begin{table}[h]
 \caption{\label{tab1} Sample code, crystallite size $S_{\rm c}$, residual stress $\epsilon$, and bulk density $D_{\rm B}$ for samples produced by different processes: AM = as milled powders, AS = as sintered bulks, MHT = bulks after transport measurements at high temperatures. The number in the code after SiC represents the mass percentages of SiC in the sample.}
\centering
 \begin{tabular}{r c c c}
\hline
\hline
 Code & HP.SiC0  & HP.SiC1 & HP.SiC3\\
\hline
 AM: $S_{\rm c}$ (nm)  & 36.2(5)  & 40.6(4) & 35.7(6)\\
 $\epsilon$ (\%)  & 0.18(1)  & 0.13(1) & 0.18(1)\\
\hline
 AS: $S_{\rm c}$ (nm)  & 69.0(5)  & 75.5(6) & 70.9(6)\\
 $\epsilon$ (\%)   & 0.006(1)  & 0.005(1) & 0.015(1)\\ 
 $D_{\rm B}$ (\%)  & 96  & 90 & 92\\
\hline
 MHT: $S_{\rm c}$ (nm)  & 88.2(8)  & 87(1) & 78(1)\\
 $\epsilon$ (\%)   & 0.035(1)  & 0.030(1) & 0.032(1)\\ 
 \hline

\hline
 \end{tabular}
\end{table}

Table\,\ref{tab1} gives the crystallite size $S_{\rm c}$, residual stress
$\epsilon$, and relative bulk density $D_{\rm B}$ of samples in different
states. $S_{\rm c}$ of the as milled nanopowders is essentially independent of
the amount of SiC. Thus the presence of SiC does not seem to influence the
crystallite size. After sintering, $S_{\rm c}$ is almost doubled but again
independent of SiC. However, after high-temperature transport measurements, the
sample with 3 wt.\% SiC has a lower $S_{\rm c}$ than the other two samples,
indicating that the SiC particles reduced the grain growth of the host
clathrate. 

\begin{figure}[ht]%
\centering
\includegraphics*[width=\columnwidth]{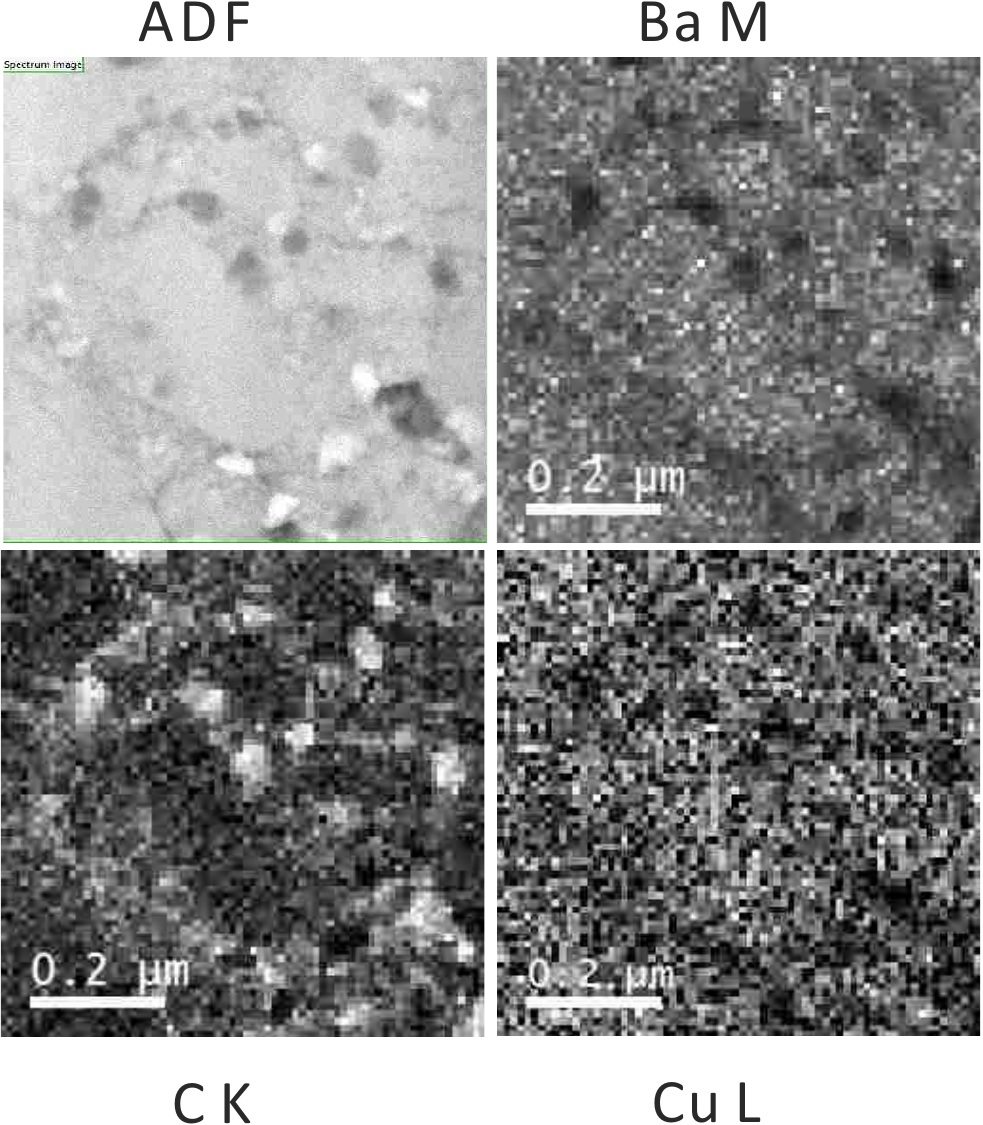}
\caption{Annular dark field (ADF) image of HP.SiC3 (top left) and corresponding
element maps of Ba, C, and Cu (see labels) measured by electron energy loss spectroscopy.}
\label{Fig:fig11}
\end{figure} 

The grain growth behavior during powder sintering and thermoelectric property
measurements at high temperatures might also depend on the amount and
distribution of secondary phases in the host material. This effect was
investigated by transmission electron microscopy (TEM). Figure\,\ref{Fig:fig11}
shows an annular dark field (ADF) image of the sample HP.SiC3 and element maps
for the green square area measured by electron energy loss spectroscopy (EELS).
As can be seen, Ba and Cu are uniformly distributed, evidenced by the quite
uniform color in the maps, except for some black spots/bands. These correspond
to white color in C (carbon) maps, indicating high C contents in these areas.
The spots/bands correspondingly have black color in the ADF image. In this way
we can identify the distributions of the C-rich phase(s) from ADF images.

\begin{figure}[h]%
\centering
\includegraphics*[width=\columnwidth]{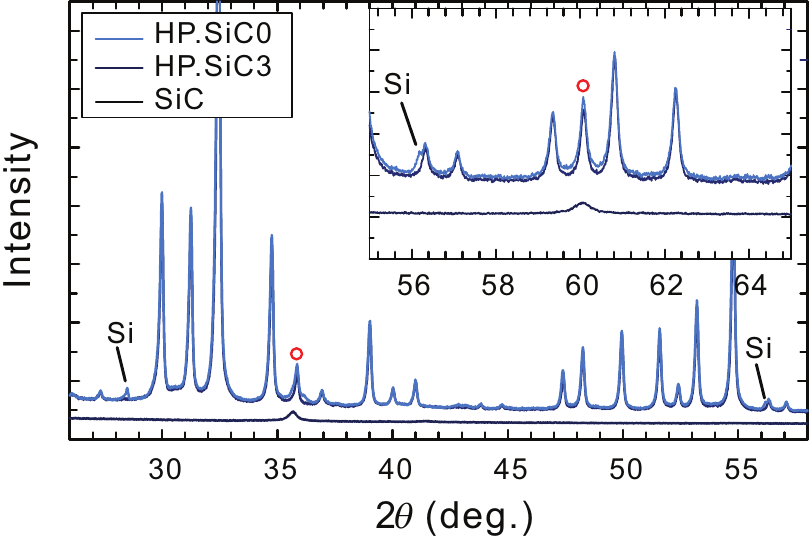}
\caption{Comparison of the XRD patterns of HP.SiC0, HP.SiC3, and SiC. The SiC in HP.SiC3 only can be confirmed by very subtle features in the picks indicated by red circles. 
}
\label{Fig:fig12}
\end{figure}

\begin{figure}[ht]%
\centering
\includegraphics*[width=\columnwidth]{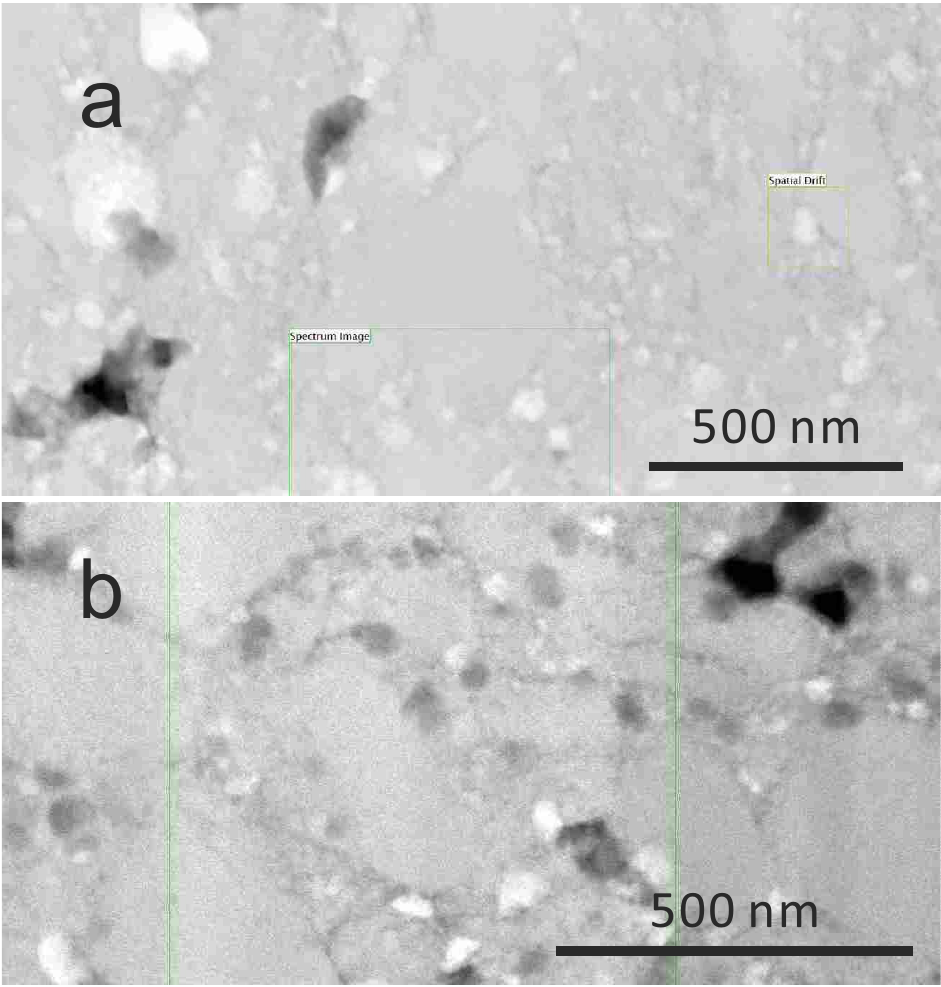}
\caption{ADF images of HP.SiC0 (a) and HP.SiC3 (b). The line/band-like black phase is seen in both HP.SiC0 and HP.SiC3 whereas the ball-like black phase is only observed in HP.SiC3.
}
\label{Fig:fig13}
\end{figure}

To measure the composition of the black phase by EDX is more difficult,
especially with the attempt to confirm if the black phase is SiC, since Si also
exists in the clathrate phase. By XRD, only from a very careful comparison of
the patterns for HP.SiC0, HP.SiC3, and SiC, one can see tiny differences in
shape and in intensity in the overlapped peaks of the clathrate phase and the
SiC phase between HP.SiC0 and HP.SiC3 (Fig.\,\ref{Fig:fig12}). This indicates
that SiC does not react with the clathrate phase and instead exists in the
HP.SiC3 sample. Comparing the ADF images of HP.SiC0 and HP.SiC3 (Fig.\,\ref{Fig:fig13}) we conclude that
the black phase of line/band-like shape, which can be also observed in the
former sample, is the stearic acid (SA) and/or a phase decomposed from SA at
high temperatures during hot pressing. The black phase in ball shape, which is
only observed for the sample HP.SiC3, is SiC. From their different phase
distributions -- the SA phase surrounds the clathrate phase while SiC are
distributed as separate grains -- one can explain why the crystallite size is
essentially independent of the SiC content. It is mostly the SA phase which
determines the behavior of grain reduction and growth. 

\begin{figure}[h]%
\centering
\includegraphics*[width=\columnwidth]{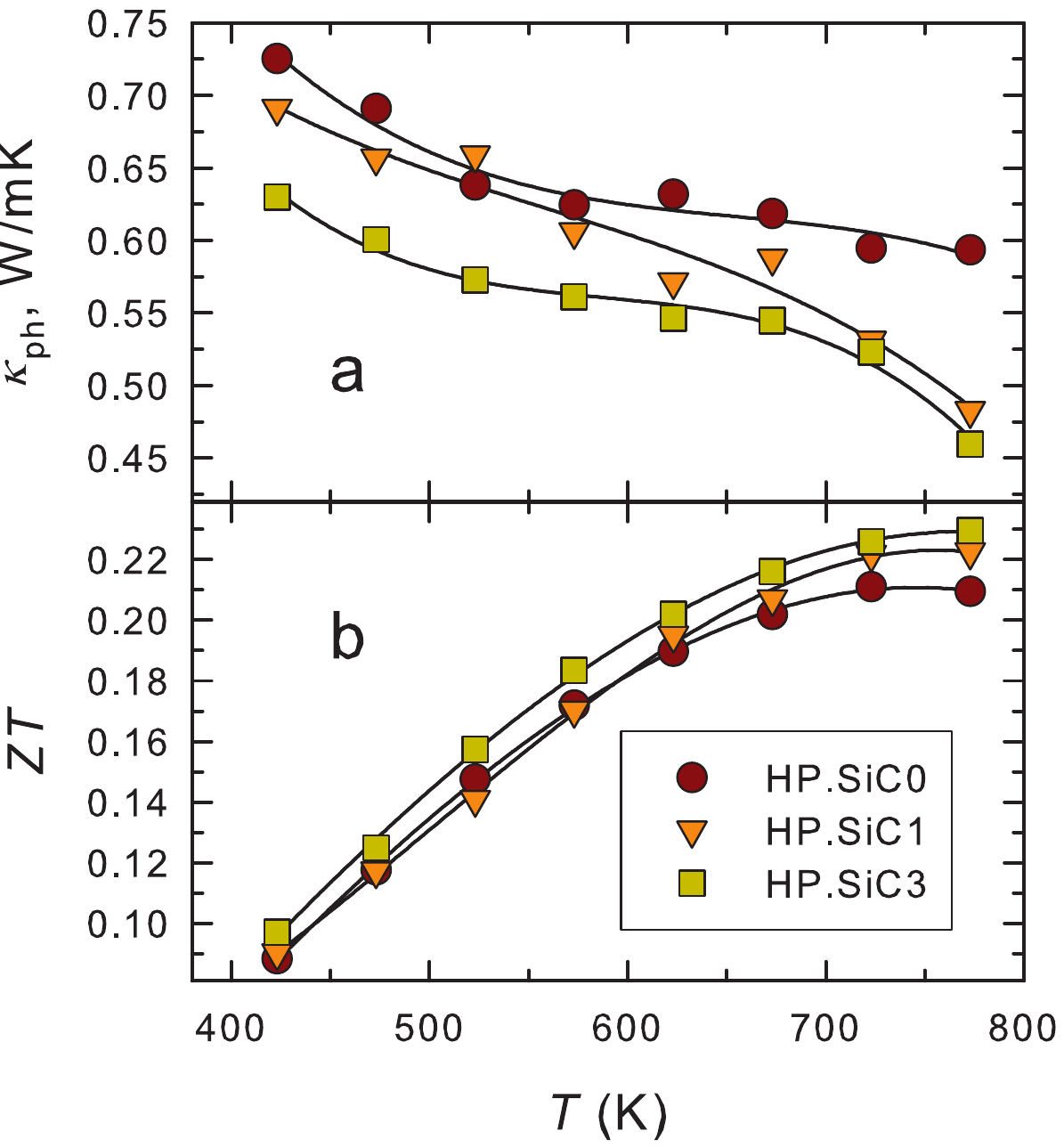}
\caption{Thermoelectric properties of nanocomposites. Lattice thermal conductivity $\kappa_{\rm ph}(T)$ (a) and $ZT(T)$ (b). 
}
\label{Fig:fig14}
\end{figure}

\begin{figure}[h]%
\centering
\includegraphics*[width=\columnwidth]{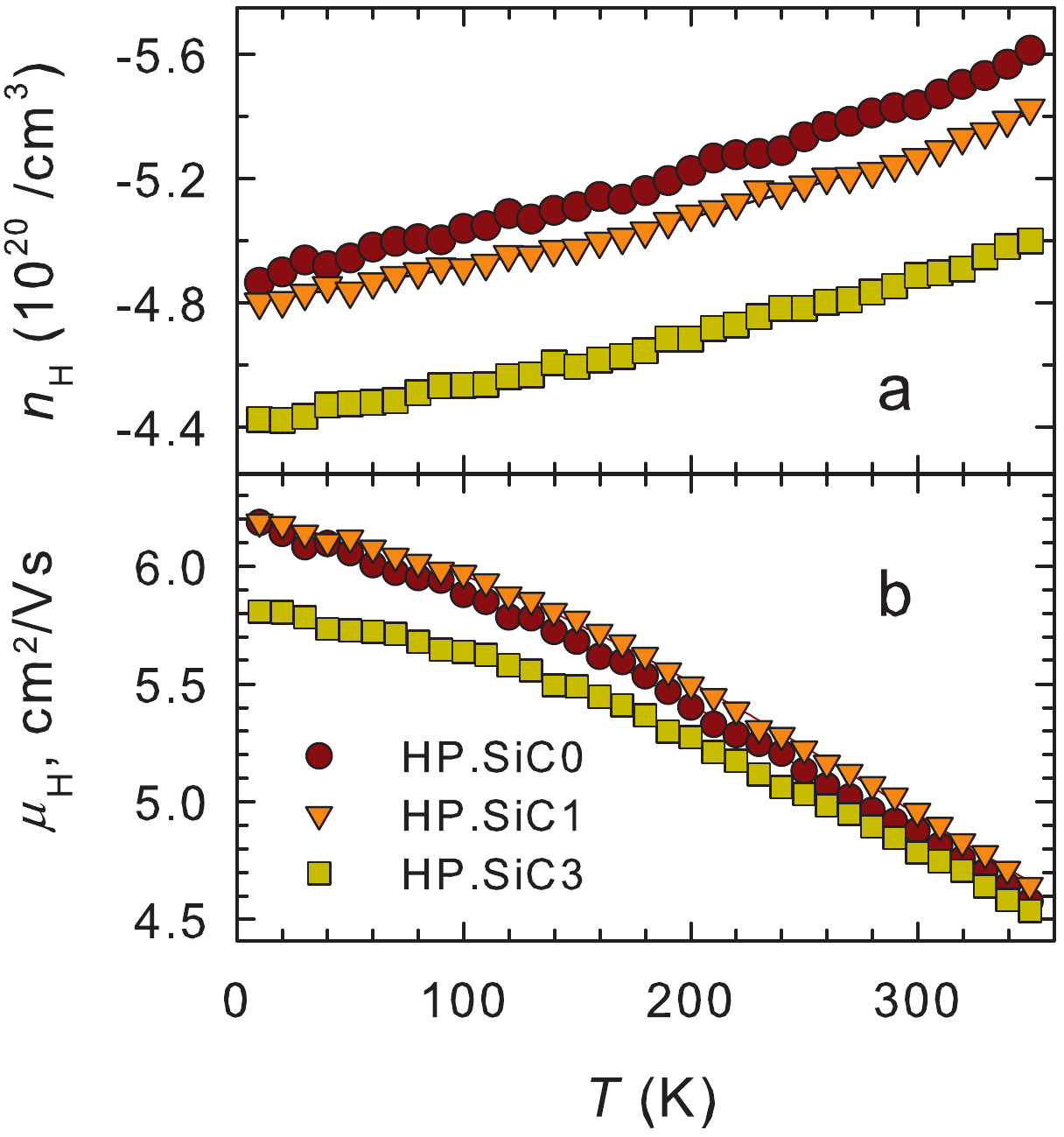}
\caption{Charge carrier concentration $n_{\rm H}$ (a) and charge carrier mobility $\mu_{\rm H}$ (b) vs $T$. 
}
\label{Fig:fig15}
\end{figure}

As a consequence of the SA phase around the nanoparticles, these samples have
high electrical resistivity and low thermal conductivity compared with the
samples prepared without SA during ball milling (see e.g., Fig.\,\ref{Fig:fig14}
and Fig.\,\ref{Fig:fig10}, HP800). The SiC nanoparticles add additional
scattering centers and further reduce the lattice thermal conductivity
(Fig.\,\ref{Fig:fig14}\,a). These nanoparticles also scatter charge carriers,
leading to a reduced charge mobility $\mu_{\rm H}$ (Fig.\,\ref{Fig:fig15}\,b),
which plays a dominant role on the increased electrical resistivity. The Hall
effect measurements revealed that the unexpectedly reduced charge carrier
concentration in the HP.SiC3 sample is also held responsible for the high
electrical resistivity. Correspondingly, the $|S(T)|$ of this sample is higher
than that of the other samples because of the lower charge carrier concentration
(Fig.\,\ref{Fig:fig15}\,a). The SiC-containing samples have slightly higher $ZT$
values due to the lower $\kappa_{\rm ph}(T)$ and higher $|S(T)|$ (not shown).

\begin{figure*}[t]
  \includegraphics*[width=\textwidth]{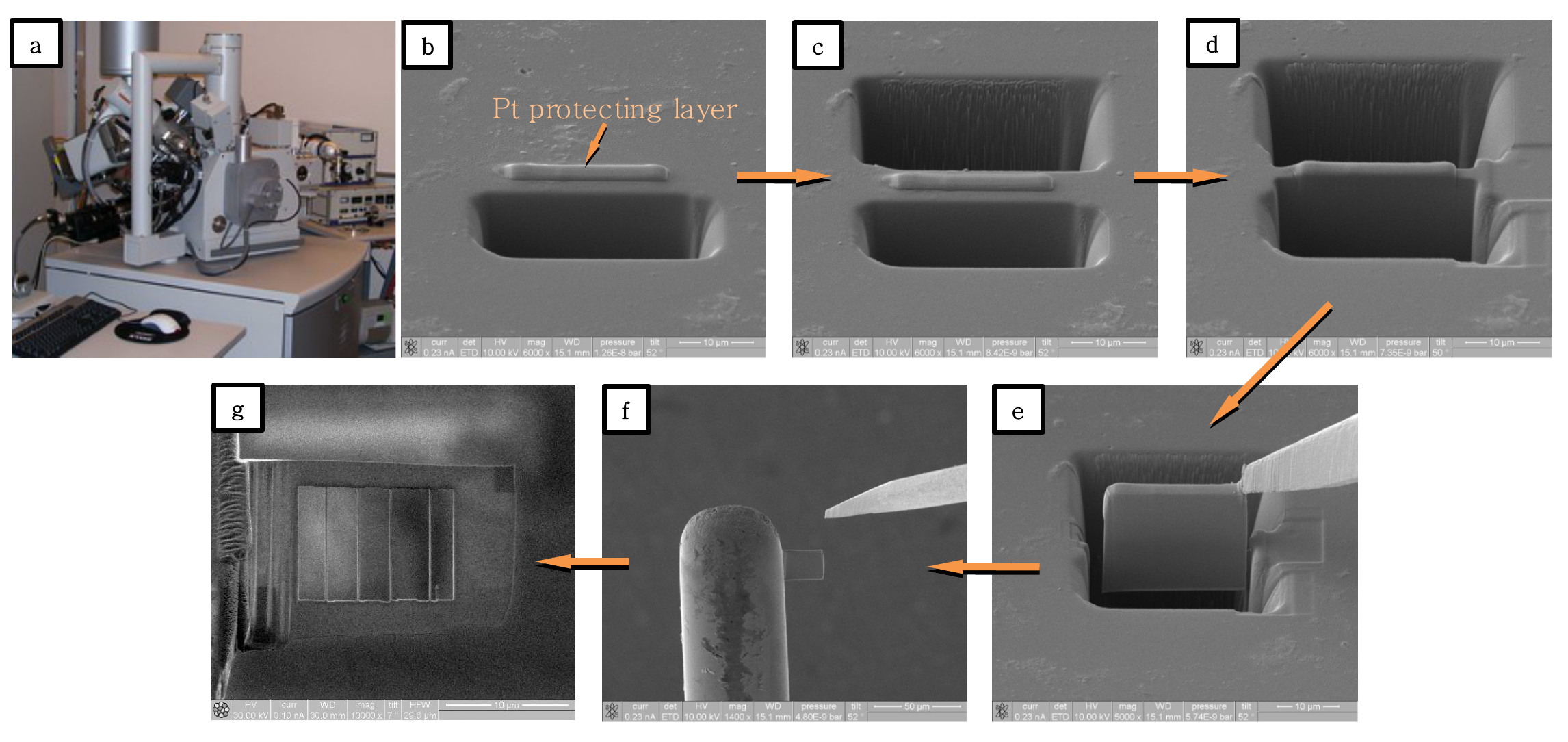}
  \caption{Quanta\,200\,3D Dual Beam-FIB at the USTEM service center at TU Wien (a), with which a lamella is cut out of a bulk intermetallic compound single crystal (b-d), transferred to a TEM grid with a nanomanipulator (e-f), and from which subsequently nanowires are defined (g).}
  \label{FIB}
\end{figure*}

\section{Nanowires}

Understanding the effects of nanostructuring by investigations on bulk nanostructured materials is challenging because usually more than one parameter (e.g., grain size, material's composition, doping level, amount of foreign phases) is changed at once. Thus it is advisable to study also simpler model systems. Here we report on our efforts to produce devices with clathrate nanowires of well-defined geometrical dimensions. Since there is no known route for the bottom-up fabrication of clathrate nanowires, we fabricate them with a top-down approach. In short, a focused ion beam (FIB) is used to sculpture nanowires out of a bulk starting material, which can be a single- or poly-crystalline sample with large enough grain size. This process is performed with the FEI Quanta 200 3D Dual Beam-FIB shown in Fig.\,\ref{FIB}a and proceeds as follows. The first step is to cut out a lamella such as is typically performed for transmission electron microscopy (TEM) sample preparation. To protect the sample surface, a platinum metal layer is deposited over the area of interest (Fig.\,\ref{FIB}b). Two trenches, one on each side of the metal line, are then carved out so as to define a lamella, which is then taken out using a lift-out technique by means of 'soldering' the lamella to a micromanipulator needle with electron-beam induced deposition (EBID) of tungston, by tilting, and by further cutting the lamella free (Fig.\,\ref{FIB}c-e). Finally the lamella is transferred to a TEM grid (Fig.\,\ref{FIB}f). While this is, in essence, a standard technique, our process is taken one step further, and from the lamella nanowires are formed as evinced in Fig.\,\ref{FIB}g.

Inevitably, an amorphous surface layer is produced during FIB-cutting by the impinging Ga$^+$ ions that have a typical energy of 30\,keV and penetrate the top of the sample surface. This layer can however be thinned down by low-energy ($~$1\,keV argon (Ar$^+$) milling. The whole process has been perfected such that we can produce micro- and nanowires with lengths of $>10\,\mu$m and diameters of $<150\,nm$.

\begin{figure*}[htb]
  \includegraphics*[width=\textwidth]{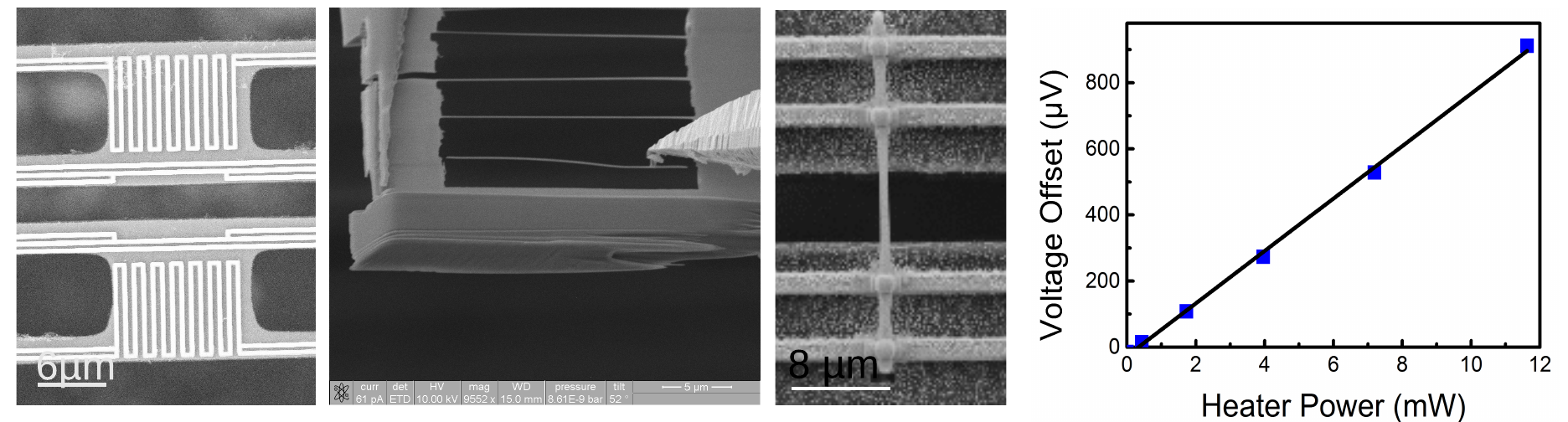}
  \caption{Measurement platform consisting of two membrane islands suspended only via thin and long beams with heaters and electrodes (left), transfer of a nanowire onto a TNCP chip with the FIB micromanipulator (middle), and raw measurement data for the determination of the Seebeck coefficient (right).}
  \label{platform}
\end{figure*}

Figure\,\ref{platform}, left, shows a microfabricated measurement platform similar to other published works \cite{Shi03.1,Karg14}. It consists of two separate silicon nitride membrane islands, across which the nanowire can be positioned. The islands are suspended by four beams that carry the input leads for the
heaters, for the temperature sensors that are directly attached to the nanowire, and for the four-probe electrical measurements of the nanowire. The innermost electrodes serve both as voltage probes in the four-point measurements of the nanowire and as thermometer. The fabrication process involves dry etching and anisotropic KOH wet etching on the back side of the substrate. Contact pads are defined by photolithography, and leads, heaters, and electrodes are defined
by electron beam lithography (EBL), as is the final etch mask that defines the islands and their supporting beams.

While we have not yet performed measurements on such a device, we managed to
transfer a nanowire onto a thermoelectric nanowire characterization platform
(TNCP) chip \cite{Wang12} as shown in Fig.\,\ref{platform}, middle, again by
means of EBID and FIB-cutting. Preliminary raw data for a room temperature
Seebeck coefficient measurement on a Ba$_8$Ga$_{16}$Ge$_{30}$ microwire are
shown in Fig.\,\ref{platform}, right. We also have built up a 3$\omega$-setup
for nanowires on which we plan to measure thermal conductivity in the near
future.

\section{Summary and conclusions}\label{5}

We have investigated the effect of nanostructuring on thermoelectric clathrates by different techniques: melt spinning, ball milling, composite formation, and the fabrication of single nanowires.

Our investigation of meltspun clathrates revealed that, at least with our
current setup, clathrate flakes with grain sizes below 1\,$\mu$m do not form
readily. They can, however, be enforced by additional actions such as severe
alloying. This is in strong contrast to our findings for the skutterudite
CoSb$_3$, where a systematic dependence of the grain size and the lattice
thermal conductivity on the cooling rate during the melt spinning process could
be demonstrated. We attribute this difference to the extremely high formation
rate of clathrates. This is also consistent with the fact that meltspun
clathrates are essentially phase pure, whereas meltspun CoSb$_3$ contains a
sizable amount of foreign phases.

Ball milling, by contrast, is a very effective means to produce nanogranular
clathrates with decreased lattice thermal conductivities. Here, the main
challenge is to prevent grain growth during the compaction process. We studied
in detail the influence of hot pressing parameters on the properties of the
compacted bulk samples. The bulk density of the samples increases with
increasing hot pressing temperature but at the same time the grains grow. Thus,
the best hot pressing temperature can only be a compromise. We also studied the
effect of a process control agent. It strongly enhances the powder yield during
ball milling and reduces grain growth during hot pressing. Overall, an
enhancement of the figure of merit by 15\,\% could be achieved with only
1\,wt.\% of stearic acid. Further improvements are expected if an agent can be
found that leaves less traces after compaction.

To prevent grain growth during compaction and improve the long-term stability of
nanostructured clathrates, we studied two different nanocomposites: a
clathrate-oxide composite and a clathrate-SiC composite. Even though the
constituent clathrate and oxide were selected because of their mutual phase
stability, the composite showed much poorer thermoelectric performance than the
clathrate alone. We speculate that this is due to a mutual doping effect at the
clathrate-oxide interfaces. SiC by contrast is entirely inert and lead to a
10\,\% improvement of the figure of merit. The long term stability remains to be
tested.

Finally, we set up a procedure to measure the thermoelectric properties of
single clathrate nanowires. A clathrate nanowire is cut out of a macroscopic
clathrate single crystal by the focused ion beam and argon milling technique,
and placed on a suspended platform with integrated heaters and thermometers. The
setup is now ready for measurements. In particular, clathrates with different nanowire diameter shall be studied. We expect the comparison of the size effect
with theoretical calculations \cite{Mad15.1} to shed further light on the intrinsic thermal transport mechanisms, which will ultimately help to further optimize bulk nanostructured clathrates.

\begin{acknowledgement} We thank A.\ Grytsiv for hot pressing and M.\ Waas for
SEM/EDX investigations. The work was supported by the German Research Foundation
(DFG project SPP 1386 nanOcla), the Austrian Science Fund (FWF project TRP 176
N22), the Christian Doppler Forschungsgesellschaft (CD-Labor f\"ur
Thermoelektrizit\"at), and the European Integrated Center for the Development of
New Metallic Alloys and Compounds (C-MAC). \end{acknowledgement}


\begin{thebibliography}{[10]}

\bibitem{Bou08.1}
 \textsc{A.\,I. Boukai},  \textsc{Y.~Bunimovich},  \textsc{J.~{Tahir-Kheli}},
  \textsc{J.\,K. Yu},  \textsc{W.\,A. {Goddard Iii}},  and  \textsc{J.\,R.
  Heath},
 \jr{{Nature}} \textbf{451}, {168} (2008).


\bibitem{Hoc08.1}
 \textsc{A.\,I. Hochbaum},  \textsc{R.~Chen},  \textsc{R.\,D. Delgado},
  \textsc{W.~Liang},  \textsc{E.\,C. Garnett},  \textsc{M.~Najarian},
  \textsc{A.~Majumdar},  and  \textsc{P.~Yang},
 \jr{{Nature}} \textbf{451}, 163 (2008).


\bibitem{2008Jos}
 \textsc{G.~Joshi},  \textsc{H.~Lee},  \textsc{Y.~Lan},  \textsc{X.~Wang},
  \textsc{G.~Zhu},  \textsc{D.~Wang},  \textsc{R.\,W. Gould},  \textsc{D.\,C.
  Cuff},  \textsc{M.\,Y. Tang},  \textsc{M.\,S. Dresselhaus},
  \textsc{G.~Chen},  and  \textsc{Z.~Ren},
 \jr{Nano Lett.} \textbf{8}, 4670 (2008).


\bibitem{2008Ma}
 \textsc{Y.~Ma},  \textsc{Q.~Hao},  \textsc{B.~Poudel},  \textsc{Y.~Lan},
  \textsc{B.~Yu},  \textsc{D.~Wang},  \textsc{G.~Chen},  and
  \textsc{Z.~Ren},
 \jr{Nano Letters} \textbf{8}(8), 2580--2584 (2008),
PMID: 18624384.


\bibitem{Pou08.1}
 \textsc{B.~Poudel},  \textsc{Q.~Hao},  \textsc{Y.~Ma},  \textsc{Y.~Lan},
  \textsc{A.~Minnich},  \textsc{B.~Yu},  \textsc{X.~Yan},  \textsc{D.~Wang},
  \textsc{A.~Muto},  \textsc{D.~Vashaee},  \textsc{X.~Chen},  \textsc{J.~Liu},
  \textsc{M.\,S. Dresselhaus},  \textsc{G.~Chen},  and  \textsc{Z.~Ren},
 \jr{Science} \textbf{320}, 634 (2008).


\bibitem{Wan08.2}
 \textsc{X.\,W. Wang},  \textsc{H.~Lee},  \textsc{Y.\,C. Lan},  \textsc{G.\,H.
  Zhu},  \textsc{G.~Joshi},  \textsc{D.\,Z. Wang},  \textsc{J.~Yang},
  \textsc{A.\,J. Muto},  \textsc{M.\,Y. Tang},  \textsc{J.~Klatsky},
  \textsc{S.~Song},  \textsc{M.\,S. Dresselhaus},  \textsc{G.~Chen},  and
  \textsc{Z.\,F. Ren},
 \jr{{Appl.\ Phys. Lett.}} \textbf{93}, 193121 (2008).


\bibitem{Min09.1}
 \textsc{A.\,J. Minnich},  \textsc{M.\,S. Dresselhaus},  \textsc{Z.\,F. Ren},
  and  \textsc{G.~Chen},
 \jr{{Energy Environ.\ Sci.}} \textbf{2}, 466 (2009).


\bibitem{Lau11.1}
 \textsc{S.~Laumann},  \textsc{M.~Ikeda},  \textsc{H.~Sassik},
  \textsc{A.~Prokofiev},  and  \textsc{S.~Paschen},
 \jr{{J.\ Mater.\ Res.}} \textbf{26}, 1861 (2011).


\bibitem{Lau12.1}
 \textsc{S.~Laumann},  \textsc{M.~Ikeda},  \textsc{H.~Sassik},
  \textsc{A.~Prokofiev},  and  \textsc{S.~Paschen},
 \jr{{Z.\ Anorg.\ Allg.\ Chem.}} \textbf{638}, 294 (2012).


\bibitem{Li08.2}
 \textsc{H.~Li},  \textsc{X.~Tang},  \textsc{X.~Su},  and
  \textsc{Q.~Zhang},
 \jr{{Appl.\ Phys.\ Lett.}} \textbf{92}, 202114 (2008).


\bibitem{Li09.1}
 \textsc{H.~Li},  \textsc{X.~Tang},  \textsc{X.~Su},  \textsc{Q.~Zhang},  and
  \textsc{C.~Uher},
 \jr{{J.\ Phys.\ D: Appl.\ Phys.}} \textbf{42}, 145409 (2009).


\bibitem{Su12.1}
 \textsc{X.~Su},  \textsc{H.~Li},  \textsc{Y.~Yan},  \textsc{G.~Wang},
  \textsc{H.~Chi},  \textsc{X.~Zhou},  \textsc{X.~Tang},  \textsc{Q.~Zhang},
  and  \textsc{C.~Uher},
 \jr{{Acta Mater.}} \textbf{60}, {3536} (2012).


\bibitem{Tan07}
 \textsc{X.~Tang},  \textsc{W.~Xie},  \textsc{H.~Li},  \textsc{W.~Zhao},
  \textsc{Q.~Zhang},  and  \textsc{M.~Niino},
 \jr{Appl. Phys. Lett.} \textbf{90}, 012102 (2007).


\bibitem{Pro13}
 \textsc{A.~Prokofiev},  \textsc{M.~Ikeda},  \textsc{E.~Makalkina},
  \textsc{R.~Svagera},  \textsc{M.~Waas},  and  \textsc{S.~Paschen},
 \jr{J. Electr. Mater.} \textbf{42}, 1628--1633 (2013).


\othercit
\bibitem{Pro08}
 \textsc{A.~Prokofiev},  \textsc{S.~Paschen},  \textsc{H.~Sassik},
  \textsc{S.~Laumann},  and  \textsc{P.~Pongratz},
utility patent AT: 10749 U1 2009-09-05, DE: 20 2008 006 946.7, patent
  applications US: 12/231,183, JP: 135994/2008 (2008).


\bibitem{Cah90.1}
 \textsc{D.\,G. Cahill},
 \jr{{Rev.\ Sci.\ Instum.}} \textbf{61}, 802 (1990).


\bibitem{Cah87.1}
 \textsc{D.\,G. Cahill} and  \textsc{R.\,O. Pohl},
 \jr{{Phys.\ Rev.\ B}} \textbf{35}, 4067 (1987).


\bibitem{Ike15.1}
 \textsc{M.~Ikeda},  \textsc{P.~Tome\v{s}},  \textsc{L.~Prochaska},
  \textsc{J.~Eilertsen},  \textsc{S.~Populoh},  \textsc{S.~L\"offler},
  \textsc{R.~Svagera},  \textsc{M.~Waas},  \textsc{H.~Sassik},
  \textsc{A.~Weidenkaff},  and  \textsc{S.~Paschen},
 \jr{{Z.\ Anorg.\ Allg.\ Chem., in press}} (2015).


\bibitem{Ayd11.1}
 \textsc{U.~Aydemir},  \textsc{C.~Candolfi},  \textsc{A.~Ormeci},
  \textsc{Y.~Oztan},  \textsc{M.~Baitinger},  \textsc{N.~Oeschler},
  \textsc{F.~Steglich},  and  \textsc{Y.~Grin},
 \jr{{Phys.\ Rev.\ B}} \textbf{84}, 195137 (2011).


\bibitem{Zei12.1}
 \textsc{I.~Zeiringer},  \textsc{M.~Chen},  \textsc{A.~Grytsiv},
  \textsc{E.~Bauer},  \textsc{R.~Podloucky},  \textsc{H.~Effenberger},  and
  \textsc{P.~Rogl},
 \jr{{Acta Mater.}} \textbf{60}, {2324} (2012).


\bibitem{Kle60}
 \textsc{K.~Klement},  \textsc{R.\,H. Willens},  and  \textsc{P.~Duwez},
 \jr{Nature} \textbf{187}, 869 (1960).


\bibitem{Gre95.1}
 \textsc{A.\,L. Greer},
 \jr{Science} \textbf{267}, 1947 (1995).


\bibitem{Nol98.1}
 \textsc{G.\,S. Nolas},  \textsc{J.\,L. Cohn},  \textsc{G.\,A. Slack},  and
  \textsc{S.\,B. Schujman},
 \jr{{Appl.\ Phys.\ Lett.}} \textbf{73}, 178 (1998).


\bibitem{Coh99.1}
 \textsc{J.\,L. Cohn},  \textsc{G.\,S. Nolas},  \textsc{V.~Fessatidis},
  \textsc{T.\,H. Metcalf},  and  \textsc{G.\,A. Slack},
 \jr{{Phys.\ Rev.\ Lett.}} \textbf{82}, 779 (1999).


\bibitem{Sal01.1}
 \textsc{B.\,C. Sales},  \textsc{B.\,C. Chakoumakos},  \textsc{R.~Jin},
  \textsc{J.\,R. Thompson},  and  \textsc{D.~Mandrus},
 \jr{{Phys.\ Rev.\ B}} \textbf{63}, 245113 (2001).


\bibitem{Pro14.1}
 \textsc{A.~Prokofiev},  \textsc{X.~Yan},  \textsc{M.~Ikeda},
  \textsc{S.~L\"offler},  and  \textsc{S.~Paschen},
 \jr{{J. Cryst. Growth}} \textbf{401}, 627 (2014).


\bibitem{Sob99}
 \textsc{F.~Sobott},  \textsc{A.~Wattenberg},  \textsc{H.\,D. Barth},  and
  \textsc{B.~Brutschy},
 \jr{Int. J. Mass Spectr.} \textbf{185-187}, 271--279 (1999).


\othercit
\bibitem{Rog05.1}
 \textsc{P.~Rogl},
{Thermoelectrics Handbook},
 ({ed.\ D. M. Rowe, CRC Press}, Boca Raton, 2006), chap. {32 (Formation and
  crystal chemistry of clathrates)}.


\bibitem{Yan13.2}
 \textsc{X.~Yan},  \textsc{E.~Bauer},  \textsc{P.~Rogl},  and
  \textsc{S.~Paschen},
 \jr{{Phys.\ Rev.\ B}} \textbf{87}, 115206 (2013).


\bibitem{Pro13.0}
 \textsc{A.~Prokofiev},  \textsc{M.~Ikeda},  \textsc{E.~Makalkina},
  \textsc{R.~Svagera},  \textsc{M.~Waas},  and  \textsc{S.~Paschen},
 \jr{{J.\ Electron.\ Mater.}} \textbf{42}, 1628 ({2013}).


\bibitem{Kaw01.2}
 \textsc{Y.~Kawaharada},  \textsc{K.~Kurosaki},  \textsc{M.~Uno},  and
  \textsc{S.~Yamanaka},
 \jr{{J.\ Alloys Compd.}} \textbf{315}, 193 (2001).


\bibitem{Koc03.1}
 \textsc{C.\,C. Koch},
 \jr{Rev.\ Adv.\ Mater.\ Sci.} \textbf{5}, 91 (2003).


\othercit
\bibitem{Fec06.1}
 \textsc{H.\,J. Fecht} and  \textsc{Y.~Ivanisenko},
in Nanostructured Materials: Processing, Properties, and Applications ({ed.\
  C.\ C.\ Koch, William Andrew Publishing: 13 Eaton Avenue Norwich, NY 13815},
  2006).


\bibitem{Eck92.1}
 \textsc{J.~Eckert},  \textsc{J.\,C. Holzer},  \textsc{C.\,E. Krill},  and
  \textsc{W.\,L. Johnson},
 \jr{{J.\ Mater.\ Res.}} \textbf{7}, 1752 (1992).


\bibitem{Koc93.1}
 \textsc{C.\,C. Koch},
 \jr{Nano.\ Mater.} \textbf{2}(2), 109 (1993),
cited By (since 1996) 408.


\bibitem{Koc97.1}
 \textsc{C.C.Koch},
 \jr{Nanostruct.\ Mater.} \textbf{9}(1-8), 13 (1997).


\bibitem{Yan15.1}
 \textsc{X.~Yan},  \textsc{E.~Bauer},  \textsc{P.~Rogl},  \textsc{J.~Bernardi},
   and  \textsc{S.~Paschen},
 \jr{{unpublished}} (2015).


\bibitem{Hun09.1}
 \textsc{T.~Hungria},  \textsc{J.~Galy},  and  \textsc{A.~Castro},
 \jr{{Adv.\ Eng.\ Mater.}} \textbf{11}(8), 615 (2009).


\othercit
\bibitem{Gro06.1}
 \textsc{J.\,R. Groza},
in Nanostructured Materials: Processing, Properties, and Applications ({ed.\
  C.\ C.\ Koch, William Andrew Publishing: 13 Eaton Avenue Norwich, NY 13815},
  2006).


\bibitem{Yan14.1}
 \textsc{X.~Yan},  \textsc{E.~Bauer},  \textsc{P.~Rogl},  and
  \textsc{S.~Paschen},
 \jr{{Phys.\ Stat.\ Solidi A}} \textbf{211}, 1282--1287 (2014).


\bibitem{Yan12.1}
 \textsc{X.~Yan},  \textsc{M.\,X. Chen},  \textsc{S.~Laumann},
  \textsc{E.~Bauer},  \textsc{P.~Rogl},  \textsc{R.~Podloucky},  and
  \textsc{S.~Paschen},
 \jr{{Phys.\ Rev.\ B}} \textbf{85}, 165127 (2012).


\bibitem{Yan10.1}
 \textsc{X.~Yan},  \textsc{G.~Giester},  \textsc{E.~Bauer},  \textsc{P.~Rogl},
  and  \textsc{S.~Paschen},
 \jr{{J.\ Electron.\ Mater.}} \textbf{39}, 1634 (2010).


\bibitem{Zol14.1}
 \textsc{A.~Zolriasatein},  \textsc{X.~Yan},  \textsc{P.~Rogl},
  \textsc{A.~Shokuhfar},  and  \textsc{S.~Paschen},
 \jr{{J.\ Nano Res.}} \textbf{29}, 121 (2014).


\bibitem{Zol15.1}
 \textsc{A.~Zolriasatein},  \textsc{X.~Yan},  \textsc{E.~Bauer},
  \textsc{P.~Rogl},  \textsc{A.~Shokuhfar},  and  \textsc{S.~Paschen},
 \jr{{to appear in Mater.\ Design}} (2014).


\bibitem{Mun06.1}
 \textsc{Z.\,A. Munir},  \textsc{U.~{Anselmi-Tamburini}},  and
  \textsc{M.~Ohyanagi},
 \jr{{J.\ Mater.\ Sci.}} \textbf{41}(3), 763 (2006).


\bibitem{Yan15.2}
 \textsc{X.~Yan},  \textsc{S.~Populoh},  \textsc{E.~Bauer},  \textsc{P.~Rogl},
  and  \textsc{S.~Paschen},
 \jr{{to appear in J.\ Electron. Mater.}} (2015).


\bibitem{Shi03.1}
 \textsc{L.~Shi},  \textsc{D.\,Y. Li},  \textsc{C.\,H. Yu},  \textsc{W.\,Y.
  Jang},  \textsc{D.\,Y. Kim},  \textsc{Z.~Yao},  \textsc{P.~Kim},  and
  \textsc{A.~Majumdar},
 \jr{{J.\ Heat Transfer}} \textbf{125}, 881 (2003).


\bibitem{Karg14}
 \textsc{S.\,F. Karg},  \textsc{V.~Troncale},  \textsc{U.~Drechsler},
  \textsc{P.~Mensch},  \textsc{P.\,D. Kanungo},  \textsc{H.~Schmid},
  \textsc{V.~Schmidt},  \textsc{L.~Gignac},  \textsc{H.~Riel},  and
  \textsc{B.~Gotsmann},
 \jr{{Nanotechnology}} \textbf{25}, 305702 (2014).


\bibitem{Wang12}
 \textsc{Z.~Wang},  \textsc{M.~Kroener},  and  \textsc{P.~Woias},
 \jr{{Sensors and Actuators A}} \textbf{188}, 417 (2012).


\othercit
\bibitem{Mad15.1}
 \textsc{G.\,K.\,H. Madsen},  \textsc{A.~Katre},  \textsc{C.~Bera},  and
  \textsc{A.~Togo},
{Phys.\ Status Solidi A, this issue}.


\end{thebibliography}
\providecommand{\WileyBibTextsc}{}
\let\textsc\WileyBibTextsc
\providecommand{\othercit}{}
\providecommand{\jr}[1]{#1}
\providecommand{\etal}{~et~al.}

\end{document}